\documentclass[11pt]{article}
\usepackage[utf8]{inputenc}
\usepackage[left=1in,right=1in]{geometry}
\usepackage{authblk}
\usepackage[pdfstartview=FitH,pdfpagemode=None]{hyperref}
\usepackage{amsmath}
\usepackage{amssymb}
\usepackage{amsfonts}
\usepackage{graphicx}
\usepackage{tabularx}
\usepackage{braket}
\usepackage{comment}
\usepackage{cite}

\title{Krylov complexity and orthogonal polynomials}
\author[1,2]{Wolfgang M\"uck}
\author[3,4,5]{Yi Yang}
\affil[1]{Dipartimento di Fisica ``Ettore Pancini", Universit\`a degli Studi di Napoli ``Federico II" \authorcr Via Cintia, 80126 Napoli, Italy}
\affil[2]{Istituto Nazionale di Fisica Nucleare, Sezione di Napoli \authorcr Via Cintia, 80126 Napoli, Italy}
\affil[3]{Department of Electrophysics, National Yang Ming Chiao Tung University, Hsinchu, ROC}
\affil[4]{Center for Theoretical and Computational Physics, \authorcr National Yang Ming Chiao Tung University, Hsinchu, ROC}
\affil[5]{National Center for Theoretical Physics, ROC}
\date{\today}

\begin{document}


\numberwithin{equation}{section}

\newcommand{\ie}{i.e.,\ }
\newcommand{\eg}{e.g.,\ }

\newcommand{\const}{\operatorname{const.}}

\newcommand{\sgn}{\operatorname{sgn}}

\newcommand{\rmd}{\,\mathrm{d}}

\newcommand{\Tr}{\operatorname{tr}}


\newcommand{\re}{\operatorname{Re}}
\newcommand{\im}{\operatorname{Im}}

\newcommand{\e}[1]{\operatorname{e}^{#1}}


\newcommand{\vev}[1]{\left\langle #1 \right\rangle}

\newcommand{\op}{\mathcal{O}}
\newcommand{\Liou}{\mathcal{L}}
\newcommand{\Hilb}{{\mathcal{H}}}

\newcommand{\Order}{\mathcal{O}}

\newcommand{\unit}[1]{\operatorname{#1}}

\newcommand{\BesselJ}[1]{\operatorname{J}_{#1}}
\newcommand{\BesselI}[1]{\operatorname{I}_{#1}}

\newcommand{\dbar}[1]{\bar{\bar{#1}}}

\newcommand{\Jacobi}[1]{P_{#1}^{(\alpha,\beta)}}
\newcommand{\hypF}[1]{\operatorname{{}_2F_1}\!\left(#1\right)}
\newcommand{\genhypF}[3]{\operatorname{{}_{#1}F_{#2}}\!\left(#3\right)}

\maketitle

\begin{abstract}
Krylov complexity measures operator growth with respect to a basis, which is adapted to the Heisenberg time evolution. The construction of that basis relies on the Lanczos algorithm, also known as the recursion method. The mathematics of Krylov complexity can be described in terms of orthogonal polynomials. We provide a pedagogical introduction to the subject and work out analytically a number of examples involving the classical orthogonal polynomials, polynomials of the Hahn class, and the Tricomi-Carlitz polynomials.

\end{abstract}
\section{Introduction}
\label{intro}

A key concept of (quantum) information theory is \emph{complexity}, which incarnates, in all of its variants, the intuitive understanding that an object is the more \emph{complex} the more ingredients are required to assemble, create, or describe it. It has been recognized in recent years that complexity embraces areas as distant as computer science, chaos, emergent phenomena in many-body systems, and black holes. Many of these efforts within quantum field theory and quantum gravity, as well as an outline of some remaining challenges, have been summarized in the snowmass white paper \cite{Faulkner:2022mlp}. 

In a classic paper, Kolmogorov \cite{Kolmogorov:1965} proposed to define as the ``relative complexity of an object $y$ with a given $x$, [\ldots] the minimal length $l(p)$ of the program $p$ for obtaining $y$ from $x$.'' Kolmogorov also pointed out that such an algorithmic approach transcends the notion of entropy for quantifying information content. The minimum number of operations necessary to implement a given task is now known as \emph{computational complexity} in computer science and information theory \cite{Aaronson:2016vto}. Similarly, quantum complexity is defined as the minimum number of elementary operations (quantum gates) needed to build a state $|\psi\rangle$ from a given reference state $|\psi_0\rangle$. A quantum circuit in operation increases the complexity of an initially simple state scrambling its information over a large number of degrees of freedom. The rate of complexity growth is expected to satisfy a physical bound proportional to the portion of energy that is available for information processing \cite{Lloyd:2000, Brown:2015lvg}. Black holes are considered the fastest scramblers in nature \cite{Hayden:2007cs, Sekino:2008he, Susskind:2014rva} and are deeply connected to quantum chaos \cite{Barbon:2011pn, Shenker:2013pqa}. Computational complexity has also been added to the holographic dictionary \cite{Susskind:2014rva, Brown:2015lvg, Susskind:2018pmk}. 

The same physical mechanism that scrambles information is responsible for the emergence of irreversible macroscopic behaviour such as thermalization or hydrodynamics \cite{Deutsch:1991, Srednicki:1994mfb, Rigol:2008, DAlessio:2015qtq}. Precisely quantifying the growth of operators \cite{vonKeyserlingk:2017dyr, Khemani:2017nda} is important to understand such emergent phenomena. A frequently used definition of operator size builds upon out-of-time-ordered correlators (OTOCs), which measure the amount a (time-dependent) operator fails to commute with other, simple, operators and provide a diagnostic of quantum chaos \cite{Larkin:1969, Maldacena:2015waa, Bhattacharyya:2019txx}. OTOC-based operator size has been discussed in \cite{Roberts:2018mnp, Qi:2018bje} for the Sachdev-Ye-Kitaev model \cite{Sachdev:1992fk, Kitaev:2015, Maldacena:2016hyu}, which has a classical gravity dual and is a benchmark model for a fast scrambler.\footnote{For a pedagogical introduction to SYK, see \cite{Trunin:2020vwy}.} 

Krylov complexity, or K-complexity for short, was introduced in \cite{Parker:2018yvk} as a measure of operator size with respect to a basis that is well adapted to Hamiltonian time evolution. In this context, the authors of \cite{Parker:2018yvk} proposed a Universal Operator Growth Hypothesis for systems with chaos, according to which the spectral density decays exponentially as $\e{-|\omega|/\omega_0}$ for large frequencies $|\omega|$, with a decay constant $\omega_0$ related to the Lyapunov exponent. Within the SYK model and when working in the thermodynamic limit, Krylov complexity is similar to other notions of operator size. However, whereas operator size fails to be a good measure of complexity beyond the thermodynamic limit, \ie at time scales much larger than the scrambling time, Krylov complexity continues to show the typical features of complexity \cite{Barbon:2019wsy, Rabinovici:2020ryf}. In particular, it grows linearly in the post-scrambling period until it saturates because of the ultimately finite dimensionality of the operator space, of order $\e{\op(S)}$.

The main advantage of Krylov complexity over OTOCs and (quantum) computational complexity is that it is free of ambiguities and provides a truly intrinsic measure of operator and complexity growth. For example, quantum complexity crucially depends on the choice of elementary gates (or operations) and a tolerance parameter $\epsilon$, which describes the accuracy with which the target state $|\psi\rangle$ is reached. Similar ambiguities exist in Nielsen's geometric approach to quantifying complexity \cite{Nielsen:2005, Nielsen:2006, Dowling:2006}, which requires a choice of an \emph{information metric} penalizing hard \emph{vs.}\ easy operations. OTOCs depend on a reference operator. In contrast, the only freedom in Krylov complexity is the choice of the inner product in operator space, but this freedom is usually limited by the physical context. Once the inner product has been defined, the construction of Krylov space proceeds by means of the Lanczos algorithm, also known as the recursion method \cite{rec-method}, and Krylov complexity may be understood as the effective dimension of Krylov space. 

The seminal work \cite{Parker:2018yvk} has spurred quite a number of applications, which we would like to mention here. Krylov complexity has been studied for the SYK model, besides in \cite{Parker:2018yvk, Rabinovici:2020ryf}, also in \cite{Jian:2020qpp}. The relation to random matrix theory and black holes has been investigated in \cite{Kar:2021nbm}. The authors of \cite{Kar:2021nbm} suggested a microcanonical inner product, which can be used to define a thermal Krylov complexity. We point out that the complexity renormalization group framework introduced in \cite{Kar:2021nbm} resembles the introduction of a square root terminator function in the recursion method \cite{rec-method}. It has also been found that quantum chaos can be understood by delocalization in Krylov space \cite{Dymarsky:2019elm}. In contrast, Krylov complexity is suppressed in integrating integrable models of finite size \cite{Rabinovici:2021qqt}, which has been called Krylov localization. 
A number of simple systems, which are symmetry generated and allow an exact treatment using generalized coherent states, were considered in \cite{Caputa:2021sib, Patramanis:2021lkx}. Further tests, numerical calculations, applications and generalizations include chaotic Ising chains and systems with many-body localization \cite{Cao:2020zls, Trigueros:2021rwj}, a variety of exemplary
systems, including 1d and 2d Ising models as well as 1d Heisenberg models \cite{Heveling:2022hth}, operator growth in conformal field theory \cite{Dymarsky:2021bjq, Caputa:2021ori}, lattice systems with local interactions under Euclidean time evolution \cite{Avdoshkin:2019trj}, the emergence of bulk Poincar\'e symmetry in systems exhibiting large-$N$ factorization \cite{Magan:2020iac}, topological phases of matter \cite{Caputa:2022eye}, integrable models with saddle-point dominated scrambling \cite{Bhattacharjee:2022vlt}, cosmological Krylov complexity \cite{Adhikari:2022oxr}, a condition for the irreversibility of operator growth \cite{Fan:2022xaa}, complexity in field theory \cite{Adhikari:2022whf}, and a fundamental and universal limit to the growth of the
Krylov complexity \cite{Hornedal:2022pkc}. Furthermore, a notion inspired by Krylov complexity is state complexity as defined in \cite{Balasubramanian:2022tpr}.   

In this paper, we will give a pedagogical introduction to Krylov complexity exploiting as much as possible the relation between the recursion method and the theory of orthogonal polynomials \cite{rec-method, Green:2001}. Our focus will lie on analytical treatments, and we resort to numerics only for producing graphs. Clearly, such an approach has limits. First, the Lanczos algorithm requires a careful numerical treatment because of inherent instabilities. Second, if one starts with typical model Hamiltonians, one must do serious numerical work. Nevertheless, we feel that a purely analytical approach is justified as an introduction and to provide a number of exactly solvable examples, which go beyond the scaling behaviours one may extract from approximations or numerical work. In fact, some of the new results included here present surprises.

The paper is organized as follows. In section~\ref{Krylov}, we shall start with a review of the Lanczos algorithm for the Heisenberg evolution of operators, emphasizing the one-to-one correspondence between the resulting operator basis and a set of orthogonal polynomials. We will continue by reviewing the associated chain of wave functions, which parameterize the operator motion within this operator basis, and their Laplace transforms. It is pointed out that the latter are essentially the functions of the second kind associated with the orthogonal polynomials. The Lanczos algorithm produces polynomials orthogonal with respect to an even measure. This has a number of simplifying consequences, which will be listed. Finally, Krylov complexity and some related measures are introduced. In particular, we will define a generating function of Krylov complexities of arbitrary degree and provide a number of useful expressions that may be helpful for the calculation of Krylov complexity in specific cases. Section~\ref{gen} is dedicated to the discussion of generic features of Krylov complexity. Some of these derive from simple mathematical statements about the spectrum, others can be understood by looking at very simple examples. In particular, we shall rederive the Krylov complexities in the symmetry generated examples of \cite{Caputa:2021sib} without explicit reference to the generalized coherent states. Moreover, we shall treat the wave functions in a continuum approximation, from which the generic behaviour of Krylov complexity in those cases, in which the Lanczos coefficients obey simple scaling laws, can be estimated. 

In sections~\ref{ex:classical}--\ref{Carlitz}, specific examples are worked out in detail. The cases discussed in section~\ref{ex:classical} involve classical orthogonal polynomials, \ie the Chebyshev, Gegenbauer, Hermite, Jacobi and Laguerre polynomials. All of these cases are known in the context of the recursion method \cite{rec-method, rec-method1985}. We add the exact calculation of complexity, whenever possible. 
Examples involving the orthogonal polynomials of the Hahn class, \ie the Charlier, Krawtchouk and Meixner polynomials, are worked out in section~\ref{hahn}. The complexities for these cases cannot be understood from a scaling regime and, somewhat surprisingly, are very similar to each other. Moreover, they appear to resemble the numerical results of systems with many-body localization \cite{Trigueros:2021rwj}. The non-standard Tricomi-Carlitz polynomials are discussed in section~\ref{Carlitz}. These are interesting, because in spite of the fact that the Lanczos coefficients obey a scaling law the continuum approximation is suspected to fail. Unfortunately, we are not able to obtain a useful analytical result, but we leave a numerical investigation for the future. 
Finally, in section~\ref{iprod}, we review and discuss several operator inner products.

\section{Operator growth in Krylov space}
\label{Krylov}

\subsection{Recursion method}
\label{La:rec.meth}

In the Heisenberg picture, an initial observable $\op$ evolves with time into 
\begin{equation}
\label{La:op.t}
	\op(t) = \e{it\Liou} \op = \e{iHt} \op \e{-iHt}~,   	
\end{equation}
where 
\begin{equation}
\label{La:Liou}
	\Liou  = [H,.]
\end{equation}
is the Liouvillian superoperator. Clearly, $\op(t)$ describes a trajectory in the vector space of all hermitean operators and becomes, generically speaking, increasingly more \emph{complex}. In order to quantify this growth of complexity it is necessary in the first place to describe in more appropriate terms the space $\op(t)$ moves in. An intrinsic approach for doing this is to construct an operator basis using the Lanczos algorithm, also called the \emph{recursion method} in physics \cite{rec-method}.\footnote{The Lanczos algorithm \cite{Lanczos:1950zz} was originally conceived to tri-diagonalize a given hermitean matrix. In the present context, the Liouvillian will become tri-diagonal.} 

The recursion method can be summarized as follows.  
A necessary preliminary step is to promote the space of operators to a Hilbert space $\hat{\Hilb}$ by endowing it with an inner product $(A|B)$ satisfying the properties
\begin{equation}
\label{La:iprod.prop}
	\forall\; |A), |B) \in \hat{\Hilb}: \quad (A|A) \geq 0~, \quad (A|B) = (B|A)^\ast~, \quad (A| a B) = a (A|B)\quad (a\in \mathbb{C})~.
\end{equation}
In addition, the Liouvillian must be hermitean with respect to the inner product, 
\begin{equation}
\label{La:Liou.hermit}
	(A|\Liou B) = (\Liou A |B)~.
\end{equation}
For simplicity, we assume $\op$ to be normalized,
\begin{equation}
\label{La:O.normalized}
	(\op|\op)=1~.
\end{equation}

The choice of the inner product is quite important and will influence the outcome of the recursion method, but we do not need to be more specific for the purpose of developing the general method. 
We will postpone the review of a variety of inner products to section~\ref{iprod}. 

Expanding the Heisenberg evolution \eqref{La:op.t} into a formal power series in time shows that the time dependent operator $|\op(t))$ is constructed out of the sequence of states $\{|\Liou^n \op), n=0,1,2,\ldots\}$. The Lanczos algorithm essentially performs a Gram-Schmidt orthogonalization on this sequence, producing an (ordered) orthogonal set of states $\{|\op_n), n=0,1,2, \ldots\}$ called the Krylov basis. The space spanned by the Krylov basis is called the Krylov (sub)space, which is a subspace of $\hat{\Hilb}$. 

In table~\ref{rec.meth.tab}, we spell out two variants of the Lanczos algorithm, which we call the \emph{orthonormal} and the \emph{monic} versions, with reference to the kinds of polynomials they produce. Both versions can be found in the literature and produce, obviously, equivalent results. Let us add a few comments. First, the positive numbers $b_n$ are called the Lanczos coefficients. The relation between $\Delta_n$ and $b_n$ is simply 
\begin{equation}
\label{La:Delta.b}
	\Delta_n = b_n^2~.
\end{equation}
Second, if the iteration stops, then the Krylov subspace is finite-dimensional. 
Third, it is known that the Lanczos algorithm is numerically unstable. Since our focus is on analytical results, we will ignore this issue and refer interested readers to \cite{Rabinovici:2020ryf} and references therein.

\begin{table}[th]
\begin{tabularx}{\textwidth}{|l|X|X|}
\hline
\multicolumn{3}{|c|}{\textbf{Lanczos algorithm (recursion method)}}\\
\hline 
 & orthonormal version & monic version \\
\hline 
start & $|\op_0)=|\op)$, $b_0=0$ & $|\op_0)=|\op)$, $\Delta_0=0$ \\
\hline
$n=1$ & $|A_1) = \Liou |\op_0)$  &$|\op_1) = \Liou |\op_0)$ \\
& $b_1=(A_1|A_1)^{\frac12}$ \hfill stop, if $b_1=0$  & $\Delta_1 = (\op_1|\op_1)$ \hfill stop, if $\Delta_1=0$ \\
& $|\op_1) = b_1^{-1} |A_1)$ &   \\
\hline 
$n>1$ & $|A_{n}) = \Liou |\op_{n-1}) -b_{n-1} |\op_{n-2})$ &
$|\op_n) = \Liou |\op_{n-1}) - \Delta_{n-1} |\op_{n-2})$ \\
&$b_n=(A_n|A_n)^{\frac12}$ \hfill stop, if $b_n=0$  & 
$\Delta_n = \frac{(\op_n|\op_n)}{(\op_{n-1}|\op_{n-1})}$ \hfill stop, if $\Delta_n=0$ \\
&$|\op_n) = b_n^{-1} |A_n)$ &   \\
\hline
\end{tabularx}
\caption{Two versions of the recursion method.\label{rec.meth.tab}}
\end{table}

In either version of the recursion method, one has by construction 
\begin{equation}
\label{La:orthogonal}
	(\op_m|\op_n) = (\op_n|\op_n)\, \delta_{mn}
\end{equation}
with $(\op_n|\op_n)=1$ in the orthonormal version. More importantly, the procedure gives 
\begin{equation}
\label{La:op.n}
	|\op_n) = |P_n(\Liou) \op)~,
\end{equation}
where $P_n(\Liou)$ denotes a certain polynomial of degree $n$ in $\Liou$, with real coefficients. It is evident from the table that these polynomials satisfy a three-term recurrence relation,\footnote{The fact that there is no term with $P_n$ on the right hand sides of \eqref{La:three.term.ortho} and \eqref{La:three.term.monic} derives from the fact that the initial operator $\op$ is assumed to be hermitean. It follows by induction that also $(i \Liou)^n \op$ is hermitean, from which one can show that $(\Liou^n\op|\Liou^m \op)=0$ for $m+n$ odd.}
\begin{align}
\label{La:three.term.ortho}
	\Liou P_n(\Liou) &= b_{n+1} P_{n+1}(\Liou) + b_{n} P_{n-1}(\Liou) && \text{(orthonormal version)}\\
\label{La:three.term.monic}
	\Liou P_n(\Liou) &= P_{n+1}(\Liou) + \Delta_{n} P_{n-1}(\Liou) &&\text{(monic version).}
\end{align}
Because the eigenvalues of $\Liou$ are real (recall that $\Liou$ is required to be hermitean with respect to the inner product), by Favard's theorem \cite{Chihara:1978, Koornwinder:2013fcq} there exists a measure $\mu(\Liou)$ on $\mathbb{R}$ such that the $P_n$ are orthogonal with respect to $\mu(\Liou)$,
\begin{equation}
\label{La:orth.poly}
	\int \rmd \mu(\Liou)\, P_m(\Liou) P_n(\Liou) = h_n \delta_{mn}\qquad (h_n>0).
\end{equation}
The support of $\mu(\Liou)$ is identified with the spectrum of $\Liou$, and integrals such as the one appearing in \eqref{La:orth.poly} are intended as integrals over the eigenvalues of $\Liou$. Consider the (normalized) measure $\mu(\Liou)$ defined by\footnote{Favard's theorem does not guarantee that the measure $\mu(\Liou)$ in \eqref{La:orth.poly} is unique, but if it is, it must coincide with \eqref{La:measure.def}. Moreover, we can assume that the measure is unique in the physically relevant cases. For example, it is unique, if the spectrum of $\Liou$ is bounded.}
\begin{equation}
\label{La:measure.def}
	\int \rmd\mu(\Liou)\, f(\Liou) = (\op|f(\Liou)\op)~.
\end{equation}
It is easy to show using \eqref{La:orthogonal} and \eqref{La:op.n} that this measure satisfies \eqref{La:orth.poly} with 
\begin{equation}
\label{La:norm}
	h_n = (\op_n|\op_n)~.
\end{equation}

\subsection{Wave functions}

Given the Krylov basis, define a semi-infinite chain of wave functions by \footnote{$h_n$ has been included to make the $\phi_n$ independent of the chosen normalization.}
\begin{equation}
\label{La:phi.n}
	\phi_n (t) = \frac{i^n}{\sqrt{h_n}} (\op(t)|\op_n)~,
\end{equation}
which are real, because both $i^n \op_n$ and $\op(t)$ are hermitean operators.\footnote{Indeed, whereas \cite{rec-method} uses \eqref{La:phi.n}, the complex conjugate definition is often found in the literature (e.g., in \cite{Parker:2018yvk, Kar:2021nbm, Caputa:2021sib, Caputa:2021ori}) and gives, obviously, the same result.}
These wave functions have the equivalent forms 
\begin{equation}
\label{La:phi.n.equiv}
	\phi_n (t) = \frac{i^n}{\sqrt{h_n}} (\op|\e{-it\Liou}P_n(\Liou)\op) 
		= i^n \int \rmd \mu(\Liou)\, \e{-it\Liou} \frac{P_n(\Liou)}{\sqrt{h_n}}~.
\end{equation}
By virtue of \eqref{La:three.term.ortho}, they satisfy the Schr\"odinger equation  
\begin{equation}
\label{La:phi.t}
	\partial_t \phi_n = -b_{n+1} \phi_{n+1} + b_n \phi_{n-1}~,\qquad \phi_n(0) = \delta_{n0}~.
\end{equation}
In other words, the wave functions $\phi_n$ constitute a semi-infinite tight-binding model, for which the Lanczos coefficients represent the hopping amplitudes \cite{Dymarsky:2019elm}. 
Moreover, they satisfy
\begin{equation}
\label{La:phi.norm}
	\sum\limits_n \phi_n(t)^2 = 1~.
\end{equation}
The easiest way to see this is to note that \eqref{La:phi.norm} holds by definition for $t=0$ and that the derivative with respect to time  
vanishes by virtue of \eqref{La:phi.t}.

In contrast to the construction of the Krylov basis, \eqref{La:phi.t} cannot be solved by iteration. A formal solution can be given as follows \cite{rec-method}. Consider the Laplace transform of $\phi_n(t)$,
\begin{equation}
\label{La:phi.Laplace}
	c_n(z) = \int\limits_0^\infty \rmd t\, \phi_n(t) \e{-zt} \qquad (\Re z >0).
\end{equation}
In terms of these, \eqref{La:phi.t} translates to
\begin{equation}
\label{La:c.z.eq}
	z c_n(z) = b_n c_{n-1}(z) - b_{n+1} c_{n+1}(z) +\delta_{n0}~. 
\end{equation}
After introducing 
\begin{equation}
\label{La:R.def}
	R_n(z)=b_n \frac{c_n(z)}{c_{n-1}(z)}\qquad  (n>0)~,
\end{equation}
\eqref{La:c.z.eq} can be rewritten as
\begin{equation}
\label{La:c.z.eq.1}
	c_0(z) = \frac1{z+R_1(z)}~, \qquad  R_n(z) = \frac{\Delta_n}{z+ R_{n+1}(z)} \quad (n>0),
\end{equation}
where \eqref{La:Delta.b} has been used. $R_1(z)$ is also called the memory function of the system \cite{rec-method}. In turn, \eqref{La:c.z.eq.1} implies the continuous fraction representation of $c_0(z)$
\begin{equation}
\label{La:c0.sol}
	c_0(z) = \frac1{z+\frac{\Delta_1}{z+\frac{\Delta_2}{z+\cdots}}}~.
\end{equation}

The functions $c_n(z)$ are closely related to the \emph{functions of the second kind} \cite{vanAsche:1990, Grinshpun:2013}
\begin{equation}
\label{La:Q.def}
	Q_n(z) = \int \rmd\mu(\Liou)\, \frac{P_n(\Liou)}{z-\Liou}~, \qquad z\in \mathbb{C} \backslash\text{supp}(\mu(\Liou))~.
\end{equation}
More precisely,
\begin{equation}
\label{La:c.Q}
	c_n(z) = \frac{i^{n+1}}{\sqrt{h_n}} Q_n(iz)~,
\end{equation}
which also provides the analytic continuation of \eqref{La:phi.Laplace} to $\Re z<0$.    
In particular, 
\begin{equation}
\label{La:resolvent}
	Q_0(z)= -ic_0(-iz) = (\op|\frac1{z-\Liou} \op)
\end{equation}
is called the \emph{resolvent}. It contains the spectral function
\begin{equation}
\label{La:spec.dens}
	\frac{\rmd \mu(\omega)}{\rmd \omega} = -\frac1\pi \lim\limits_{\epsilon\to 0^+} \Im \left[Q_0(\omega+i\epsilon)\right]
	= \frac1\pi \lim\limits_{\epsilon\to 0^+} \Re \left[c_0(\epsilon -i\omega)\right]~.
\end{equation}
The functions of the second kind satisfy, except for $n=0$, the same three-term recurrence relation as the polynomials $P_n$, which is easily proven from \eqref{La:Q.def} and \eqref{La:three.term.monic}. In the case of monic $P_n$, one has
\begin{equation}
\label{La:Q.rec}
	z Q_n(z) = Q_{n+1}(z) + \Delta_n Q_{n-1}(z) + \delta_{n0}~.
\end{equation}

We conclude this subsection by remarking that the entire information about the operator $\op(t)$ is contained equivalently in any of the following: 
\begin{itemize}
	\item the set of Lanczos coefficients ($b_n$ or $\Delta_n$),
	\item the fluctuation function $\phi_0(t)$ or its Laplace transform $c_0(z)$, which is equivalent to the function of the second kind $Q_0(z)$,
	\item the spectral density, which is related to $Q_0(z)$ by \eqref{La:spec.dens},
	\item the moments of the spectral density, which coincide with the coefficients of the expansion of $\phi_0(t)$ in a series in $t^2$. 
	There are recursive algorithms translating the set of moments into the set of Lanczos coefficients and \emph{vice versa}. These are useful for numerical calculations with known spectral functions, but we shall not need them here. Interested readers are referred to \cite{rec-method}.
\end{itemize}

\subsection{Consequences of the even spectrum}
\label{La:spectrum}

In this subsection, we will list some simple, but important, properties, which derive from the fact that the recurrence relations \eqref{La:three.term.ortho} and \eqref{La:three.term.monic} do not contain a term with $P_n$ on their right hand sides. This implies 
that the measure $\mu(\Liou)$ is even, $\mu(\Liou)=\mu(-\Liou)$. To be specific, let us work with the monic polynomials.  

A general property of orthogonal polynomials with respect to an even measure is that $P_{2n}(\Liou)$ and $P_{2n+1}(\Liou)$ can be expressed in terms of orthogonal polynomials in $\Liou^2$ with appropriate measures. Explicitly, we have\footnote{There is a subtlety in \eqref{La:p.even}, if the measure $\rmd \mu(\Liou)$ contains a delta function at $\Liou=0$, $\rmd \mu(\Liou)= a \delta(\Liou) \rmd \Liou+\cdots$. In this case, one formally gets a contribution $\rmd \bar{\mu}(\Liou^2) = 2a\delta(\Liou^2)\rmd\Liou^2 + \cdots$. In order to avoid double counting, one can adopt the convention \[ \int\limits_0^u \rmd x\,\delta(x) f(x) = \frac12 f(0)\qquad (u>0).\]\label{La:footnote.delta}} 
\begin{align}
\label{La:p.even}
	P_{2n}(\Liou) &= \bar{P}_n(\Liou^2)~,& \rmd \bar{\mu}(\Liou^2) &= 2 \rmd \mu(\Liou)~,\\ 
\label{La:p.odd}
	P_{2n+1}(\Liou) &= \Liou \dbar{P}_n(\Liou^2)~,& \rmd \dbar{\mu}(\Liou^2) &= \frac{2}{\Delta_1} \rmd \mu(\Liou)\Liou^2~. 
\end{align}
Moreover, with 
\begin{align}
\label{La:orth.poly.even}
	\int \rmd \bar{\mu}(x)\, \bar{P}_m(x) \bar{P}_n(x) &= \bar{h}_n \delta_{mn}~,\\
\label{La:orth.poly.odd}
	\int \rmd \dbar{\mu}(x)\, \dbar{P}_m(x) \dbar{P}_n(x) &= \dbar{h}_n \delta_{mn}~,
\end{align}
we also have 
\begin{equation}
\label{La:h.even.odd}
	h_{2n} = \bar{h}_n~,\qquad h_{2n+1} = \Delta_1 \dbar{h}_n~. 
\end{equation}
We remark that the measures $\bar{\mu}$ and $\dbar{\mu}$ are normalized, \ie $\bar{h}_0=\dbar{h}_0=1$. 

Both, $\bar{P}_n$ and $\dbar{P}_n$, satisfy three-term recurrence relations, which can be obtained from \eqref{La:three.term.monic}. Multiplying \eqref{La:three.term.monic} by $\Liou$ and using it once more one finds
\begin{align}
\label{La:rec.even}
	\Liou^2  \bar{P}_n &= \bar{P}_{n+1} +\bar{b}_n \bar{P}_n + \bar{c}_n \bar{P}_{n-1}~, &
	\bar{b}_n &= \Delta_{2n} + \Delta_{2n+1}~, & \bar{c}_n &=  \Delta_{2n} \Delta_{2n-1}~,\\
\label{La:rec.odd}
	\Liou^2 \dbar{P}_{n} &= \dbar{P}_{n+1} + \dbar{b}_n \dbar{P}_n + \dbar{c}_n \dbar{P}_{n-1}~,&
	\dbar{b}_n &= \Delta_{2n+1} + \Delta_{2n+2}~,& \dbar{c}_n &=  \Delta_{2n} \Delta_{2n+1}~.
\end{align}
Moreover, they are interrelated by 
\begin{equation}
\label{La:even.fom.odd}
	\bar{P}_n = \dbar{P}_n + \Delta_{2n} \dbar{P}_{n-1}~.
\end{equation}

The functions of the second kind \eqref{La:Q.def} are 
\begin{align}
\label{La:Q.even}
	Q_{2n}(z) &= z \bar{Q}_n(z^2)~,\\
\label{La:Q.odd}
	Q_{2n+1}(z) &= \Delta_1 \dbar{Q}_n(z^2)~,
\end{align}
where the definition of $\bar{Q}_n$ and $\dbar{Q}_n$ is analogous to \eqref{La:Q.def}. They are interrelated by 
\begin{equation}
\label{La:Q.odd.from.even}
	\Delta_1 \dbar{Q}_n(z^2) = \bar{Q}_{n+1} (z^2) + \Delta_{2n+1} \bar{Q}_n(z^2)~.
\end{equation}

Finally, consider the wave functions \eqref{La:phi.n.equiv}. With \eqref{La:p.even} and \eqref{La:p.odd}, they become
\begin{align}
\label{La:phi.even}
	\phi_{2n}(t) &= (-1)^n\bar{h}_{n}^{-\frac12} \int\rmd\bar{\mu}(x) \cos(t\sqrt{x}) \bar{P}_n(x)~,\\
\label{La:phi.odd}
	\phi_{2n+1}(t) &= (-1)^n\Delta_1^{\frac12}\dbar{h}_{n}^{-\frac12}   \int\rmd\dbar{\mu}(x) \frac{\sin(t\sqrt{x})}{\sqrt{x}} \dbar{P}_n(x)~,
\end{align}
where \eqref{La:h.even.odd} has been used.

\subsection{Measures of operator growth}
\label{La:measures}

Krylov- or K-complexity has been defined in \cite{Parker:2018yvk} as the time-dependent mean position on the semi-infinite chain of wave functions,
\begin{equation}
\label{La:K.complexity}
	K_\op(t) = \vev{n}_t = \sum\limits_n n \phi_n(t)^2~.
\end{equation}
This definition can be generalized to an entire class of measures of complexity growth. We define Krylov complexity of degree $k$ as
\begin{equation}
\label{La:Kk.complexity}
	K^{(k)}_\op(t) = \sum\limits_n n^k \phi_n(t)^2~,
\end{equation}
provided the sum is convergent. Related quantities are the Krylov variance \cite{Caputa:2021ori}
\begin{equation}
\label{La:K.variance}
	\Delta K_\op = \frac{K^{(2)}_\op}{[K_\op]^2} -1
\end{equation}
and the Krylov entropy \cite{Barbon:2019wsy, Rabinovici:2020ryf}, 
\begin{equation}
\label{La:K.entropy}
	S_\op = -\sum\limits_n \phi_n(t)^2 \ln \phi_n(t)^2~.
\end{equation}

The recurrence relation of the wave functions \eqref{La:phi.t} allows to obtain 
\begin{equation}
\label{La:K.diff.eq}
	\partial_t^2 K_\op(t) = 2 \sum\limits_{n=0}^\infty (\Delta_{n+1}-\Delta_n)\phi_n(t)^2~,
\end{equation}
which, taking into account the boundary conditions $K_\op(0)=\partial_t K_\op(0) = 0$, may be used to solve for $K_\op(t)$.

It is useful to define the generating function
\begin{equation}
\label{La:K.complexity.gen.func}
	K_\op(\lambda;t) = \sum\limits_n \e{n\lambda} \phi_n(t)^2~,
\end{equation}
from which the Krylov complexities of various degrees can be obtained by
\begin{equation}
\label{La:K.complexity.from.gen.func}
	K^{(k)}_\op(t) = \left. \frac{\partial^k}{\partial \lambda^k} K_\op(\lambda;t)\right|_{\lambda=0}~.
\end{equation}
$\lambda$ is formal parameter, which one can take negative in order to ensure that the sum in \eqref{La:K.complexity.gen.func} is convergent.   

Instead of $K_\op(\lambda;t)$, it is often more useful to consider its Laplace transform,
\begin{equation}
\label{La:K.gen.func.laplace1}
	K_\op(\lambda;z) = \int\limits_0^\infty \rmd t \e{-zt} K_\op(\lambda;t) = \sum\limits_{n=0}^\infty \e{n\lambda} \int\limits_0^\infty \rmd t \e{-zt} \phi_n(t)^2~.
\end{equation}
With the help of \eqref{La:phi.n.equiv}, this can be expressed as
\begin{equation}
\label{La:K.gen.func.laplace2}
	K_\op(\lambda;z) = \sum\limits_{n=0}^\infty \frac{(-\e{\lambda})^n}{h_n} \int\rmd \mu(\Liou) \int \rmd \mu(\Liou') \frac{P_n(\Liou)P_n(\Liou')}{z+i(\Liou+\Liou')}~,
\end{equation}
which, after using \eqref{La:Q.def}, becomes
\begin{equation}
\label{La:K.gen.func.laplace3}
	K_\op(\lambda;z) = \sum\limits_{n=0}^\infty \frac{i(-\e{\lambda})^n}{h_n} \int\rmd \mu(\Liou) P_n(\Liou) Q_n(iz-\Liou)~.
\end{equation}

From \eqref{La:K.complexity.from.gen.func} and \eqref{La:K.gen.func.laplace3} one gets the following formal expression for the Laplace transform of the complexity,
\begin{equation}
\label{La:K.laplace.formal}
	K_\op(z) = \sum\limits_{n=0}^\infty \frac{i(-1)^nn}{h_n} \int\rmd \mu(\Liou) P_n(\Liou) Q_n(iz-\Liou)~,
\end{equation}
but this is not very useful in most cases. Instead, we rewrite \eqref{La:K.gen.func.laplace3} with the help of the recurrence relations \eqref{La:three.term.monic} and \eqref{La:Q.rec} as follows,
\begin{align}
\notag
	K_\op(\lambda;z) &= \frac1{z} \sum\limits_{n=0}^\infty \frac{(-\e{\lambda})^n}{h_n} \int\rmd \mu(\Liou) P_n(\Liou) Q_n(iz-\Liou) (iz-\Liou+\Liou)\\
\notag 
	&= \frac1{z} \sum\limits_{n=0}^\infty \frac{(-\e{\lambda})^n}{h_n} \int\rmd \mu(\Liou) \left\{ P_n(\Liou) \left[ Q_{n+1}(iz-\Liou) +\Delta_n Q_{n-1}(iz-\Liou)+\delta_{n0} \right]\right.\\
\notag 
	&\quad \left.+\left[ P_{n+1}(\Liou)+\Delta_n P_{n-1}(\Liou)\right] Q_n(iz-\Liou)\right\}\\
\label{La:K.gen.func.laplace5}
	&=	
	\frac1{z} +\frac{1-\e{\lambda}}z \sum\limits_{n=0}^\infty \frac{(-\e{\lambda})^n}{h_n} \int\rmd \mu(\Liou) \left[ P_n(\Liou) Q_{n+1}(iz-\Liou) +P_{n+1} Q_n(iz-\Liou)\right]~.
\end{align}
The term independent of $\lambda$ is just the Laplace transform of unity, \ie the norm \eqref{La:phi.norm}. The term linear in $\lambda$ gives the Laplace transform of the complexity,
\begin{equation}
\label{La:K.laplace1}
	K_\op(z) = \frac1z \sum\limits_{n=0}^\infty \frac{(-1)^{n+1}}{h_n} \int\rmd \mu(\Liou) \left[ P_n(\Liou) Q_{n+1}(iz-\Liou) +P_{n+1} Q_n(iz-\Liou)\right]~.
\end{equation}
Applying to \eqref{La:K.laplace1} the same trick as above, one gets  
\begin{equation}
\label{La:K.laplace2}
	K_\op(z) = \frac{2i}{z^2} \sum\limits_{n=0}^\infty \frac{(-1)^{n}}{h_n} \int\rmd \mu(\Liou) P_n(\Liou) Q_{n}(iz-\Liou) (\Delta_{n+1}-\Delta_n)~.
\end{equation}
This is, of course, nothing but the Laplace transform of \eqref{La:K.diff.eq}.

\section{Generic features of operator growth}
\label{gen}

In this section, we shall discuss a number of generic features, which follow directly from general properties of orthogonal polynomials or can be derived easily from the results of the previous section.

\subsection{Measures with finite, bounded and unbounded support}
\label{gen:spectra}

We must distinguish three types of systems according to whether the support of the measure $\mu(\Liou)$ is finite, bounded, or unbounded. 

\paragraph{Finite spectrum}
If the Lanczos algorithm terminates at some $N$, \ie $\Delta_N=0$, then the support of $\mu(\Liou)$ consists of $N$ discrete points $\omega_k$, $k=0,1,\ldots,N-1$, which are simply the zeros of the polynomial $P_N(\Liou)$. It is a trivial statement that the Krylov complexity cannot exceed $N$. Depending on the system, it can oscillate back and forth between $0$ and $N$, approach a limiting value, or continue to fluctuate around some mean value. We shall meet all of these possibilities in the examples.

In the case of infinite spectra, which are associated with a semi-infinite chain, the late-time complexity depends crucially on the asymptotic behaviour of the recursion coefficients $\Delta_n$ for large $n$. Several mathematical facts are known, some of which we will state here. 

\paragraph{Bounded spectrum}
Assume that the Liouvillian $\Liou$ has a bounded spectrum, $\text{supp}(\mu(\Liou))\subseteq [-2E,2E]$. Instead of $\mu(\Liou)$, it is more useful to consider $\bar{\mu}(\Liou^2)$, which is the measure associated with the even polynomials $\bar{P}_n(\Liou^2)=P_{2n}(\Liou)$, as discussed in subsection~\ref{La:spectrum}. Clearly, $\text{supp}(\bar{\mu}(\Liou^2))\subseteq [0,4E^2]$. The same discussion could be applied to $\dbar{\mu}(\Liou^2)$, which is associated with the odd polynomials.

Let $I=[\xi_1,\eta_1]$ be the \emph{true domain of orthogonality} of the polynomials $\bar{P}_n$, which is defined as the smallest closed interval containing all zeroes $y_{n,i}$ of all $\bar{P}_n$ \cite{Chihara:1978, Koornwinder:2013fcq}. It is also the smallest closed interval containing the support of $\bar{\mu}(\Liou^2)$, \ie $\text{supp}(\bar{\mu}(\Liou^2))\subseteq I \subseteq [0,4E^2]$. 
We recall from \eqref{La:rec.even} that the polynomials $\bar{P}_n$ satisfy a three-term recurrence relation with coefficients 
\begin{equation}
\label{gen:bc.recall}
	\bar{b}_n= \Delta_{2n}+\Delta_{2n+1}~, \qquad \bar{c}_n= \Delta_{2n}\Delta_{2n-1}~.
\end{equation} 
Then, the following mathematical statements hold \cite{Chihara:1978, Koornwinder:2013fcq}. 
\begin{itemize}
\item Because $I$ is bounded, the sets of both coefficients, $\bar{b}_n$ and $\bar{c}_n$, are bounded. Therefore, also the set of $\Delta_n$ is bounded. 
\item If, for $n\to \infty$, $\bar{b}_n\to \bar{b}$ and $\bar{c}_n\to \bar{c}$, then $\text{supp}(\bar{\mu})$ is bounded with at most countably many points outside of the interval $J=[\bar{b}-2\sqrt{\bar{c}},\bar{b}+2\sqrt{\bar{c}}]$, and $\bar{b}\pm 2\sqrt{\bar{c}}$ are limit points of $\text{supp}(\bar{\mu})$. Moreover, because $\bar{b} -2\sqrt{\bar{c}}\geq \xi_1\geq0$ and $\bar{b} +2\sqrt{\bar{c}}\leq \eta_1\leq4E^2$, we must have $\bar{b}\geq 2\sqrt{\bar{c}}$ and $\sqrt{\bar{c}}\leq E^2$.
\end{itemize} 

Consider the simplest case in which $\Delta_n\to \Delta$ in the large $n$ limit. It follows from \eqref{gen:bc.recall} that $\bar{b}_n\to 2\Delta$ and $\bar{c}_n \to \Delta^2$, so that $J=[0,4\Delta]$ with $\Delta \leq E^2$. Moreover, $0$ and $4\Delta$ are limit points of $\text{supp}(\bar{\mu})$. 
Hence, because $0$ is a limit point, there cannot be a ``gap'', if by ``gap'' we intend the absence of states with arbitrarily small positive energy.  

A single ``gap'' occurs, if the Lanczos coefficients for even and odd $n$ approach different limiting values, \ie $\Delta_{2n}\to \Delta_e$ and $\Delta_{2n+1} \to \Delta_o$. In this case, $J=[ (\sqrt{\Delta_e} - \sqrt{\Delta_o})^2, (\sqrt{\Delta_e} + \sqrt{\Delta_o})^2]$, with the limitation $\sqrt{\Delta_e} + \sqrt{\Delta_o}\leq 2E$. We remark, however, that there remains the possibility of having countably many points  outside of $J$. A typical example would be a delta function at $\Liou=0$ in addition to the continuous spectrum. We shall encounter such cases in the examples in section~\ref{ex:classical}.

More generally, if the support of $\Liou^2$ contains several disconnected intervals (``bands'') within $[0,4E^2]$, then with increasing $n$ the coefficients $\Delta_n$ will continue to oscillate in a predictable fashion \cite{rec-method1985}.

\paragraph{Unbounded spectrum}
An unbounded set of coefficients $\bar{c}_n$ is sufficient to have a spectrum with unbounded support \cite{Chihara:1978, Koornwinder:2013fcq}. Then, because of \eqref{gen:bc.recall}, also $\Delta_n$ and $\bar{b}_n$ grow without bounds.
We may characterize the average growth using a power law \cite{rec-method},
\begin{equation}
\label{gen:delta.growth}
	\Delta_n \sim n^\lambda~,
\end{equation}
where $0<\lambda \leq2$. $\lambda=2$ is associated with chaos, whereas systems with $0<\lambda<2$ are integrable\cite{Parker:2018yvk}.
In physical applications, $\Delta_n$ may oscillate considerably around the mean behaviour. The lower limit $\lambda=0$ corresponds to a bounded spectrum, while the upper limit $\lambda=2$ arises from physical considerations on the spectral function \cite{rec-method, Parker:2018yvk}, which decays as 
\begin{equation}
\label{gen:spec.decay}
	\frac{\rmd\mu(\omega)}{\rmd\omega} \sim \exp\left(-\left|\frac{\omega}{\omega_0}\right|^{\frac2{\lambda}}\right)
\end{equation}  
for large frequencies. With \eqref{gen:spec.decay} one can associate an exponential growth of $\phi_0$ for imaginary times \cite{rec-method}, 
\begin{equation}
\label{gen:phi0.growth}
	\phi_0(i\tau) \sim \exp(\tau^\rho)~,\qquad \rho =\frac2{2-\lambda}~.
\end{equation}

Mathematically, everything that can go wrong in the general formalism goes wrong for $\lambda\geq2$. The fluctuation function $\phi_0(t)$ is not an entire function (of complex time), the continuous fraction representation of $c_0(z)$ \eqref{La:c0.sol} does not converge, and the measure cannot be uniquely determined from the sequence of $\Delta_n$. Therefore, with the exception of the limiting case $\lambda=2$, we shall exclude these cases from our considerations.

We shall explicitly see in the next subsections and in the examples of section~\ref{ex:classical} that the special cases $\lambda=0$, $\lambda=1$ and $\lambda=2$ are associated with linear, quadratic and exponential late-time growth of complexity, respectively.

\subsection{Simple solutions for the complexity}
\label{gen:simple}
There are only a few cases, in which one can provide an explicit and exact formula for the complexity $K_\op(t)$. In this subsection, we shall present three exceptionally simple cases, which coincide with the three cases related to generalized coherent states discussed in \cite{Caputa:2021sib}. Our new derivations of $K_\op(t)$ are based on the Laplace transform $K_\op(z)$ introduced in subsection~\ref{La:measures}. 

Before discussing these examples, let us briefly recall how the structure of the poles of $K_\op(z)$ in the complex plane encodes the generic late-time behaviour of $K_\op(t)$. It is sufficient to consider the Laplace transform of the simple function $f(t) = t^\nu \e{\gamma t}$, with $\nu>-1$ and complex $\gamma$. Obviously, $\gamma>0$ means exponential growth, $\gamma<0$ exponential decay, and imaginary $\gamma$ implements an oscillating behaviour. The Laplace transform of $f(t)$ is 
\begin{equation}
\label{gen:f.Laplace} 
	f(t) = t^\nu \e{\gamma t}\quad \Rightarrow \quad F(z) = \frac{\Gamma(\nu+1)}{(z-\gamma)^{\nu+1}}~.
\end{equation}    
Therefore, the location of a pole of $K_\op(z)$ encodes exponential or oscillating behaviour, and for the latter the poles must come in complex conjugate pairs. Moreover, the degree of that pole is related to a power law correction, or a pure power law, if the pole is at $z=0$. 

The simplest example is the purely linear sequence $\Delta_n = n\Delta$. Let us use \eqref{La:K.laplace2} to calculate $K_\op(z)$, which gives
\begin{equation}
\label{gen:ex1}
	K_\op(z) = \frac{2i\Delta}{z^2} \sum\limits_{n=0}^\infty \frac{(-1)^{n}}{h_n} \int\rmd \mu(\Liou) P_n(\Liou) Q_{n}(iz-\Liou)~.
\end{equation}
The sum in \eqref{gen:ex1} is the same as in \eqref{La:K.gen.func.laplace3} with $\lambda=0$, so that the result can be read off from  \eqref{La:K.gen.func.laplace5}. Thus, one finds
\begin{equation}
\label{gen:K.linear}
	\Delta_n = n\Delta\quad \Rightarrow \quad K_\op(z)= \frac{2\Delta}{z^3}\quad \Rightarrow \quad K_\op(t) = \Delta t^2~.
\end{equation}
This is the Heisenberg algebra case of \cite{Caputa:2021sib}. We shall meet it again in subsection \eqref{ex:hermite} in connection with the Hermite polynomials. 

The next case is the limiting case of a quadratic plus linear growth, $\Delta_n= \alpha^2 n^2 +\beta n$. Now, \eqref{La:K.laplace2} becomes
\begin{equation}
\notag
	K_\op(z) = \frac{2 i}{z^2} \sum\limits_{n=0}^\infty \frac{(-1)^{n}}{h_n} \int\rmd \mu(\Liou) P_n(\Liou) Q_{n}(iz-\Liou)
	(2\alpha^2n +\alpha^2+\beta)~.
\end{equation}
The sum involving the term $(\alpha^2+\beta)$ can be done as in the example above, while the sum with $\alpha^2n$ becomes again $K_\op(z)$ by virtue of \eqref{La:K.laplace.formal}. Hence, one obtains
\begin{equation}
\notag
	K_\op(z) = \frac{4\alpha^2}{z^2} K_\op(z) + \frac{2(\alpha^2+\beta)}{z^3}~,
\end{equation} 
which is solved by
\begin{equation}
\notag
	K_\op(z) = \frac{2(\alpha^2+\beta)}{z(z^2-4\alpha^2)} = \frac{\alpha^2+\beta}{4\alpha^2} \left( \frac1{z-2\alpha}+\frac1{z+2\alpha}-\frac2z \right)~. 
\end{equation}
Performing the inverse Laplace transform yields
\begin{equation}
\label{gen:K.quadratic}
	\Delta_n= \alpha^2 n^2 +\beta n \quad \Rightarrow \quad 	K_\op(t) = \left(1+\frac{\beta}{\alpha^2}\right) \sinh^2 (\alpha t)~.
\end{equation}
Clearly, with $\beta=\alpha^2 (2h-1)$, the result of the $SU(1,1)$ case of \cite{Caputa:2021sib} is reproduced. We remark that the orthogonal polynomials for this case are Meixner-Pollaczek polynomials with an even measure \cite{NIST:DLMF,Koornwinder:1988},
\begin{equation}
\label{gen:Hahn.P}
	P_n(\Liou) = n! \alpha^n \,P_n^{(h)}\left(x;\frac{\pi}2\right)~, 
\end{equation}
where $\Liou = 2\alpha x$. These are orthogonal in $x\in (-\infty,\infty)$ with respect to the (normalized) measure
\begin{equation}
\label{gen:Hahn.measure}
	\rmd \mu(x) = \frac{2^{2h-1}}{\pi\Gamma(2h)}
	\left|\Gamma(h+ix)\right|^2 \rmd x~. 
\end{equation}
For the case $h=\frac12$ see \cite{Bender:1986ht,Bender:1987}.

Finally, we may consider $\Delta_n= -\alpha^2 n^2 +\beta n$, which is consistent only if the spectrum is discrete, \ie $\Delta_N=0$ for some $N\geq1$, which implies $\beta=\alpha^2 N$. The calculation of $K_\op(z)$ goes as
in the previous case after replacing $\alpha\to i\alpha$. Hence, 
\begin{equation}
\label{gen:K.quadratic2}
	\Delta_n= \alpha^2 (-n^2 + N n) \quad \Rightarrow \quad 	K_\op(t) = (N-1) \sin^2 (\alpha t)~.
\end{equation}
Here, we recognize the $SU(2)$ case of \cite{Caputa:2021sib} with $2j=N-1$. The polynomials associated with this simple case are (shifted) Krawtchouk polynomials \cite{NIST:DLMF},
\begin{equation}
\label{gen:Kraw.P}
	P_n(\Liou) = (-\alpha)^n (2j+1-n)_n\,K_n\left(x+\frac{j}2;\frac12,j\right)\qquad (n=0,1,\ldots,j)
\end{equation}
with $x=\Liou/(2\alpha)\in \{-j/2,-j/2+1,\ldots,j/2\}$.

\subsection{Continuum limit}
\label{cont}

If the Lanczos coefficients $b_n$ (or $\Delta_n$) can be approximated by sufficiently simple functions of $n$, then it is possible to estimate the late-time behaviour of the complexity using a continuum limit. In this subsection, we shall illustrate this approach for some simple cases following \cite{Barbon:2019wsy}. 

To start, let us introduce 
\begin{equation}
\label{cont:x.def}
	x= \epsilon n~, \qquad b(x) = b_n~, \qquad \phi(x,t) = \phi_n(t)
\end{equation}
with $\epsilon\ll 1$, where we assume that both, $b(x)$ and $\phi(x,t)$ are sufficiently smooth functions of $x$. It is clear that this assumption does not hold for the wave function $\phi(x,t)$ at $t=0$, but it is sufficient for our purposes that it holds for some $t>t_0$. For the rest of this subsection, we will shift $t-t_0\to t$.  
With \eqref{cont:x.def}, the discrete Schr\"odinger equation \eqref{La:phi.t} can be approximated by the first-order partial differential equation
\begin{equation}
\label{cont:phi.eq}
	\partial_t \phi + 2\epsilon b \partial_x\phi = -\epsilon (\partial_x b) \phi~,
\end{equation}
where we omitted the $\Order(\epsilon^2)$ terms. Equation \eqref{cont:phi.eq} can be solved by means of the method of characteristics. Using $s=t$ to parameterize the characteristic curves, \eqref{cont:phi.eq} translates into the system
\begin{equation}
\label{cont:char}
	\frac{\rmd x}{\rmd t} = 2 \epsilon b~, \quad \frac{\rmd \phi}{\rmd t} = -\epsilon (\partial_x b) \phi~.
\end{equation}
Formally, the general solution can be expressed as 
\begin{equation}
\label{cont:gen.sol}
	\phi(x,t) = \frac1{\sqrt{b}} \psi(\tau(x,t))~,
\end{equation} 
where $\tau(x,t)$ is an integration constant arising in the first equation of \eqref{cont:char}, \eg
\begin{equation}
\label{cont:tau}
	\int\limits^{x_\tau(t)}_0 \frac{\rmd x}{2\epsilon b} = t +\tau~.
\end{equation}

To take some simple specific examples, consider the scaling behaviour
\begin{equation}
\label{cont:b.scaling}
	b_n = \alpha n^{\frac{\lambda}2} \quad \Rightarrow \quad b(x) = \alpha \left(\frac{x}{\epsilon}\right)^{\frac{\lambda}2}~.
\end{equation} 
In the special case $\lambda=2$, which represents a fast scrambler, the two equations in \eqref{cont:char} are solved by $x=x_0\e{2\alpha t}$ and $\phi = \psi\e{-\alpha t}$, respectively, where $\psi$ is now an arbitrary function of the integration constant $x_0$. The constant $\alpha$ is related to the Lyapunov exponent. Thus, we can write down the solution 
\begin{equation}
\label{cont:phi.lam.2}
	\phi(x,t) = \e{-\alpha t} \psi\left( \frac{x}{\epsilon}\e{-2\alpha t} \right)~.
\end{equation} 
Now, $\phi(x,t)$ is subject to the norm constraint 
\begin{equation}
\label{cont:norm}
	1 = \sum\limits_n \phi_n^2 \approx \frac1{\epsilon} \int\limits_0^\infty \rmd x\, \phi^2(x,t)~,
\end{equation}
\ie 
\begin{equation}
\label{cont:norm.lam.2}
	\frac1{\epsilon} \int\limits_0^\infty \rmd x\, \e{-2\alpha t} \psi^2\left( \frac{x}{\epsilon}\e{-2\alpha t} \right)
	= \int\limits_0^\infty \rmd y\, \psi^2(y) \approx 1~.
\end{equation} 
Therefore, the complexity is, approximately,
\begin{align}
\label{cont:K.def}
	K_\op(t) &\approx \frac1{\epsilon^2} \int\limits_0^\infty \rmd x\, x \phi^2(x,t) \\
\notag
	&\approx \frac1{\epsilon^2} \int\limits_0^\infty \rmd x\, x \e{-2\alpha t} \psi^2\left( \frac{x}{\epsilon}\e{-2\alpha t} \right)\\
\label{cont:K.lam.2}
	&\approx \e{2\alpha t} \int\limits_0^\infty \rmd y\, y \psi^2(y) = \e{2\alpha t} K_0~.
\end{align} 
This agrees with the exponential growth at late times in \eqref{gen:K.quadratic}.

For $\lambda<2$, the solution of the first equation in \eqref{cont:char} is\footnote{Eqn.~\eqref{cont:x.sol} tells us that, if $\lambda>2$, there can be only solutions with a limited time domain. This is related to the discussion at the end of subsection \eqref{gen:spectra}.}  
\begin{equation}
\label{cont:x.sol}
	\frac1{(2-\lambda)\alpha}\left(\frac{x}{\epsilon} \right)^{1-\frac{\lambda}2} = t + \tau~,
\end{equation}
where $\tau$ is a non-negative constant. Thus, the general solution \eqref{cont:gen.sol} reads 
\begin{equation}
\label{cont:phi.lam.gen}
	\phi(x,t) = \left(\frac{x}{\epsilon}\right)^{-\frac{\lambda}4} \psi(\tau)~.
\end{equation}
Let us assume that, for $t=0$, the support of $\phi$ is restricted to $x\in [0,X)$ with $X=\epsilon[(2-\lambda)\alpha T]^{2/(2-\lambda)}$, which implies $\psi(\tau)=0$ for $\tau \geq T>0$. Therefore, the norm constraint \eqref{cont:norm} becomes, after a change of variable,
\begin{equation}
\label{cont:norm.lam.gen}
	2\alpha \int\limits_{-t}^T \rmd \tau\, \psi^2(\tau) \approx 1~.
\end{equation} 
Because the norm must be time-independent, one has the additional constraint $\psi(\tau)=0$ for $\tau<0$.
Taking this into account, the complexity \eqref{cont:K.def} is easily found to be 
\begin{equation}
\label{cont:K.lam.gen}
	K_\op(t) \approx 2\alpha \int\limits_0^T \rmd \tau \left[(2-\lambda)\alpha (t+\tau) \right]^{\frac2{2-\lambda}} \psi^2(\tau)~.
\end{equation} 
For large enough $t$, this behaves as 
\begin{equation}
\label{cont:K.lam.gen.t.large}
	K_\op(t) \approx \left[ (2-\lambda) \alpha t \right]^{\frac2{2-\lambda}}~,
\end{equation} 
where \eqref{cont:norm.lam.gen} has been used. 

Let us briefly comment on these results. One can observe from \eqref{cont:x.sol} that the length of the interval on which $\phi(x,t)$ has support grows with time, if $0<\lambda <2$, whereas it shrinks for $\lambda<0$. More precisely, if $\phi(x,0)$ has support in $x\in [0,X)$ then, at late times, the size of the domain of support at time $t$ is   
$$
	\Delta x(t) \approx \frac{2X}{2-\lambda} \left(\frac{t}{T} \right)^\frac{\lambda}{2-\lambda}
$$
to leading order in $T/t$. This does not shrink for $\lambda\geq 0$, which implies that we can trust the continuum approximation. As a nice check, for $\lambda=1$, \eqref{cont:K.lam.gen.t.large} agrees precisely with the exact result \eqref{gen:K.linear}. In the limiting case $\lambda=0$, $\phi(x,t)$ describes a wave moving with constant velocity $2\epsilon\alpha$, which implies $K_\op(t) \approx 2 \alpha t$. As we shall see in the subsections \ref{ex:Chebyshev.U}, \ref{ex:Chebyshev.T}, and \ref{ex:Gegenbauer}, this reproduces only the qualitative linear growth of the complexity with time, whereas the velocity is not generic. The slope $2\alpha$ appears to be an upper limit. The situation is different for $\lambda<0$. Although the solution of the differential equation remains formally correct, the continuum approximation must fail at a certain point, and we should not trust the result \eqref{cont:K.lam.gen.t.large}.

The arguments used in these simple examples can be extended to the general solution \eqref{cont:gen.sol}--\eqref{cont:tau}. With $\psi(\tau)$ having support in $\tau\in[0,T)$, one has the norm constraint
\begin{equation}
\label{cont:norm.gen}
	2 \int\limits_0^T\rmd \tau\, \psi^2(\tau) \approx 1
\end{equation} 
and the complexity  
\begin{equation}
\label{cont:K.gen}
	K_\op(t) \approx \frac{2}{\epsilon} \int\limits_0^T\rmd \tau\, x_\tau(t) \psi^2(\tau) = \frac{2}{\epsilon} \int\limits_0^T\rmd \tau\, x_0(t+\tau) \psi^2(\tau)
	\approx \frac1\epsilon x_0(t+\tau_0)= \frac1\epsilon x_{\tau_0}(t)~.
\end{equation} 
The last approximation holds for $t\gg T$, and $\tau_0 \in(0,T)$. Remember, however, that the time variable has been shifted in the continuum approximation with respect to the exact initial value problem. This suggests that $\tau_0$ simply describes this shift. Extending \eqref{cont:K.gen} to model the complexity also at early times, we set $\tau_0=0$, such that $x_{0}(0)=0$ from \eqref{cont:tau}.

\begin{figure}[th]
\begin{center}
\includegraphics[width=0.7\textwidth]{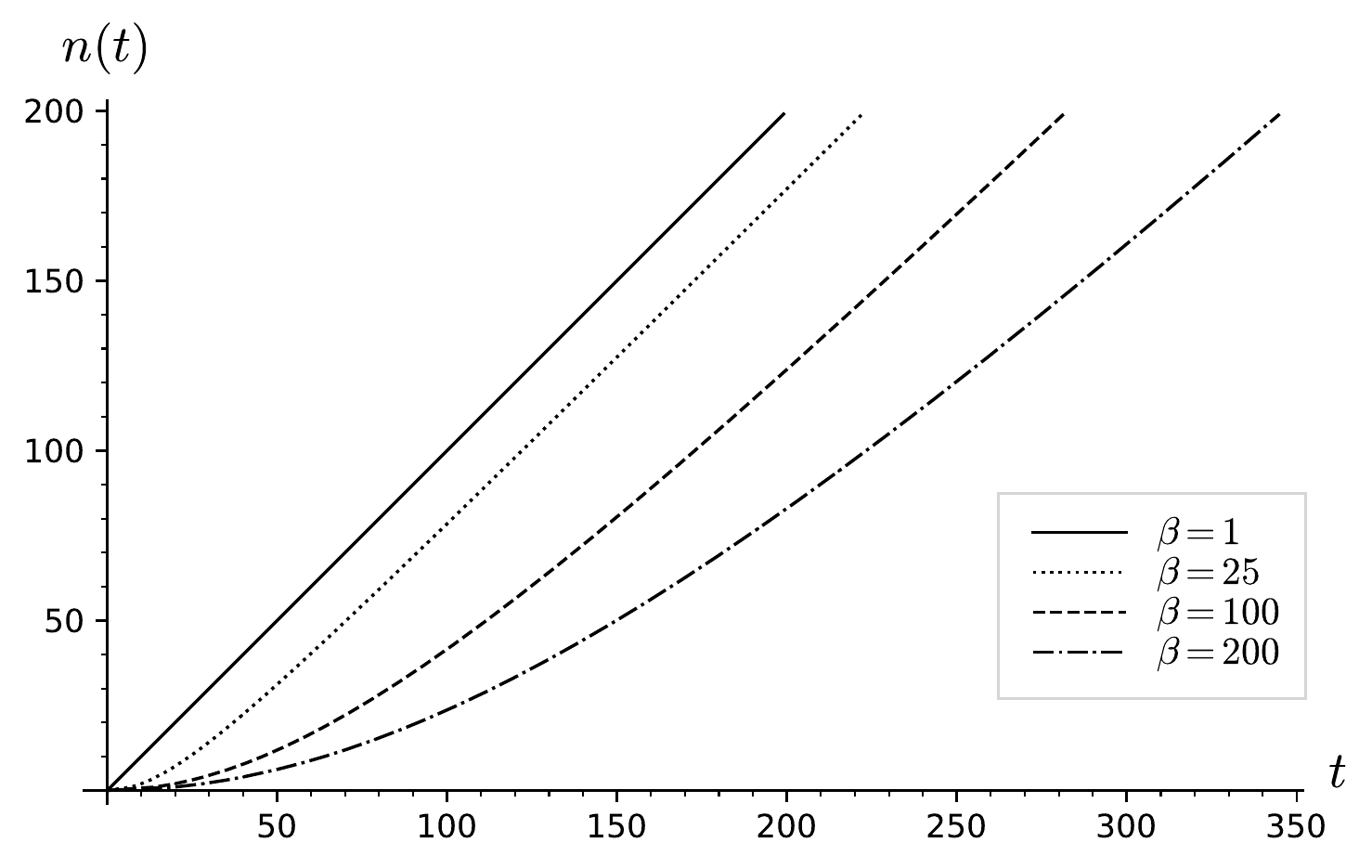}
\caption{The characteristic curves corresponding to the Lanczos coefficients \eqref{cont:Gegenbauer.b}. \label{cont:Gegenbauer.fig}} 
\end{center}
\end{figure}

Let us illustrate this idea with two examples, which will be treated in more detail in later sections. First, consider
\begin{equation}
\label{cont:Gegenbauer.b}
	b_{n}= \sqrt{\frac{n\left(n+2\beta-1\right)}{4\left(n+\beta\right)\left(n+\beta-1\right)}}~,
\end{equation}
which gives rise to the Gegenbauer polynomials, see subsection~\ref{ex:Gegenbauer}. In this case, \eqref{cont:tau} yields an implicit solution of $n(t)=\frac1\epsilon x(t)$, 
\begin{equation}
\label{cont:geg.implicit}
 	t= \int\limits_0^{n(t)}\frac{\rmd n}{2 b}= \int\limits_0^{n(t)} \rmd n\, \sqrt{1+\frac{\beta(\beta-1)}{n(n+2\beta-1)}}~.
\end{equation}
Clearly, we must restrict to $\beta\geq 1$, as otherwise the integrand would be imaginary for small $x$.  
For large $n$, the $b_n$ satisfy the scaling relation \eqref{cont:b.scaling} with $\alpha=\frac12$. In agreement with the previous discussion, \eqref{cont:geg.implicit} shows that $n(t)\approx t$ for $t$ large, but one would expect that this overestimates the velocity of complexity growth. Indeed, this will be confirmed in subsection~\ref{ex:Gegenbauer}. The integral in \eqref{cont:geg.implicit} can be solved in terms of elliptic integrals, but numerical integration is sufficient for our purposes. Some cases of $\beta$ are shown in figure~\ref{cont:Gegenbauer.fig}.

\begin{figure}[th]
\begin{center}
\includegraphics[width=0.7\textwidth]{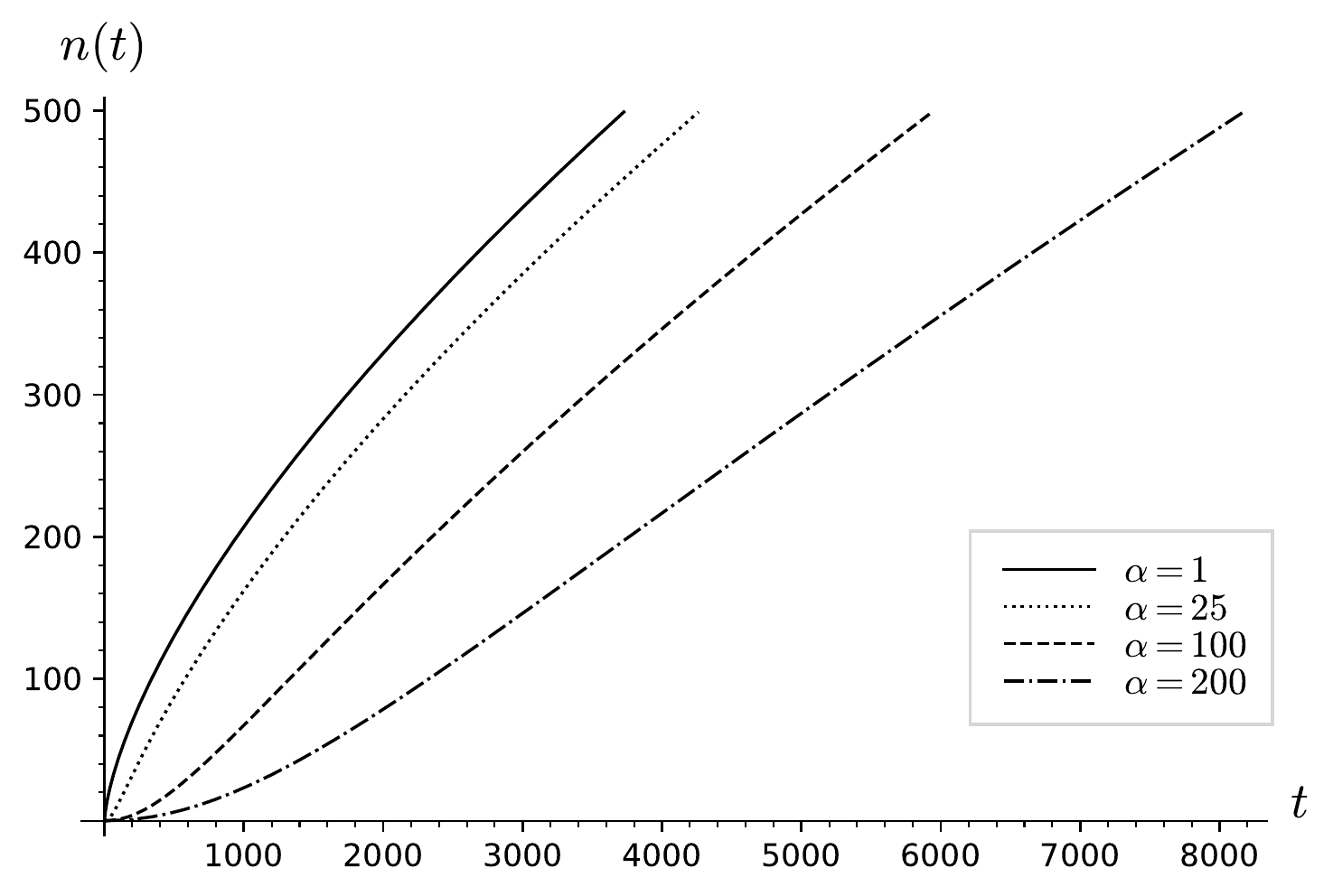}
\caption{The characteristic curves corresponding to the Lanczos coefficients \eqref{cont:Tricomi.b}. \label{cont:Tricomi.fig}} 
\end{center}
\end{figure}

Another example is the sequence of Lanczos coefficients
\begin{equation}
\label{cont:Tricomi.b}
	b_{n}=\sqrt{\frac{n}{(n+\alpha)(n+\alpha-1)}}~,
\end{equation}
which gives rise to the Tricomi-Carlitz polynomails, see section~\ref{Carlitz}. In this case, the characteristic curves $n(t)$ have the implicit form  
\begin{equation}
\label{cont:Tri.implicit}
 	t= \int\limits_0^{n(t)}\frac{\rmd n}{2 b}= \int\limits_0^{n(t)} \rmd n\, \sqrt{\frac{(n+\alpha)(n+\alpha-1)}{4n}}~,
\end{equation}
where $\alpha$ must be restricted to $\alpha\geq1$. As before, \eqref{cont:Tri.implicit} can be expressed in terms of elliptic integrals, but numerical integration suffices for our purposes. Some cases are shown in figure~\ref{cont:Tricomi.fig}. We remark that the characteristics in this case behave like $n(t)\sim t^{\frac23}$ for large $t$, but we should not trust the continuum approximation, as discussed above. 

As one last example, let us consider the case in which the Lanczos coefficients $b_n$ approach two different constants, depending on whether $n$ is even or odd. This case is clearly not included in the simple scaling behaviour discussed above, but it can be analyzed analytically, as we shall see in subsection~\ref{ex:gap}. Consider the Schr\"odinger equation \eqref{La:phi.t} with
\begin{equation}
\label{cont:two.b}
	b_{2n} = b_e \quad (n>0)~, \qquad  b_{2n+1}= b_o~.
\end{equation}
From \eqref{La:phi.t}, one easily finds the second-order differential equation
\begin{equation}
\label{cont:two:pde}
	-\partial_t^2 \phi_n + b_eb_o  \left( \phi_{n+2} -2 \phi_n + \phi_{n-2} \right) - (b_e-b_o)^2 \phi_n =0~,
\end{equation}
which holds for both, even and odd, $n$ ($n>1$). The continuum approximation of \eqref{cont:two:pde} is the massive Klein-Gordon equation
\begin{equation}
\label{cont:two:massive.KG}
	\left[ -\partial_t^2 + 4 b_eb_o \partial_n^2 \phi - (b_e-b_o)^2 \right] \phi=0~.
\end{equation}
Without further calculation, we can conclude that the ``pulse'' $\phi_n(t)$ propagates like a massive particle, with constant velocity smaller than the ``speed of light''. Hence, we obtain an approximate upper bound for the late-time growth,
\begin{equation}
\label{cont:two:K}
	K_\op(t) \lesssim 2 \sqrt{b_eb_o}\, t~.
\end{equation}

\section{Examples with classical orthogonal polynomials}
\label{ex:classical}

In this section, we will illustrate the recursion method and the calculation of Krylov complexity using systems described by the classical orthogonal polynomials. Following a pedagogical spirit, we present the examples in order of increasing complexity. All of the examples we present here have appeared elsewhere, see \cite{rec-method,rec-method1985,Kar:2021nbm} and references therein, to which we only add the calculation of the Krylov complexity. 
Analytic examples are also important for the construction of \emph{terminator functions}, which provide suitable approximations of the continued fraction representation of the function $c_0(z)$ \cite{rec-method}. Such terminator functions have a physical interpretation as describing the complex inaccessible environment, into which the operator dynamics dissolves in the late-time limit \cite{Kar:2021nbm}. 
The position $N_c$, at which the terminator function is put sets a renormalization scale. One may then study the renormalization group flow followed by the complexity as a function of $N_c$.

\subsection{Chebyshev polynomials of the second kind}
\label{ex:Chebyshev.U}

The simplest case is given by the Chebyshev polynomials of the second kind, which are associated with constant $\Delta_n=\Delta$. They are a special case of the Gegenbauer polynomials, which shall be discussed in subsection~\ref{ex:Gegenbauer}. Here, we will present two derivations of the Krylov complexity. 

Our starting point is a set of constant Lanczos coefficients. For simplicity, we use 
\begin{equation}
\label{ex:cheU.delta}
	\Delta_n=\frac14 \qquad (n>0)~,
\end{equation}
which one can achieve by re-scaling $\Liou$ to a dimensionless variable $x$. Here and in the following examples, we will use $x$ in place of $\Liou$ in order to remind ourselves of this fact.

The recurrence relation \eqref{La:three.term.monic} now reads
\begin{equation}
\label{ex:cheU.rec}
	 x  P_n(x)  = P_{n+1}(x) +\frac14 P_{n-1}(x)~,
\end{equation}
which is solved by the (monic) Chebyshev Polynomials of the second kind \cite{NIST:DLMF}
\begin{equation}
\label{ex:cheU.U}
	P_n(x) = 2^{-n} U_n(x)~.
\end{equation}
They are orthogonal with respect to the measure
\begin{equation}
\label{ex:cheU.measure}
	\rmd \mu(x) = \frac2{\pi} \sqrt{1- x ^2}\rmd x~, \qquad x \in (-1,1)
\end{equation}
and have the norms $h_n = 4^{-n}$. 

The wave functions \eqref{La:phi.n} are 
\begin{equation}
\label{ex:cheU.phi.n}
	\phi_n(\tau) = \frac{2(n+1)}{\tau}\BesselJ{n+1}(\tau)~.
\end{equation}
The first way to calculate the complexity is by solving the differential equation \eqref{La:K.diff.eq}. The sum on the right hand side of that equation simplifies to the single summand with $n=0$, so that the problem to be solved is
\begin{equation}
\label{ex:cheU.d2K}
	\partial_\tau^2 K_\op(\tau) = \frac2{\tau^2}\BesselJ{1}(\tau)^2~,\qquad K_\op(0)=0~,\quad \partial_\tau K_\op(0)=0~.
\end{equation}
The solution is 
\begin{equation}
\label{ex:cheU.Ksol}
	K_\op(\tau) = \frac13\left[(4\tau^2+3)\BesselJ{0}(\tau)^2 + (4\tau^2+1)\BesselJ{1}(\tau)^2 -4\tau\BesselJ{0}(\tau)\BesselJ{1}(\tau) -3 \right]~. 
\end{equation}
This has the late-time asymptotic behaviour
\begin{equation}
\label{ex:cheU.K.t}
	K_\op(\tau) = \frac{8}{3\pi} \tau - 1 +\cdots~.
\end{equation}

Another way to calculate the complexity is via its Laplace transform \eqref{La:K.laplace2}. For this, we need the functions of the second kind, $Q_n(z)$, or equivalently $c_n(z)$. 
These are easily obtained from the recursion \eqref{La:c.z.eq.1} by noting that one can set $R_n(z)=R(z)$ by virtue of \eqref{ex:cheU.delta}. The recursion leads to the quadratic equation
\begin{equation}
\label{ex:cheU.Req}
	R(z)^2 +zR(z) -\frac14 =0~,
\end{equation}
which is solved by\footnote{We only write the physical solution.}
\begin{equation}
\label{ex:cheU.R}
	R(z) = \frac12\left( \sqrt{z^2+1} -z \right)~.
\end{equation}
Then, \eqref{La:c.z.eq.1} and \eqref{La:R.def} give 
\begin{equation}
\label{ex:cheU.c}
	c_n(z) = 2 [2R(z)]^{n+1}~.
\end{equation}
It is easy to check the measure \eqref{ex:cheU.measure} from $c_0(z)$. The functions of the second kind are obtained from \eqref{La:c.Q}, 
\begin{equation}
\label{ex:cheU.Q}
	Q_n(iz) = 4 [-iR(z)]^{n+1}~.
\end{equation}

Consider now the Laplace transform of the complexity \eqref{La:K.laplace2}. As before, only the $n=0$ term contributes to the sum, so that
\begin{equation}
\label{ex:cheU.K1}
	K_\op(z) = \frac{2i}{z^2} \int \rmd \mu(x) \frac14 Q_0(iz-x) = \frac{2}{z^2} \int \rmd \mu(x) R(z+ix)~.
\end{equation}
After substituting the measure \eqref{ex:cheU.measure} and using the fact that it is even, this becomes 
\begin{align}
\notag
	K_\op(z) &= 
	\frac2{\pi z^2} \int\limits_{-1}^1 \rmd  x  \sqrt{1- x ^2} \left[R(z+i x ) + R(z-i x ) \right]\\
\label{ex:cheU.K3}
 	&= \frac2{\pi z^2} \int\limits_{-1}^1 \rmd  x  \sqrt{1- x ^2} \left[ -z + \frac12 \sqrt{1- x ^2 +z^2 +2iz x } + \frac12 \sqrt{1- x ^2 +z^2 -2iz x } \right]~.
\end{align}
We note that \eqref{ex:cheU.K3} should be the Laplace transform of \eqref{ex:cheU.Ksol}, but this is not obvious to us. 
For small $z$, \eqref{ex:cheU.K3} reduces to\footnote{The $\op(z^2)$ terms give rise to divergent integrals, which is a hint that the expansion should continue with $z^2\ln z$, but we shall not pursue this issue.} 
\begin{align}
\notag
	K_\op(z) &= \frac2{\pi z^2} \int\limits_{-1}^1 \rmd  x  \left[ (1- x ^2) -z \sqrt{1- x ^2} +\op(z^2) \right]\\
\label{ex:cheU.K}
 	&= \frac{8}{3\pi z^2} -\frac1z +\cdots~.
\end{align}
This agrees with the late-time behaviour \eqref{ex:cheU.K.t}.

We end this subsection by noting that the function $R(z)$ \eqref{ex:cheU.R} is nothing but the square root terminator that is often used for sequences of $\Delta_n$ that approach a constant $\Delta$ for large $n$. This terminator function was advocated in \cite{Kar:2021nbm} as the universal terminator function for chaotic systems with a parametrically long Lanczos plateau. With this terminator function implemented at level $k$, the operator ``environment'', \ie the basis states $|\op_n)$ with $n>k$, become irrelevant for the Krylov complexity. This follows directly from \eqref{La:K.laplace2}, in which the sum terminates at $n=k$, because the square root terminator corresponds to constant $\Delta_n$. In the example discussed in this subsection, this simplification was maximal.

\subsection{Chebyshev polynomials of the first kind}
\label{ex:Chebyshev.T}

The second simplest case are the Chebyshev polynomials of the first kind. They correspond to an parameter exceptional value, which is excluded from the Gegenbauer polynomials that will be discussed in the next subsection. The Chebyshev polynomials of the first kind are orthogonal on $x\in (-1,1)$ with the (normalized) weight 
\begin{equation}
\label{ex:che.w}
	w(x) = \frac{1}{\pi} \left(1-x^2\right)^{-\frac12}~.
\end{equation}
The monic polynomials are 
\begin{equation}
\label{ex:che:monic}
	P_0(x) =1~,\qquad P_n(x) = 2^{1-n} T_n(x) \quad (n>0)~,
\end{equation}
and we have\footnote{For the Chebyshev polynomials of the second kind (Gegenbauer polynomials with $\beta=1$) one has $\Delta_n=\frac14 \,\forall n$.}
\begin{equation}
\label{ex:che.hn} 
	h_0 = 1~, \quad h_n = 2^{1-2n}\quad (n>0)~,\qquad 
	\Delta_1 = \frac12~,\quad \Delta_n =\frac14\quad (n>1)~.
\end{equation} 
The wave functions \eqref{La:phi.n} are
\begin{equation}
\label{ex:cheT.phi}
	\phi_0(\tau) = \BesselJ{0}(\tau)~,\qquad \varphi_n(\tau) = \sqrt{2} \BesselJ{n}(\tau) \quad (n>0)~.
\end{equation}
The unitarity relation \eqref{La:phi.norm} is simply represented by the known sum formula \cite{Gradshteyn, NIST:DLMF}
\begin{equation}
\label{ex:cheT.phi.complete}
	\BesselJ{0}^2(\tau) + 2 \sum\limits_{n=1}^\infty \BesselJ{n}^2(\tau) = 1~.
\end{equation} 

To calculate the Krylov complexity we solve again the differential equation \eqref{La:K.diff.eq}, which in this case reads
\begin{equation}
\label{ex:che.d2K}
	\partial_\tau^2 K_\op(\tau) = \BesselJ{0}(\tau)^2 -\BesselJ{1}(\tau)^2~, \qquad K_\op(0)=0~,\quad \partial_\tau K_\op(0)=0~.
\end{equation} 
The solution is 
\begin{equation}
\label{ex:che.K}
	K_\op(\tau) = \tau^2[\BesselJ{0}(\tau)^2 +\BesselJ{1}(\tau)^2 ] -\tau \BesselJ{0}(\tau)\BesselJ{1}(\tau)~,
\end{equation} 
which has the asymptotic behaviour $K_\op(\tau)\approx \frac2{\pi}\tau$. 
We remark that \eqref{ex:che.K} establishes the following sum
\begin{equation}
\label{ex:che.new.sum}
	\sum\limits_{n=1}^\infty n \BesselJ{n}^2(\tau) = \frac12 \tau^2[\BesselJ{0}(\tau)^2 +\BesselJ{1}(\tau)^2 ] -\frac12 \tau \BesselJ{0}(\tau)\BesselJ{1}(\tau)~.
\end{equation}

It is also possible to calculate $K^{(2)}$ using Neumann's addition theorem \cite{NIST:DLMF},
\begin{equation}
\label{ex:che.K2}
	K_\op^{(2)} = 2 \sum\limits_{n=1}^\infty n^2\BesselJ{n}^2(\tau) 
	= \frac12 \tau^2\sum\limits_{n=1}^\infty \left[\BesselJ{n-1}(\tau)+ \BesselJ{n+1}(\tau)\right]^2 = \frac12 \tau^2~.
\end{equation} 

For completeness, we consider also the Laplace transform of the complexity \eqref{La:K.laplace2}. As above, two terms contribute to the sum in  \eqref{La:K.laplace2}, because $\Delta_1-\Delta_0=\frac12$ and  $\Delta_2-\Delta_1=-\frac14$. Therefore, we need the functions $Q_0(z)$ and $Q_1(z)$. These are obtained after solving the recursion \eqref{La:c.z.eq.1}, which is quite similar to the case of the previous subsection. Indeed, one simply has $R_n(z)=R(z)$ for $n\geq2$ and $R_1(z)=2R(z)$ with $R(z)$ given again by \eqref{ex:cheU.R}. Hence, 
\begin{equation}
\label{ex:che.cn}
	c_0(z) = \frac1{z+2R(z)}~,\qquad c_n(z) = \frac2{\sqrt{h_n}}R(z)^n c_0(z)~,
\end{equation}
so that
\begin{equation}
\label{ex:che.Q}
	Q_0(iz) = \frac{-i}{z+2R(z)}~,\qquad Q_1(iz) = \frac{-2R(z)}{z+2R(z)}~.
\end{equation}

With this information, \eqref{La:K.laplace2} becomes
\begin{align}
\notag
	K_\op(z) &= \frac{i}{z^2} \int\rmd \mu(x)\left[ Q_0(iz-x) + x Q_1(iz-x) \right] \\
\notag
	&= \frac{1}{z^2}  \int\rmd \mu(x) \frac{1-2ixR(z+ix)}{z+2R(z+ix)}~.
\end{align}
This should be the Laplace transform of \eqref{ex:che.K}, which is again not obvious to us.
To leading order in $z$, the integral is 
$$\int \rmd \mu(x) \frac{1-2ixR(z+ix)}{z+2R(z+ix)} = \frac1{\pi} \int\limits_{-1}^1 \frac{\rmd x}{\sqrt{1-x^2}}       \frac{1-ix(\sqrt{1-x^2}-ix)}{\sqrt{1-x^2}-ix} +\op(z)= \frac2\pi+\op(z)~. $$
This value translates into the late-time complexity $K_\op(\tau) \approx\frac2{\pi}\tau$ in agreement with the result obtained above.

\subsection{Gegenbauer (or ultraspherical) polynomials}
\label{ex:Gegenbauer}

The Gegenbauer polynomials are orthogonal on $(-1,1)$ with respect to the even weight function 
\begin{equation}
\label{ex:geg.w}
	w(x) = \frac{\Gamma(\beta+1)}{\sqrt{\pi} \Gamma(\beta+\frac12)} \left(1-x^2\right)^{\beta-\frac12}~, \qquad \left(\beta>-\frac12, \beta\neq0\right)
\end{equation}
which we have normalized for convenience. The cases $\beta=0$ and $\beta=1$ correspond to the Chebyshev polynomials of the first and second kinds, respectively, and have been discussed in the preceding subsections. Other special cases are the Legendre polynomials ($\beta=\frac12$). 

The monic Gegenbauer polynomials are 
\begin{equation}
\label{ex:geg.monic}
	P_n(x) = \frac{n!}{2^n (\beta)_n} C_n^{(\beta)}(x)~,
\end{equation}
where $(a)_n =\Gamma(a+n)/\Gamma(a)$ denotes the Pochhammer symbol. For these, we have \cite{NIST:DLMF}
\begin{equation}
\label{ex:geg.hn} 
	h_n = \frac{n! (2\beta)_n}{4^n (\beta)_n (\beta+1)_n}~,\qquad 
	\Delta_n = \frac{n (n+2\beta-1)}{4(n+\beta)(n+\beta-1)}~.
\end{equation} 

The wave functions \eqref{La:phi.n} can be easily obtained from the Gegenbauer formula \cite[10.23.9]{NIST:DLMF}
\begin{equation}
\label{ex:geg.gegform}
	\e{iz x} = \frac{\Gamma(\beta)}{(\frac{z}2)^{\beta}} \sum\limits_{n=0}^\infty (n+\beta) i^n \BesselJ{n+\beta}(z) C_n^{(\beta)}(x)~,
\end{equation}
where $\BesselJ{\nu}$ denotes a Bessel function. One finds
\begin{equation}
\label{ex:geg.phi}
	\phi_n(\tau) = \left[ \frac{(2\beta)_n(\beta+1)_n}{n! (\beta)_n}\right]^\frac12 \Gamma(\beta+1) 
		\left(\frac{2}{\tau}\right)^\beta \BesselJ{n+\beta}(\tau)~.
\end{equation}
With the help of sum formulas \cite{NIST:DLMF, Gradshteyn}, it can be checked that \eqref{La:phi.norm} holds. 

To calculate the Krylov complexity, we need the following sum \cite{Bailey1930},
\begin{equation}
\label{ex:geg.Bailey}
	\sum\limits_{n=0}^\infty
	\frac{(2\nu)_n(\nu+1)_n(a)_n(b)_n}{n!(\nu)_n(1+2\nu-a)_n(1+2\nu-b)_n} \frac{\BesselJ{n+\nu}(z)^2}{z^{2\nu}} 
	= \frac{2^{-2\nu}}{[\Gamma(\nu+1)]^2}\operatorname{{}_2F_3}\left(\substack{\nu+\frac12,1+2\nu-a-b\\ \nu+1,1+2\nu-a,1+2\nu-b}; -z^2 \right)~,
\end{equation}
where $\operatorname{{}_2F_3}$ denotes a generalized hypergeometric series. For $a=2\nu+1-b$, \eqref{ex:geg.Bailey} can be used to check the unitarity relation \eqref{La:phi.norm}. The complexity is
\begin{equation}
\label{ex:geg.K1}
	K_\op(\tau) = 2^{2\beta}[\Gamma(\beta+1)]^2 \sum\limits_{n=0}^\infty\frac{n(2\beta)_n(\beta+1)_n}{n!(\beta)_n}
	\frac{\BesselJ{n+\beta}(\tau)^2}{\tau^{2\beta}}~. 
\end{equation}
Writing 
$$ n = \beta \left[\frac{(\beta+1)_n}{(\beta)_n}-1 \right] $$
and using \eqref{ex:geg.Bailey} with $\nu=\beta$, $a=\beta+1$, $b=\beta+\frac12$ gives\footnote{The function $\operatorname{{}_2F_3}$ has reduced to $\operatorname{{}_1F_2}$ because $b=\beta+\frac12$.}
\begin{equation}
\label{ex:geg.K2}
	K_\op(\tau) = \beta \left[ \operatorname{{}_1F_2}\left( -\frac12;\beta,\beta+1;-\tau^2\right) -1 \right]~.
\end{equation}
One can check with the help of \cite{Wolfram1} that \eqref{ex:geg.K2} agrees with \eqref{ex:che.K} for $\beta=1$. Similarly, one can check \eqref{ex:cheU.Ksol} for $\beta=0$, but first one needs to apply the contiguity relation \cite{Wolfram2}.

It is interesting to note that the limiting case $\beta=-\frac12$ gives 
$$\lim\limits_{\beta\to-\frac12}K_\op(\tau) = \sin^2 \tau~,$$
which is \eqref{gen:K.quadratic2} for $N=2$. This agrees with the fact that the system collapses to the trivial finite system with $N=2$, because $\Delta_2=0$ for $\beta=-\frac12$.  
 
\begin{figure}[t]
	\includegraphics[width=0.48\textwidth]{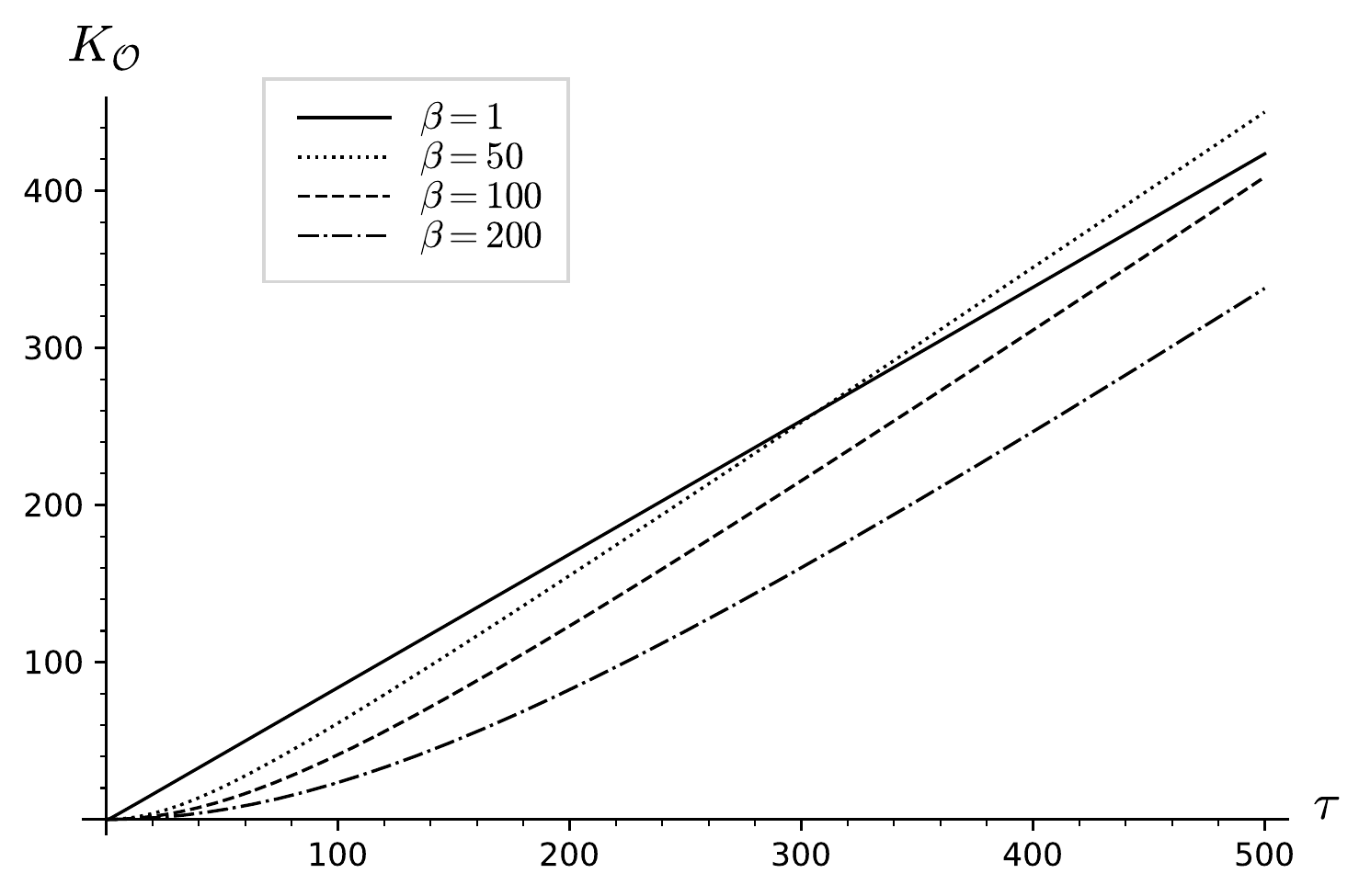}
	\hfill
	\includegraphics[width=0.48\textwidth]{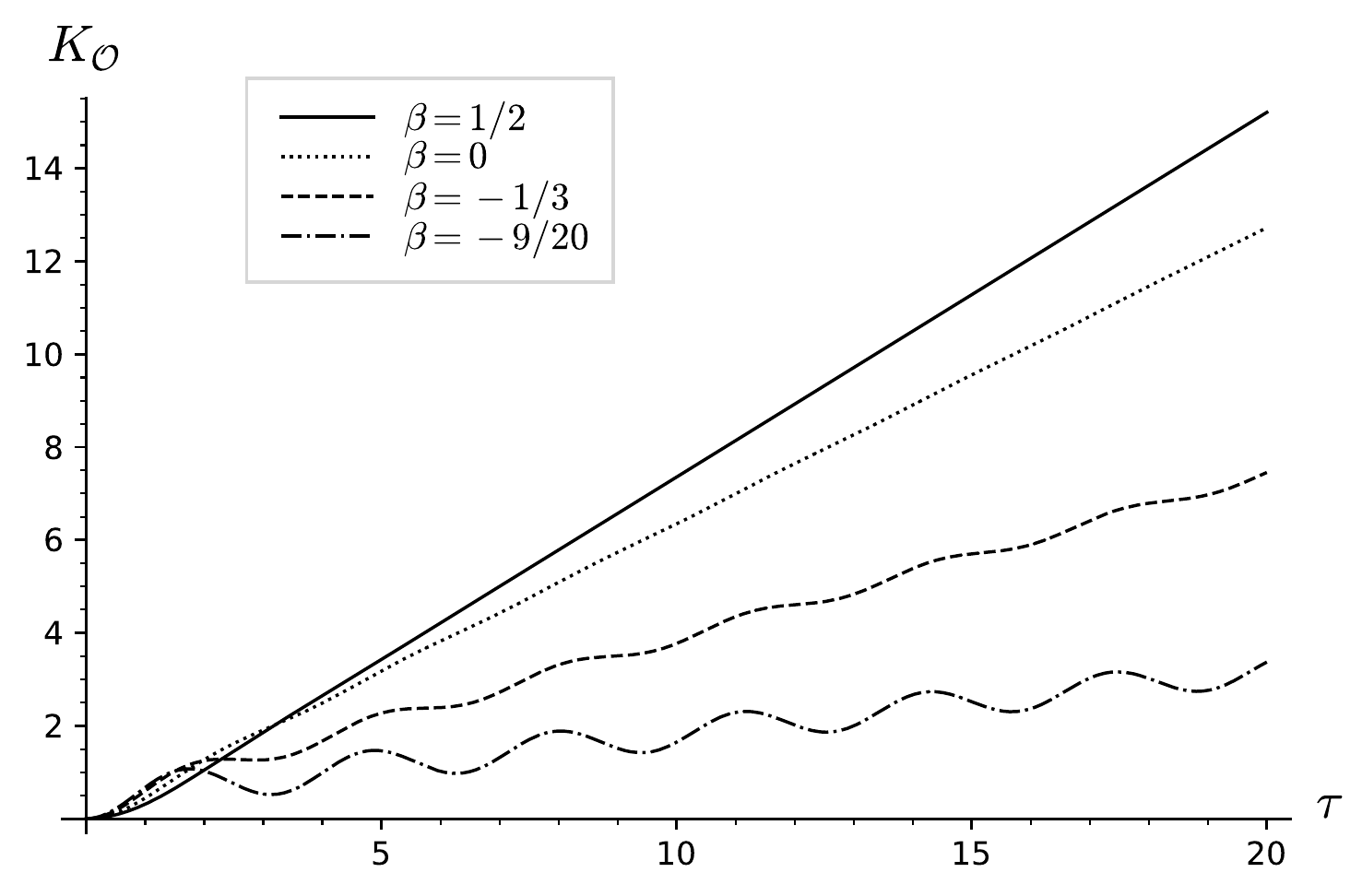}
\caption{Complexity \eqref{ex:geg.K2} as a function of time for a variety of values of $\beta$. Note the different scales. \label{ex:geg.K.fig}}  
\end{figure}

At late times, the asymptotic expansion of \eqref{ex:geg.K2} gives \cite{Wolfram3}
\begin{equation}
\label{ex:geg.K.asy}
	K_\op(\tau) \approx \frac{\Gamma(\beta+1)^2\,\tau}{\Gamma(\beta+\frac12)\Gamma(\beta+\frac32)}  - \beta +\op(\tau^{-1}) 
	- \frac{\Gamma(\beta+1)^2}{2\pi\,\tau^{2\beta+1}} \left[ \sin(2\tau-\pi\beta) + \op(\tau^{-1}) \right]~.
\end{equation}
For $\beta\geq0$, the last term can be omitted, and we can rewrite \eqref{ex:geg.K.asy} as
\begin{equation}
\label{ex:geg.K.asy2}
	K_\op(\tau) \approx  (\tau-\tau_0)\dot{K}_\infty~, \qquad 
	\dot{K}_\infty= \frac{\Gamma(\beta+1)^2}{\Gamma(\beta+\frac12)\Gamma(\beta+\frac32)}~,\qquad 
	\tau_0 = \frac{\beta}{\dot{K}_\infty}~.
\end{equation}
The late-time growth rate $\dot{K}_\infty$ is not universal. It is a monotonously increasing function of $\beta$, starting with zero at $\beta=-\frac12$ and reaching unity at $\beta=\infty$. However, the larger $\beta$ the later the linear growth regime is reached ($\tau_0$ grows with $\beta$). 
For $-\frac12<\beta<0$ the last term in \eqref{ex:geg.K.asy} gives an oscillating sub-dominant contribution.
The complexity \eqref{ex:geg.K2} is plotted in figure~\ref{ex:geg.K.fig} for various values of the parameter $\beta$.

\subsection{Hermite polynomials}
\label{ex:hermite}

The last example of classical orthogonal polynomials with for an even measure are the Hermite polynomials.  In contrast to the previous examples, the associated spectrum is not bounded. This is, actually, the simplest case and corresponds to the Heisenberg algebra case considered in \cite{Caputa:2021sib}. We have already met this case among the examples discussed in subsection~\ref{gen:simple}.

Hermite polynomials are orthogonal with respect to the (normalized) weight function
\begin{equation}
\label{ex:herm.measure}
	w(x) = \frac1{\sqrt{\pi}} \e{-x^2}~, \qquad x\in (-\infty,\infty)~.
\end{equation}
The monic polynomials are 
\begin{equation}
\label{ex:herm.ploy}
	P_n(x) = 2^{-n} H_n(x)
\end{equation}
with
\begin{equation}
\label{ex:herm.h}
	h_n = 2^{-n} n!~,\qquad \Delta_n = 2n~.
\end{equation}
The wave functions \eqref{La:phi.n} are 
\begin{equation}
\label{ex:herm.phi.n}
	\phi_n(\tau) = \frac{i^n}{\sqrt{\pi 2^n n!}}  \int\limits_{-\infty}^\infty \rmd x\, \e{-x^2-i\tau x} H_n(x) 
	= \frac{\tau^n}{\sqrt{2^n n!}} \e{-\frac{\tau^2}4}~.
\end{equation}
To characterize the operator growth, consider the complexity generating function \eqref{La:K.complexity.gen.func},
\begin{equation}
\label{ex:herm.gen.func}
	K_\op(\lambda;\tau) = \sum\limits_{n=0}^\infty \frac1{n!} \left( \frac{\e{\lambda}\tau^2}{2} \right)^n \e{-\frac12 \tau^2}
	= \e{\frac12 (\e{\lambda}-1)\tau^2}~.
\end{equation} 
It follows that the complexities (of degree one and two) are 
\begin{equation}
\label{ex:herm.complex}
	K^{(1)}_\op (\tau) = \frac12 \tau^2~,\qquad  K^{(2)}_\op (\tau) = \frac14 \tau^2 (\tau^2+2)~,
\end{equation}
and the Krylov variance \eqref{La:K.variance} is 
\begin{equation}
\label{ex:herm.variance}
	\Delta K_\op (\tau) = \frac{2}{\tau^2}~.
\end{equation}

\subsection{The simplest case with a gapped spectrum}
\label{ex:gap}

As discussed in subsection~\ref{gen:spectra}, a gapped spectrum can be associated with a sequence of Lanczos coefficients $\Delta_n$ that  approach different limits depending on whether $n$ is even or odd. The simplest case, which can be treated analytically \cite{rec-method}, is given by
\begin{equation}
\label{ex:gap.Delta}
	\Delta_{2n} = \Delta_e\quad  (n>0)~, \qquad \Delta_{2n+1}=\Delta_o~.
\end{equation}
For completeness, we first report the solution from \cite{rec-method} and then add some further details in order to prepare the ground for the more general case involving the Jacobi polynomials, which will be the subject of the next subsection.

Our first goal is to obtain the functions $c_n(z)$. For this purpose, consider the second equation of \eqref{La:c.z.eq.1} for even and odd $n$, 
\begin{equation}
\label{ex:gap.c.eq}
	R_e(z) = \frac{\Delta_e}{z +R_o(z)}~,\qquad 
	R_o(z) = \frac{\Delta_o}{z +R_e(z)}~.
\end{equation}
This system is solved by\footnote{The fact that $R_e$ and $R_o$ must vanish for large $z$ fixes the sign in the solution of the quadratic equation.}
\begin{align}
\label{ex:gap.Ro}
	R_o(z) &= \frac{1}{2} \left[\sqrt{z^2 +2(\Delta_o+\Delta_e) + \frac{(\Delta_o-\Delta_e)^2}{z^2}} 
		-z + \frac{\Delta_o-\Delta_e}{z} \right]~,  \\	
\label{ex:gap.Re}
	R_e(z) &= \frac12 \left[\sqrt{z^2 +2(\Delta_o+\Delta_e) + \frac{(\Delta_o-\Delta_e)^2}{z^2}} 
	-z - \frac{\Delta_o-\Delta_e}{z} \right]~.
\end{align}
The functions $c_n$ then follow from the first equation of \eqref{La:c.z.eq.1} and by applying \eqref{La:R.def} recursively, with the result 
\begin{equation}
\label{ex:gap.cn}
	c_0(z) =\frac{R_e}{\Delta_e}~,\qquad 
	c_{2n} = T(z)^n c_0(z)~,\qquad
	c_{2n+1}(z) = \frac1{\sqrt{\Delta_e}} T(z)^{n+1}~,
\end{equation}
where we have introduced 
\begin{align}
\notag
	T(z) &= \frac{R_e(z)R_o(z)}{\sqrt{\Delta_e\Delta_o}} \\
\label{ex:gap.R.prod}
	&= \frac1{2\sqrt{\Delta_e\Delta_o}} \left[ z^2 +\Delta_o+\Delta_e -\sqrt{z^4+2(\Delta_e+\Delta_o)z^2 +(\Delta_o-\Delta_e)^2} \right]~. 
\end{align}
 
The spectral density can be obtained from $c_0(z)$ using \eqref{La:spec.dens}. One finds  
\begin{equation}
\label{ex:gap.spec.dens}
	\frac{\rmd \mu}{\rmd \omega} = \frac1{2\pi \Delta_e |\omega|} \sqrt{(\omega^2-\omega_-^2)(\omega_+^2-\omega^2)} \Theta(\omega_+^2-\omega^2) \Theta(\omega^2-\omega_-^2) +\frac{\Delta_e-\Delta_o}{\Delta_e} \delta(\omega) \Theta(\Delta_e-\Delta_o)~,
\end{equation}
with
\begin{equation}
\label{ex:gap.omegas}
	\omega_- = \left|\sqrt{\Delta_o} - \sqrt{\Delta_e}\right|~,\qquad \omega_+ = \sqrt{\Delta_o} + \sqrt{\Delta_e}~,
\end{equation}
Notice that the delta function at $\omega=0$ appears only if $\Delta_e<\Delta_o$. 

Alternatively, one may consider separately the spectral densities for the even and odd polynomials. For the even ones, first use \eqref{La:Q.even} to obtain\footnote{We retain the factors of $i^2$ under the square root in order to keep track of the phase.}
\begin{align}
\notag
	\bar{Q}_0(z) &= \frac{-i}{\sqrt{z}}c_0(-i\sqrt{z}) \\
\label{ex:gap.Q.even}
	&=
	\frac1{2\Delta_e} \left[ -i \sqrt{(-i)^2 +\frac{2(\Delta_e+\Delta_o)}{z} +\frac{i^2(\Delta_o-\Delta_e)^2}{z^2}} 
	+1 -\frac{\Delta_o-\Delta_e}{z} \right]~.
\end{align}
From \eqref{ex:gap.Q.even} we can read off the spectral density\footnote{The comment made in footnote~\ref{La:footnote.delta} applies here. Since $\omega^2\in[0,\infty)$, there is an extra factor of 2 in the term with the delta function. This factor must be dropped, if we one defines $x=\omega^2$ and extends its domain to $x\in (-\infty,\infty)$.}
\begin{equation}
\label{ex:gap.spec.dens.even}
	\frac{\rmd \bar{\mu}}{\rmd \omega^2} = \frac1{2\pi \Delta_e \omega^2} \sqrt{(\omega^2-\omega_-^2)(\omega_+^2-\omega^2)}\, \Theta(\omega_+^2-\omega^2) \Theta(\omega^2-\omega_-^2) +2\frac{\Delta_e-\Delta_o}{\Delta_e} \delta(\omega^2) \Theta(\Delta_e-\Delta_o)~.
\end{equation}
It is convenient to introduce 
\begin{equation}
\label{ex:gap.y.def}
	y = \gamma \omega^2 -\bar{y}~,\qquad \gamma= \frac1{2\sqrt{\Delta_e\Delta_o}}~,\qquad \bar{y} = \gamma(\Delta_o+\Delta_e)~,
\end{equation}
in terms of which \eqref{ex:gap.spec.dens.even} reads
\begin{equation}
\label{ex:gap.spec.dens.even.y}
	\frac{\rmd \bar{\mu}}{\rmd y} = \frac1\pi \sqrt{\frac{\Delta_o}{\Delta_e}} \frac{\sqrt{1-y^2}}{y+\bar{y}} \Theta(1-y^2)
	+\frac{\Delta_e-\Delta_o}{\Delta_e}\, \Theta(\Delta_e-\Delta_o)\,\delta(y+\bar{y})~.
\end{equation}
It is easy to check that \eqref{ex:gap.spec.dens.even.y} is normalized on $y\in(-\infty,\infty)$. 

Similarly, to get the measure for the odd polynomials, we calculate 
\begin{align}
\notag
	\dbar{Q}_0(z) &= -\frac{c_1(-i\sqrt{z})}{\sqrt{\Delta_o}} \\
\label{ex:gap.Q.odd}
	&= \frac1{2\Delta_e \Delta_o} \left[ -i \sqrt{(-i)^2 z^2+2(\Delta_e+\Delta_o)z +i^2 (\Delta_o-\Delta_e)^2} +z -\Delta_e -\Delta_o \right]~.
\end{align}
This yields the spectral density
\begin{equation}
\label{ex:gap.spec.dens.odd}
	\frac{\rmd \dbar{\mu}}{\rmd \omega^2} = \frac1{2\pi \Delta_e \Delta_o} \sqrt{(\omega^2-\omega_-^2)(\omega_+^2-\omega^2)}\, \Theta(\omega_+^2-\omega^2) \Theta(\omega^2-\omega_-^2)~,
\end{equation}
which becomes
\begin{equation}
\label{ex:gap.spec.dens.odd.y}
	\frac{\rmd \dbar{\mu}}{\rmd y} = \frac2\pi \sqrt{1-y^2}\, \Theta(1-y^2)
\end{equation}
upon using \eqref{ex:gap.y.def}. Obviously, \eqref{ex:gap.spec.dens}, \eqref{ex:gap.spec.dens.even} and \eqref{ex:gap.spec.dens.odd} are related by \eqref{La:p.even} and \eqref{La:p.odd}, while the delta function at $\omega=0$ is irrelevant for the odd polynomials.

Because the measure \eqref{ex:gap.spec.dens.odd.y} is nothing but the orthogonality measure for the Chebyshev polynomials of the second kind, we can immediately write down 
\begin{equation}
\label{ex:gap.p.odd}
	\dbar{P}_n(\omega^2) = \gamma^{-n} 2^{-n} U_n(y) = (\Delta_e \Delta_o)^{\frac{n}2} U_n(y)~.
\end{equation}
These polynomials are monic in $\omega^2$. One easily obtains $\dbar{h}_n = (\Delta_e\Delta_o)^n$, which with \eqref{La:h.even.odd} yields $h_{2n+1}= \Delta_e^n\Delta_o^{n+1}$. 

The even polynomials follow immediately from \eqref{La:even.fom.odd}, 
\begin{equation}
\label{ex:gap.p.even}
	\bar{P}_n = (\Delta_e \Delta_o)^{\frac{n}2} \left[ U_n(y) + \sqrt{\frac{\Delta_e}{\Delta_o}} U_{n-1}(y)\right]~,
\end{equation}
By consistency, we have $\bar{h}_n=h_{2n}= \Delta_e h_{2n-1} = (\Delta_e\Delta_o)^n$.

Let us try to find an analytic expression for the complexity. Because we do not know the wave functions $\phi_n(t)$, but we have their Laplace transforms $c_n(z)$ as well as the polynomials $P_n$, we shall look at the generating function $K_\op(\lambda;z)$ in the form \eqref{La:K.gen.func.laplace3}. Hence, consider
\begin{equation}
\label{ex:gap.K.gen.func1}
	K_\op(\lambda;z) = \int\rmd \mu(\Liou) \sum\limits_{n=0}^\infty \frac{(i\e{\lambda})^n}{\sqrt{h_n}} P_n(\Liou) c_n(z+i\Liou)
\end{equation}
and look separately at the terms with odd and even $n$. For odd $n$, using \eqref{ex:gap.p.odd} and \eqref{ex:gap.cn}, the summands become 
\begin{equation}
\label{ex:gap.summand.odd}
	\frac{(i\e{\lambda})^{2n+1}}{\sqrt{h_{2n+1}}} P_{2n+1}(\Liou) c_{2n+1}(z+i\Liou)
	= \frac{i\e{\lambda} \Liou}{\sqrt{\Delta_e\Delta_o}} T(z+i\Liou)U_n(y) \left[-\e{2\lambda}T(z+i\Liou)\right]^n~,
\end{equation}
where we recall the definition of $T(z)$ from \eqref{ex:gap.R.prod}. The terms \eqref{ex:gap.summand.odd} can be added up using the definition of the generating function of the Chebyshev polynomials \cite{NIST:DLMF}. This results in
\begin{equation}
\label{ex:gap.sum.odd}
	\sum\limits_{n\; \mathrm{odd}} \frac{(i\e{\lambda})^n}{\sqrt{h_n}} P_n(\Liou) c_n(z+i\Liou)
	=\frac{i \e{\lambda}\Liou}{\sqrt{\Delta_e\Delta_o}}\frac{T(z+i\Liou)}{1+2y\e{2\lambda}T(z+i\Liou) + \e{4\lambda}T(z+i\Liou)^2}~. 
\end{equation}
Similarly, for even $n$, we rewrite the summands in \eqref{ex:gap.K.gen.func1} using \eqref{ex:gap.p.even} and \eqref{ex:gap.cn}, which gives, for $n>0$,
\begin{equation}
\label{ex:gap.summand.even}
	\frac{(i\e{\lambda})^{2n}}{\sqrt{h_{2n}}} P_{2n}(\Liou) c_{2n}(z+i\Liou)
	= c_0(z+i\Liou) \left[ U_n(y) + \sqrt{\frac{\Delta_e}{\Delta_o}} U_{n-1}(y) \right] \left[-\e{2\lambda}T(z+i\Liou)\right]^n~,
\end{equation}
while one simply gets $c_0(z+i\Liou)$ for $n=0$. Then, summing up the terms with even $n$ yields
\begin{equation}
\label{ex:gap.sum.even}
	\sum\limits_{n\; \mathrm{even}} \frac{(i\e{\lambda})^n}{\sqrt{h_n}} P_n(\Liou) c_n(z+i\Liou)
	= c_0(z+i\Liou) \frac{1-\e{2\lambda}\sqrt{\frac{\Delta_e}{\Delta_o}}T(z+i\Liou)}{1+2y\e{2\lambda}T(z+i\Liou) + \e{4\lambda}T(z+i\Liou)^2}~. 
\end{equation}
Thus, after adding \eqref{ex:gap.sum.odd} and \eqref{ex:gap.sum.even} one has
\begin{align}
\notag
	\Sigma(\lambda;z,\Liou) &= \sum\limits_{n=0}^\infty \frac{(i\e{\lambda})^n}{\sqrt{h_n}} P_n(\Liou) c_n(z+i\Liou)\\
\label{ex:gap.sum}
	&= \frac{c_0(z+i\Liou) \left[ 1-\e{2\lambda}\sqrt{\frac{\Delta_e}{\Delta_o}}T(z+i\Liou)\right] 
	+ \frac{i \e{\lambda}\Liou}{\sqrt{\Delta_e\Delta_o}} T(z+i\Liou)}{1+2y\e{2\lambda}T(z+i\Liou) + \e{4\lambda}T(z+i\Liou)^2}~. 
\end{align}
To make progress, we can use the identities
\begin{equation}
\label{ex:gap.c.id}
	zc_0(z) = 1-\sqrt{\frac{\Delta_o}{\Delta_e}} T(z)~,\qquad 
	\sqrt{\frac{\Delta_e}{\Delta_o}} T(z) c_0(z) = c_0(z) -\frac{z}{\sqrt{\Delta_e\Delta_o}} T(z) 
\end{equation}
and
\begin{equation}
\label{ex:gap.T.id}
	T(z)^2 = -1 + \frac{T(z)}{\sqrt{\Delta_e\Delta_o}} (z^2+\Delta_e+\Delta_o)~,
\end{equation}
which follow from the recurrence relations. It is then easy to check that setting $\lambda=0$ gives the expected result $\Sigma(0;z,\Liou)=1/z$ after using \eqref{ex:gap.c.id}, \eqref{ex:gap.T.id} and \eqref{ex:gap.y.def}.

To find the (Laplace transform of) complexity, we need 
\begin{equation}
\label{ex:gap.K.def}
	K_\op(z) = \int \rmd\mu(\Liou)\, \left. \partial_\lambda \Sigma(\lambda;z,\Liou)\right|_{\lambda=0}~.
\end{equation}
With the help of the identities \eqref{ex:gap.c.id} and \eqref{ex:gap.T.id} one gets
\begin{equation}
\label{ex:gap.dsigma}
	\left. \partial_\lambda \Sigma(\lambda;z,\Liou)\right|_{\lambda=0} = \frac1{z^2} \left[ i\Liou +2\Delta_e c_0(z+i\Liou) +\frac{2(\Delta_o-\Delta_e)}{z+2i\Liou} \right]~.
\end{equation}
Substituting \eqref{ex:gap.dsigma} into \eqref{ex:gap.K.def}, the first term in the brackets does not contribute, because it is odd. The easiest way to find the integral of the third term is by writing 
\begin{equation}
\label{ex:gap.con.int}
	\int \rmd \mu(\Liou)\, \ldots = \frac1{2\pi i} \oint \rmd \omega\, Q_0(\omega)\, \ldots ~,
\end{equation}
where the contour surrounds only the support of the measure $\mu(\Liou)$, and then deforming the contour to pick up the residue at $\omega= \frac{iz}2$. (There is no contribution from $\infty$.) Finally, the integral of the second term can be written explicitly using the measure \eqref{ex:gap.spec.dens}. This yields
\begin{align}
\label{ex:gap.K.int}
	K_\op(z) &= \frac1{z^2} \Bigg\{ (\Delta_o-\Delta_e) c_0\left(\frac{z}2\right) +2 (\Delta_e-\Delta_o) \Theta(\Delta_e-\Delta_o) c_0(z) 
	\\
\notag
	&\quad + \int\limits_{\omega_-}^{\omega_+} \frac{\rmd \omega}{\pi\omega}  \sqrt{(\omega_+^2-\omega^2)(\omega^2-\omega_-^2)}
	\left[c_0(z+i\omega)+c_0(z-i\omega)\right] \Bigg\}~.
\end{align}
To find the late-time asymptotics of the complexity, consider the dominant term in \eqref{ex:gap.K.int} for small $z$. Observe that
$$ c_0(z) = \frac{\Delta_e-\Delta_o}{\Delta_e z} \Theta(\Delta_e-\Delta_o) + \op(z)~,$$
so that the $1/z$ term precisely cancels between the first two terms on the right hand side of \eqref{ex:gap.K.int}. Therefore, the dominant contribution to $K_\op(z)$ is found by taking the limit $z\to 0$ in the integral, which gives
\begin{align}
\notag 
	K_\op(z) &= \frac1{\pi z^2} \int\limits_{\omega_-}^{\omega_+} \frac{\rmd \omega}{\omega} \sqrt{(\omega_+^2-\omega^2)(\omega^2-\omega_-^2)}
	\left[c_0(i\omega)+c_0(-i\omega)\right]+ \op\left(\frac1z\right)\\
\notag
	&= \frac1{\pi\Delta_ez^2} \int\limits_{\omega_-}^{\omega_+} \frac{\rmd \omega}{\omega^2}(\omega_+^2-\omega^2)(\omega^2-\omega_-^2)+\op\left(\frac1z\right)\\
\label{ex:gap.K.final}
	&= \frac{2(\omega_+-\omega_-)^3}{3\pi\Delta_e z^2}  +\op\left(\frac1z\right)~.
\end{align}
Note that 
\begin{equation}
\label{ex:gap.domega}
	\omega_+ - \omega_- = 
	\begin{cases}
		2 \sqrt{\Delta_e} \quad &\text{for $\Delta_o>\Delta_e$}, \\
		2 \sqrt{\Delta_o} &\text{for $\Delta_o<\Delta_e$}. 
	\end{cases}
\end{equation}
Finally, \eqref{ex:gap.K.final} translates into the late-time complexity 
\begin{equation}
\label{ex:gap.K.t}
	K_\op(t) = \frac{2(\omega_+-\omega_-)^3}{3\pi\Delta_e}\, t +\op(1)~.
\end{equation}
It can be checked that \eqref{ex:gap.K.t} satisfies the bound \eqref{cont:two:K}. Moreover, if we set $\Delta_e=\Delta_o=\frac14$, we reproduce \eqref{ex:cheU.K.t}.

\subsection{Jacobi polynomials}
\label{ex:Jacobi}

Jacobi polynomials \cite{Gradshteyn, NIST:DLMF} are, in general, related to a non-symmetric measure. They can be used to model a more general system with a gap than that of the previous subsection while maintaining a bounded spectrum. Let us assume a measure with support in $\omega_-<|\omega|<\omega_+$ plus possibly a delta function at $\omega=0$. To be concrete, define  
\begin{equation}
\label{ex:jac.y.def}
	y = \gamma \left(\omega^2 - \bar{\omega}^2 \right) 
\end{equation}
with 
\begin{equation}
\label{ex:jac.gamma.def}
	\gamma = \frac2{\omega_+^2-\omega_-^2}~,\qquad \bar{\omega}^2 = \frac12 \left(\omega_+^2+\omega_-^2\right)~,
\end{equation}
so that $y\in (-1,1)$. 

Because the odd polynomials are insensitive to a delta function at $\omega=0$, let us postulate 
\begin{equation}
\label{ex:jac.odd.p}
	\dbar{P}_n(\omega^2) = \frac{n!\, 2^n\gamma^{-n}}{(n+\alpha+\beta+1)_n} \Jacobi{n}(y)~,
\end{equation}
where $(a)_n$ denotes the Pochhammer symbol, and $\Jacobi{n}(y)$ are the standard Jacobi polynomials ($\alpha,\beta>-1$). The prefactor has been chosen such that the $\dbar{P}_n$ are monic in $\omega^2$. They are orthogonal with respect to the (normalized) measure
\begin{equation}
\label{ex:jac.measure.odd}
	\rmd \dbar{\mu} = \frac{\Gamma(\alpha+\beta+2)}{2^{\alpha+\beta+1}\Gamma(\alpha+1)\Gamma(\beta+1)} (1-y)^\alpha (1+y)^\beta \rmd y
\end{equation}
and have the norms
\begin{equation}
\label{ex:jac.norm.odd}
	\dbar{h}_n = \frac{n!\,4^n\gamma^{-2n} (\alpha+1)_n(\beta+1)_n}{(n+\alpha+\beta+1)_n(\alpha+\beta+2)_{2n}}~.
\end{equation}

The three-term recurrence relation of the Jacobi polynomials,
\begin{align}
\label{ex:jac.rec.odd}
	2(n+1)&(n+\alpha+\beta+1)(2n+\alpha+\beta)\Jacobi{n+1}(y) =\\
\notag
	&(2n+\alpha+\beta+1)[(2n+\alpha+\beta)(2n+\alpha+\beta+2)y +\alpha^2 -\beta^2] \Jacobi{n}(y)\\
\notag
	&-2(n+\alpha)(n+\beta)(2n+\alpha+\beta+2) \Jacobi{n-1}(y)
\end{align}
corresponds to \eqref{La:rec.odd} with 
\begin{align}
\label{ex:jac.rec.odd.b}
	\dbar{b}_n &= \Delta_{2n+1}+\Delta_{2n+2} 
	= \bar{\omega}^2 - \frac{\alpha^2-\beta^2}{\gamma(2n+\alpha+\beta)(2n+\alpha+\beta+2)}~,\\
\label{ex:jac.rec.odd.c}
	\dbar{c}_n &= \Delta_{2n}\Delta_{2n+1} 
	= \frac{4n(n+\alpha)(n+\beta)(n+\alpha+\beta)}{\gamma^2(2n+\alpha+\beta+1)(2n+\alpha+\beta)^2(2n+\alpha+\beta-1)}~.
\end{align}
For large $n$, the right hand sides of \eqref{ex:jac.rec.odd.b} and \eqref{ex:jac.rec.odd.c} reduce to $\bar{\omega}^2$ and $1/(4\gamma^2)$, respectively, which suggests that the $\Delta_n$ with even and odd $n$ approach certain limiting values $\Delta_e$ and $\Delta_o$, respectively. These two values satisfy the relations of the simpler case discussed in the previous subsection \eqref{ex:gap.omegas},
\begin{equation}
\label{ex:gap.delta.eo}
	\Delta_{e/o} = \frac14 \left( \omega_+ \pm \omega_-\right)^2~, 
\end{equation} 
but we cannot tell yet which one is which.

Let us continue with the even part. The polynomials are determined from \eqref{La:even.fom.odd} and \eqref{ex:jac.odd.p},
\begin{equation}
\label{ex:jac.even.p}
	\bar{P}_n(\omega^2) = \frac{n!\, 2^n\gamma^{-n}}{(n+\alpha+\beta+1)_n} \left[ \Jacobi{n}(y) 
	+\frac{\gamma \Delta_{2n} (2n+\alpha+\beta)(2n+\alpha+\beta-1)}{2n(n+\alpha+\beta)} \Jacobi{n-1}(y) \right]~.
\end{equation}

Except for the delta function at $\omega=0$, the measure $\rmd \bar{\mu}$ is simply given by combining the measures in \eqref{La:p.even} and \eqref{La:p.odd}. Thus, 
\begin{equation}
\label{ex:jac.measure.even}
	\rmd \bar{\mu} = \frac{\Delta_1}{\omega^2} \rmd \dbar{\mu} + 2A \delta(\omega^2) \rmd \omega^2~,
\end{equation}
with some $A\in [0,1)$, which will be determined shortly.\footnote{Recall the convention in footnote~\ref{La:footnote.delta}.} With \eqref{ex:jac.measure.odd}, this becomes
\begin{equation}
\label{ex:jac.measure.even.2}
 	\rmd \bar{\mu} = \frac{\Delta_1 \gamma\, \Gamma(\alpha+\beta+2)}{2^{\alpha+\beta+1}\Gamma(\alpha+1)\Gamma(\beta+1)} 
 	\frac{(1-y)^\alpha (1+y)^\beta}{y + \gamma\bar{\omega}^2} \rmd y
 	+ 2A \delta(\omega^2) \rmd \omega^2~.
\end{equation}
To determine $A$, one imposes that the measure \eqref{ex:jac.measure.even.2} be normalized,
\begin{equation}
\label{ex:jac.A}
		\int \rmd \bar{\mu} = \frac{\Delta_1}{\omega_+^2} \hypF{1,\alpha+1;\alpha+\beta+2;1-\frac{\omega_-^2}{\omega_+^2}} + A = 1~.
\end{equation}

\begin{figure}[t]
	\includegraphics[width=0.48\textwidth]{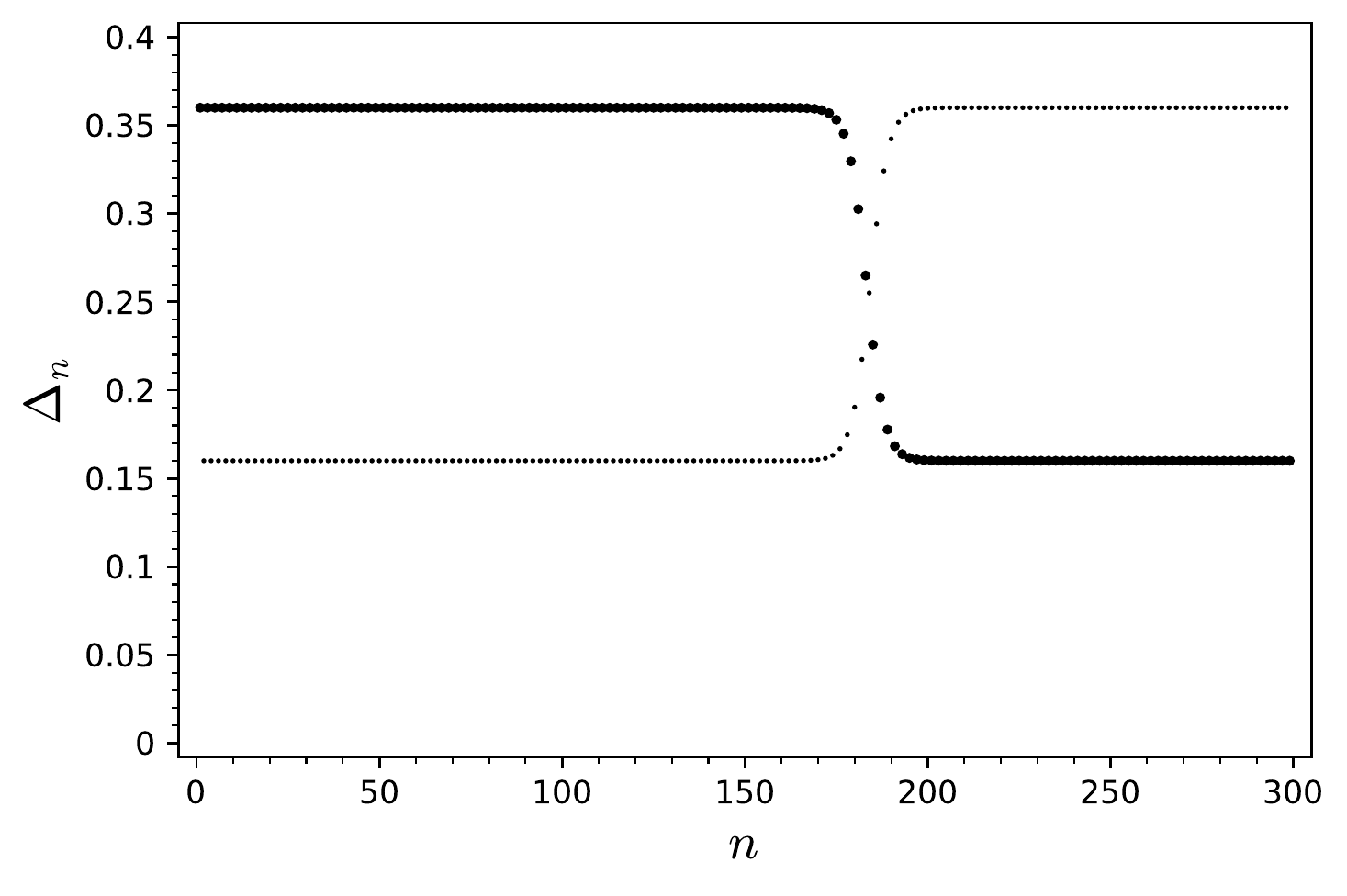}
	\hfill
	\includegraphics[width=0.48\textwidth]{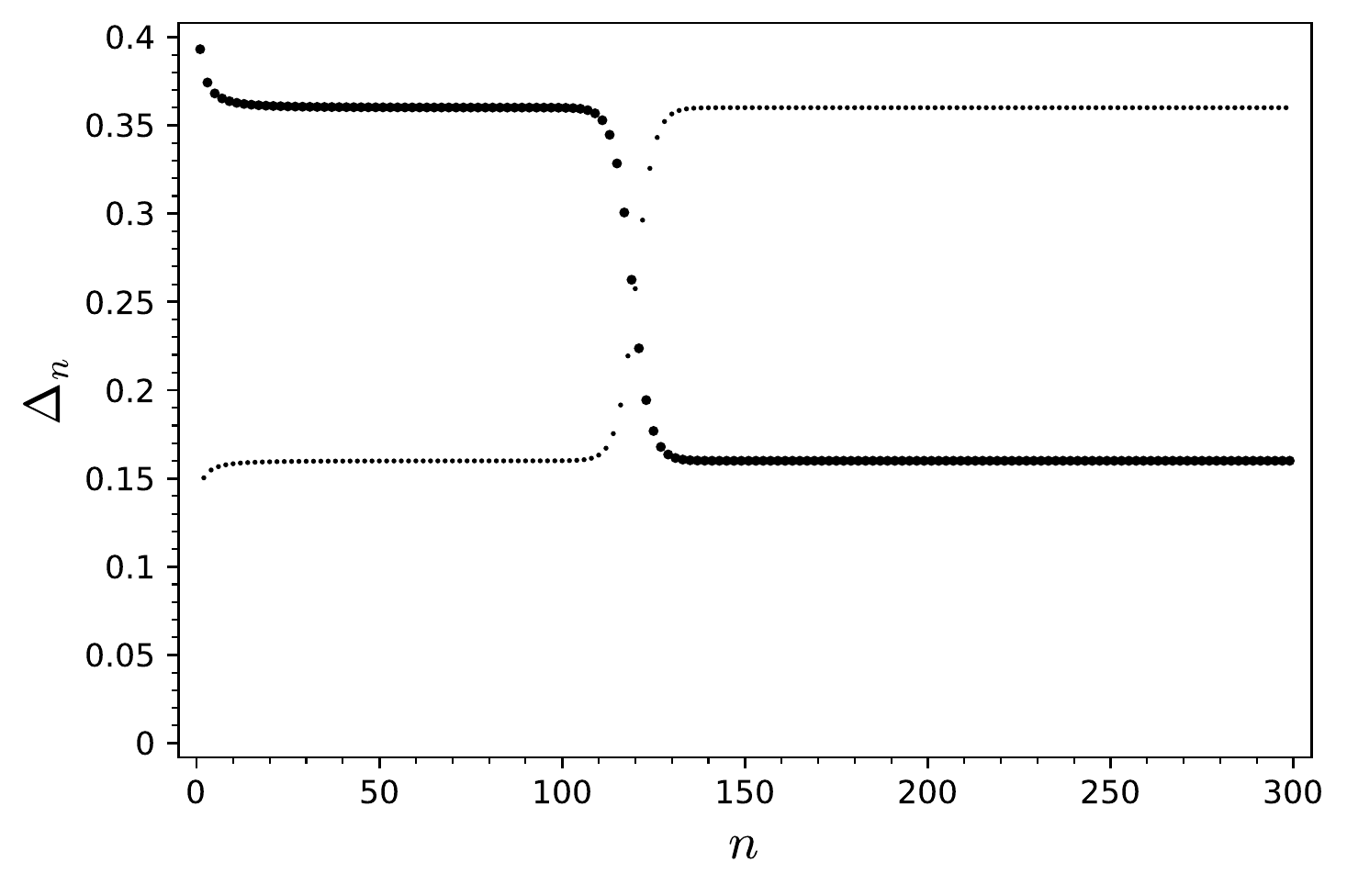}
\caption{Two examples of sequences $\{\Delta_n\}$ that illustrate the generic behaviour $\Delta_o<\Delta_e$ with parameters $\omega_+=1$, $\omega_-=0.2$. The small dots depict $\Delta_{2n}$, the larger dots $\Delta_{2n+1}$. The recursion starts close to the unstable limit solution at $n=99$ and is performed forwards and backwards. Left: $\alpha=\beta=1/2$, $\Delta_1=0.36$, which corresponds to the special case discussed in subsection~\ref{ex:gap}. In this case, the numerical error is sufficient to drive the cross-over to the stable solution. Right: $\alpha=-0.3$, $\beta=0.8$, $\Delta_1=0.39312029$.\label{ex:jac.fig}}  
\end{figure}

$\Delta_1$ remains as a free parameter. All the other $\Delta_n$ follow by recursively applying \eqref{ex:jac.rec.odd.b} and \eqref{ex:jac.rec.odd.c},
$$ \Delta_2 = \dbar{b}_0 -\Delta_1~, \qquad 
	\Delta_3 = \frac{\dbar{c}_1}{\Delta_2}~,\qquad  \text{etc.} $$
However, not any $\Delta_1$ gives rise to a sensible sequence of positive $\Delta_n$. If it does, numerical evidence shows that the sequence converges for large $n$ to 
\begin{equation}
\label{ex:jac.limit}
	\Delta_{2n} \to \Delta_e = \frac14 (\omega_+ +\omega_-)^2~,\qquad 
	\Delta_{2n+1} \to \Delta_o = \frac14 (\omega_+ -\omega_-)^2~. 	 
\end{equation}
If $\omega_->0$, then these limiting values are approached with terms of order $1/n^2$. Notice that $\Delta_e>\Delta_o$, which implies that the central delta function in the measure \eqref{ex:jac.measure.even.2} is a generic feature. Indeed, the limit solution in which $\Delta_e<\Delta_o$ is unstable, as we illustrate in figure~\ref{ex:jac.fig}.

Cross-over solutions such as those illustrated in figure~\ref{ex:jac.fig} are models for systems with two different regimes of linear operator growth. Using the results of the previous subsection, if the sequence of $\Delta_n$ remains long enough close to the unstable fixed point then, before the cross-over, the early-time complexity is  
\begin{equation}
\label{ex:jac.K.early}
	K_\op(t)_\text{early}  =\frac{8(\omega_+ -\omega_-)}{3\pi}\, t +\op(1)~.
\end{equation}
At late times, \ie after the cross-over to the stable fixed point, the complexity grows as 
\begin{equation}
\label{ex:jac.K.late}
	K_\op(t)_\text{late}  =\frac{8(\omega_+ -\omega_-)^3}{3\pi(\omega_+ +\omega_-)^2}\, t +\op(1)~.
\end{equation}
Obviously, $\dot{K}_\op(t)_\text{early} >\dot{K}_\op(t)_\text{late}$.

The case $\omega_-=0$ is special. In this case, there is no gap, but there might be a central singularity, and the $\Delta_n$ with even and odd $n$ approach the common limit $\frac14 \omega_+^2$ from opposite sides with terms of order $1/n$. For further details, we refer the reader to \cite{rec-method, rec-method1985}. 

The systems discussed in this subsection can be further generalized to spectral densities with delta function singularities at the lower and upper bounds of the spectrum. The orthogonal polynomials for these systems were derived from the Jacobi polynomials in \cite{Koornwinder:1984}.

\subsection{Laguerre polynomials}
\label{ex:laguerre}

Laguerre polynomials can be used to model a system with an unbounded spectrum, with or without a gap, generalizing the case of Hermite polynomials discussed in subsection~\ref{ex:hermite}. Consider a spectrum with support $|\omega| \in (\omega_-,\infty)$ with $\omega_-\geq0$. Let us define 
\begin{equation}
\label{ex:lag.omega.def}
	\omega^2 = \omega_-^2 + \kappa x~,
\end{equation}
and postulate that the odd polynomials are given by \eqref{La:p.odd} with
\begin{equation}
\label{ex:lag.P.odd}
	\dbar{P}_n(\omega^2) =  (-\kappa)^n n!L_n^{(\alpha)}(x)~, 
\end{equation}
where $L_n^{(\alpha)}(x)$ denotes a Laguerre polynomial. The polynomials \eqref{ex:lag.P.odd} are monic in $\omega^2$, orthogonal with respect to the measure
\begin{equation}
\label{ex:lag.measure.odd}
	\rmd \dbar{\mu} = \frac{\e{-x} x^\alpha}{\Gamma(1+\alpha)}  \rmd x~, 
\end{equation}
and possess the norms $\dbar{h}_n = \frac{h_{2n+1}}{\Delta_1} = \kappa^{2n}n!(\alpha+1)_n$.
The three-term recurrence relation of the Laguerre polynomials \cite{Gradshteyn}, 
\begin{equation}
\label{ex:lag.rec.odd}
	(n+1)L_{n+1}^{(\alpha)}(x) - (2n+1+\alpha-x)L_{n}^{(\alpha)}(x) +(n+\alpha) L_{n-1}^{(\alpha)}(x) =0~,
\end{equation}
translates into the general three-term recurrence relation \eqref{La:rec.odd} with
\begin{align}
\label{ex:lag.b.odd}
	\dbar{b}_n &= \Delta_{2n+1}+\Delta_{2n+2} = \omega_-^2 + \kappa(2n+\alpha+1)~,\\
\label{ex:lag.c.odd}
	\dbar{c}_n &= \Delta_{2n}\Delta_{2n+1} = \kappa^2 n(n+\alpha)~.
\end{align}
Given $\Delta_1$ as input, all $\Delta_n$ for $n>1$ can be calculated recursively from these two equations.

The even polynomials are given by \eqref{La:even.fom.odd}. They are orthogonal with respect to the measure
\begin{equation}
\label{ex:lag.measure.even}
	\rmd \bar{\mu} = \frac{\Delta_1}{\omega_-^2+\kappa x} \frac{\e{-x} x^\alpha}{\Gamma(1+\alpha)} \rmd x +2A\delta(\omega^2) \rmd \omega^2~, 
\end{equation} 
where the constant $A$ is determined by imposing that \eqref{ex:lag.measure.even} be normalized,
\begin{equation}
\label{ex:lag.A}
	\int \rmd \bar{\mu} =  \frac{\Delta_1}{\alpha\kappa} \left[ \Phi\left(1,1-\alpha;\frac{\omega_-^2}{\kappa}\right) 
	- \Gamma(1-\alpha) \e{\frac{\omega_-^2}{\kappa}} \left(\frac{\omega_-^2}{\kappa}\right)^{\alpha} \right] + A = 1~. 
\end{equation}
Here, $\Phi(\alpha,\gamma;x)$ denotes the Kummer confluent hypergeometric function. 

First, let us discuss the case without a gap, $\omega_-=0$. In this case, \eqref{ex:lag.A} simplifies to 
\begin{equation}
\label{ex:lag.nogap.A}
	\frac{\Delta_1}{\alpha\kappa} +A =1~,
\end{equation}
which restricts $\Delta_1$ to $\Delta_1\in (0,\alpha\kappa]$. Therefore, as mentioned above, given any $\Delta_1\in (0,\alpha\kappa]$, all other $\Delta_n$ are determined recursively by \eqref{ex:lag.b.odd} and \eqref{ex:lag.c.odd}. As one can easily see from \eqref{ex:lag.b.odd} and \eqref{ex:lag.c.odd}, the limiting case $\Delta_1=\kappa\alpha$ gives rise to the following exact sequence 
\begin{equation}
\label{ex:lag.exact}
	\Delta_{2n}=\kappa n~,\qquad \Delta_{2n+1} = \kappa(n+\alpha)~.
\end{equation}
Because $A=0$ from \eqref{ex:lag.nogap.A} in this case, for this sequence the central central delta function in the measure is absent. 
We shall calculate the complexity for the sequence \eqref{ex:lag.exact} further below, but before doing so let us return to sequences with $0<\Delta_1<\alpha\kappa$. 

\begin{figure}[t]
	\includegraphics[width=0.48\textwidth]{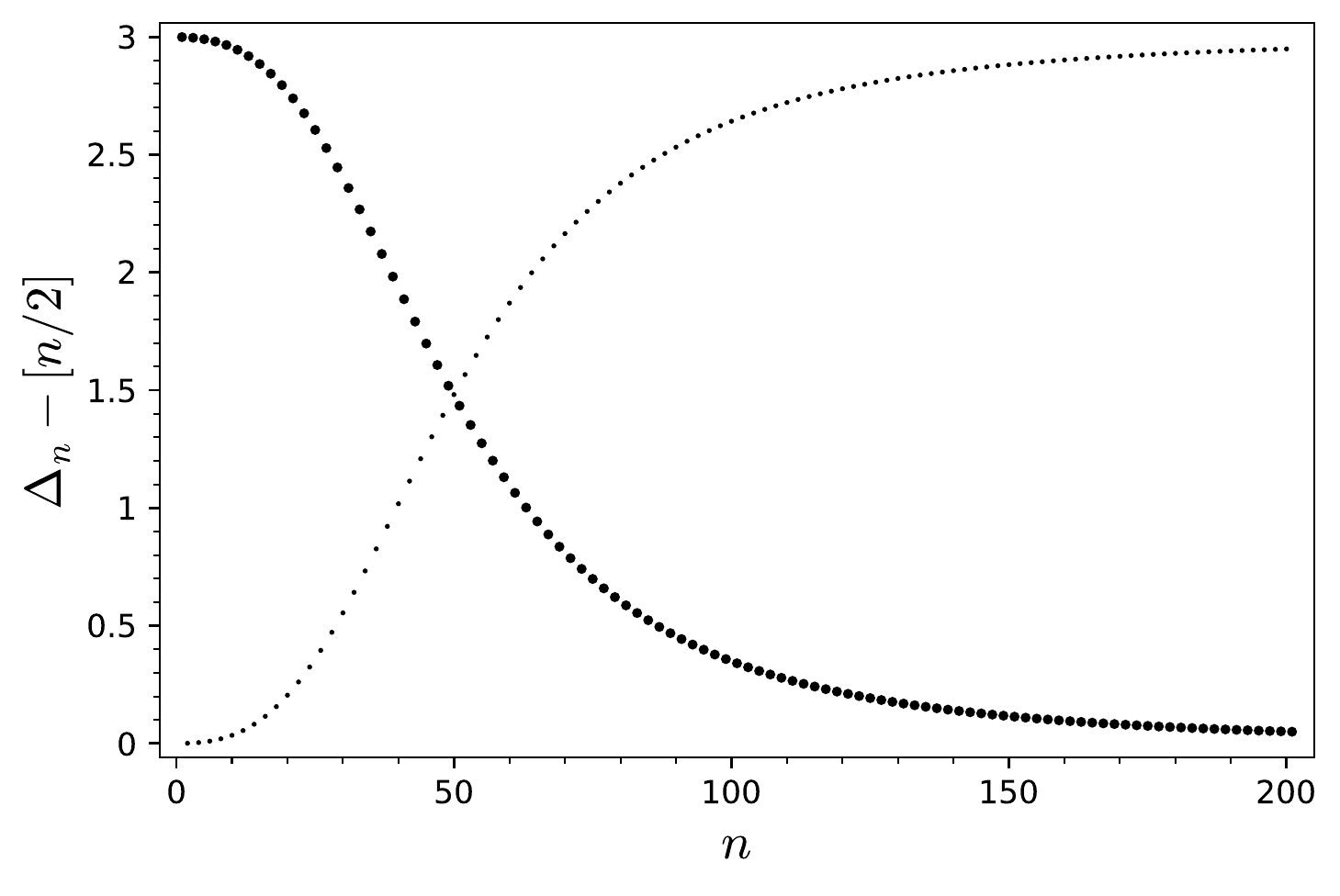}
	\hfill
	\includegraphics[width=0.48\textwidth]{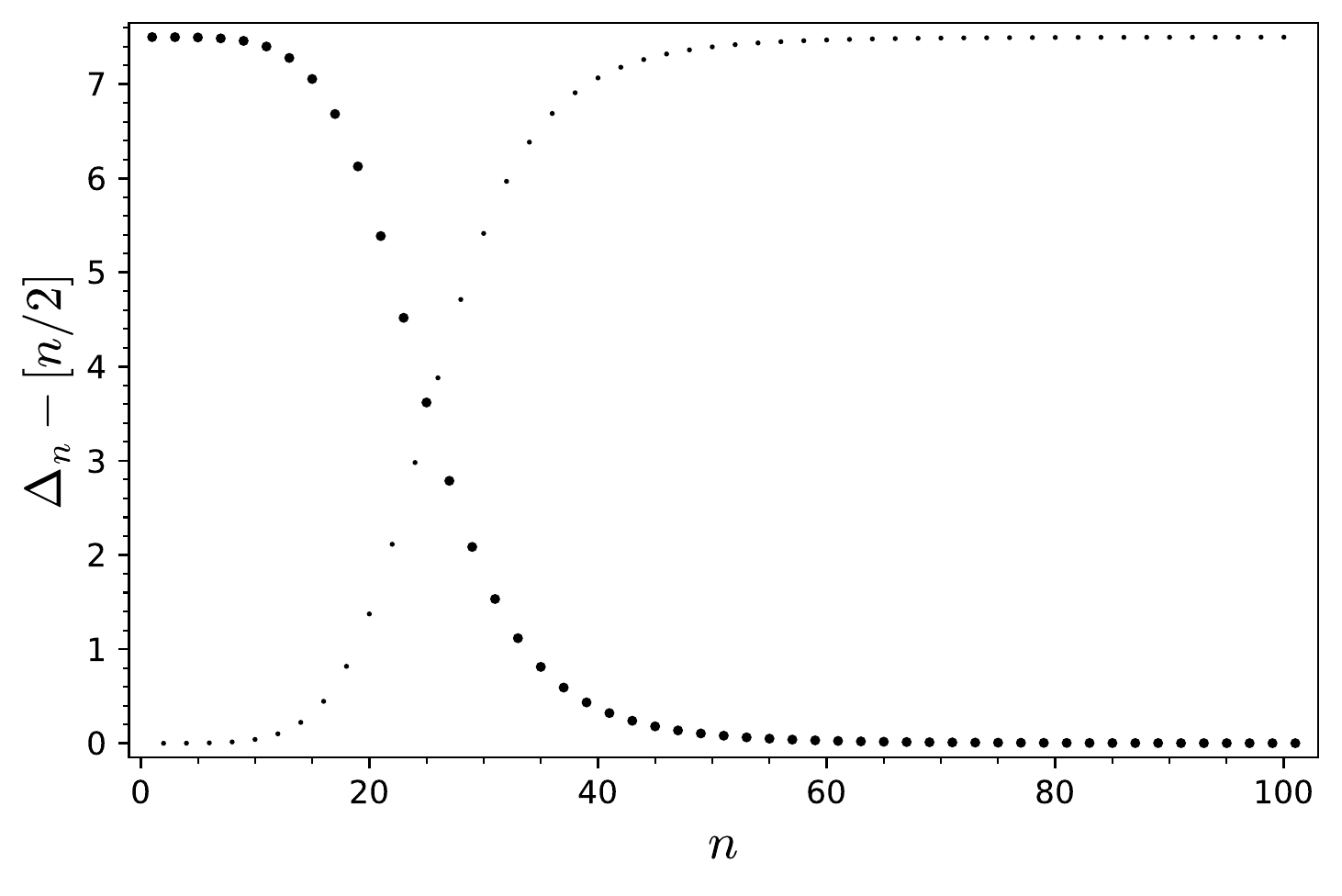}
\caption{Two examples of sequences $\{\Delta_n\}$ that illustrate the convergence towards \eqref{ex:lag.large.n} for $\omega_-=0$, $\kappa=1$. The small dots depict $\Delta_{2n}$, the larger dots $\Delta_{2n+1}$. Left: $\alpha=3$, $\Delta_1=2.999$. Right: $\alpha=7.5$, $\Delta_1=7.4999$.\label{ex:lag.fig}}  
\end{figure}

Another formal solution of \eqref{ex:lag.b.odd} and \eqref{ex:lag.c.odd} when $\omega_-=0$ is
\begin{equation}
\label{ex:lag.large.n}
	\Delta_{2n}=\kappa (n+\alpha)~,\qquad \Delta_{2n+1} = \kappa n~.
\end{equation}
This cannot be an exact solution, because it would imply $\Delta_1=0$. Nevertheless, \eqref{ex:lag.large.n} provides a good large-$n$ asymptotic solution. In fact, numerical evidence suggests that all sequences with $0<\Delta_1< \alpha \kappa$ converge towards \eqref{ex:lag.large.n}, with corrections of order $1/n$. This behaviour is illustrated in the numerical examples in figure~\ref{ex:lag.fig}.

Let us return to the special sequence \eqref{ex:lag.exact}. In this case, the even polynomials are simply 
\begin{equation}
\label{ex:lag.P.even.special}
	\bar{P}_n(\omega^2) =  (-\kappa)^n n!L_n^{(\alpha-1)}(x)~,  
\end{equation}
and the measure \eqref{ex:lag.measure.even} reduces to 
\begin{equation}
\label{ex:lag.measure.even.special}
	\rmd \bar{\mu} = \frac{\e{-x} x^{\alpha-1}}{\Gamma(\alpha)} \rmd x~. 
\end{equation} 
The norm of \eqref{ex:lag.P.even.special} is $h_{2n} = \kappa^{2n} n! (\alpha)_n$. The polynomials $P_n(\omega)$ ($n$ even or odd) are also called \emph{generalized Hermite polynomials} \cite{Chihara:1978}. 

To calculate the complexity, let us consider the generating function $K_\op(\lambda;z)$ in the form \eqref{La:K.gen.func.laplace2}. The even and odd contributions to the sum over $n$ can be considered separately. For the even part, one has
\begin{align}
\label{ex:lag.K.1}
	K_\op(\lambda;z)_\text{even} &= \int\rmd \mu(\omega) \int \rmd \mu(\omega') \sum\limits_{n=0}^\infty \frac{(\e{2\lambda})^n}{h_{2n}}  \frac{P_{2n}(\omega)P_{2n}(\omega')}{z+i(\omega+\omega')}\\
\notag 
	&= \int\rmd \bar{\mu}(\omega^2) \int \rmd \bar{\mu}({\omega'}^2) 
	\frac{\Gamma(\alpha) z(z^2+\omega^2+{\omega'}^2)}{(z^2+\omega^2+{\omega'}^2)^2 -4 \omega^2{\omega'}^2}\\
\notag &\quad \times \sum\limits_{n=0}^\infty 
	\frac{n!L_n^{(\alpha-1)}(\omega^2/\kappa)L_n^{(\alpha-1)}({\omega'}^2/\kappa)(\e{2\lambda})^n}{\Gamma(\alpha+n)}\\
	&= \frac{z}{\Gamma(\alpha)} \int\limits_0^\infty \rmd x \int\limits_0^\infty \rmd y 
	\frac{[z^2+\kappa(x+y)] (xy)^{\frac{\alpha-1}{2}} \e{\lambda(1-\alpha)}}{\{[z^2+\kappa(x+y)]^2-4\kappa^2 xy\}(1-\e{2\lambda})}\\
\notag &\quad \times
	\exp\left(-\frac{x+y}{1-\e{2\lambda}}\right) \BesselI{\alpha-1}\left(\frac{2\sqrt{xy}\e{\lambda}}{1-\e{2\lambda}} \right)~, 
\end{align}
where the summation has been performed using the formula \cite[8.976.1]{Gradshteyn}, and $\BesselI{\alpha}$ denotes a modified Bessel function. Recall that $\lambda$ is a formal parameter, and one should take $\lambda<0$ for the sum to be convergent.
Similarly, for the odd part one obtains
\begin{align}
\label{ex:lag.K.2}
	K_\op(\lambda;z)_\text{odd} &= \int\rmd \mu(\omega) \int \rmd \mu(\omega') \sum\limits_{n=0}^\infty \frac{(-\e{\lambda})(\e{2\lambda})^n}{h_{2n+1}}  \frac{P_{2n+1}(\omega)P_{2n+1}(\omega')}{z+i(\omega+\omega')}\\
\notag 
	&=  \int\rmd \dbar{\mu}(\omega^2) \int \rmd \dbar{\mu}({\omega'}^2) 
	\frac{2\kappa z \e{\lambda} \alpha^2\Gamma(\alpha)}{(z^2+\omega^2+{\omega'}^2)^2 -4 \omega^2{\omega'}^2}\\
\notag &\quad \times \sum\limits_{n=0}^\infty 
	\frac{n!L_n^{(\alpha)}(\omega^2/\kappa)L_n^{(\alpha)}({\omega'}^2/\kappa)(\e{2\lambda})^n}{\Gamma(\alpha+n+1)}\\
	&= \frac{z}{\Gamma(\alpha)} \int\limits_0^\infty \rmd x \int\limits_0^\infty \rmd y 
	\frac{2\kappa (xy)^{\frac{\alpha}{2}} \e{\lambda(1-\alpha)}}{\{[z^2+\kappa(x+y)]^2-4\kappa^2 xy\}(1-\e{2\lambda})}\\
\notag &\quad \times
	\exp\left(-\frac{x+y}{1-\e{2\lambda}}\right) \BesselI{\alpha}\left(\frac{2\sqrt{xy}\e{\lambda}}{1-\e{2\lambda}} \right)~. 
\end{align}
Thus, after adding \eqref{ex:lag.K.1} and \eqref{ex:lag.K.2}, the complexity generating function is 
\begin{align}
\label{ex:lag.K.3}
	K_\op(\lambda;z) &= \frac{z}{\Gamma(\alpha)} \int\limits_0^\infty \rmd x \int\limits_0^\infty \rmd y 
	\frac{(xy)^{\frac{\alpha-1}{2}} \e{\lambda(1-\alpha)}}{\{[z^2+\kappa(x+y)]^2-4\kappa^2 xy\}(1-\e{2\lambda})} \exp\left(-\frac{x+y}{1-\e{2\lambda}}\right)\\
\notag 
	&\quad \times \left\{ \left[z^2+\kappa(x+y)\right]  \BesselI{\alpha-1}\left(\frac{2\sqrt{xy}\e{\lambda}}{1-\e{2\lambda}} \right) 
	+2\kappa \sqrt{xy}  \BesselI{\alpha}\left(\frac{2\sqrt{xy}\e{\lambda}}{1-\e{2\lambda}} \right)  \right\}~.
\end{align}

To make progress, consider the asymptotic behaviour of the modified Bessel function \cite{Gradshteyn}
\begin{equation}
\label{ex:lag.I.asympt}
	I_\alpha (z) = \frac{\e{z}}{\sqrt{2\pi z}} \left[ 1- \frac{(\alpha+\frac12)(\alpha-\frac12)}{2z} +\cdots \right]
	+ \frac{\e{-z}(-1)^{\alpha+\frac12}}{\sqrt{2\pi z}} \left[ 1+ \frac{(\alpha+\frac12)(\alpha-\frac12)}{2z} +\cdots \right]~,
\end{equation}
where we have omitted the higher order terms, which give contributions of order $\lambda^2$ in \eqref{ex:lag.K.3}. After substituting \eqref{ex:lag.I.asympt} into \eqref{ex:lag.K.3} and changing an integration variable by $y=w^2$, one realizes that the contributions arising from the terms with $\e{-z}$ in \eqref{ex:lag.I.asympt} are identical to those arising from the terms with $\e{z}$ upon replacing $w\to -w$. Thus, the result of these operations is 
\begin{align}
\label{ex:lag.K.4}
	K_\op(\lambda;z) &= \frac{z}{\sqrt{\pi}\Gamma(\alpha)} \int\limits_0^\infty \rmd x\, \e{-x} x^{\frac{2\alpha-3}{4}} 
	\int\limits_{-\infty}^\infty \rmd w \frac{w^{\alpha-\frac12} \e{\lambda(\frac12-\alpha)}
	\e{-\frac{(\e{\lambda} \sqrt{x}-w)^2}{1-\e{2\lambda}}}}{\sqrt{1-\e{2\lambda}} [z^2+\kappa(\sqrt{x}-w)^2]}\\
\notag &\quad \times 
	\left\{ 1 - \frac{(1-\e{2\lambda})(\alpha-\frac12)}{2\e{\lambda} [z^2+\kappa(\sqrt{x}+w)^2]} 
	\left[ \frac{z^2+\kappa(x+w^2)}{2w\sqrt{x}}\left(\alpha-\frac32\right) + \kappa  \left(\alpha+\frac12\right) \right] +\cdots\right\}~.
\end{align}
After another change of variable, $w= \e{\lambda}\sqrt{x} +\sqrt{1-\e{2\lambda}} v$, we expand the integrand for small $\lambda$ and find
\begin{align}
\label{ex:lag.K.5}
	K_\op(\lambda;z) &= \frac{1}{\sqrt{\pi}\Gamma(\alpha)z} \int\limits_0^\infty \rmd x\,\e{-x} x^{\alpha-1} \int\limits_{-\infty}^\infty \rmd v
	\, \e{-v^2} \\
\notag &\quad \times
	\left\{ 1 + \lambda \left[ \frac{2\kappa v^2}{z^2} +  \frac{2\kappa (\alpha-\frac12)}{z^2+4\kappa x} + \frac{(\alpha-\frac12)(\alpha-\frac32) (1-2v^2)}{2x} \right] +\cdots\right\}~,
\end{align}
where the ellipses denote terms of higher order in $\lambda$. We remark that the expansion of the integrand also contains a term of order $\sqrt{-\lambda}$, but this term is odd in $v$ and, therefore, vanishes upon integration. We can now perform the integral in $v$ in \eqref{ex:lag.K.5}, which gives
\begin{equation}
\label{ex:lag.K.6}
	K_\op(\lambda;z) = \frac1z + \lambda K_\op(z)+ \cdots~,
\end{equation}
with the complexity
\begin{align}
\label{ex:lag.K.7}
	K_\op(z) &= \frac{1}{\Gamma(\alpha)z} \int\limits_0^\infty \rmd x\,\e{-x} x^{\alpha-1} \left[ \frac{\kappa}{z^2} + 
	\frac{2\kappa (\alpha-\frac12)}{z^2+4\kappa x} \right]\\
\notag 
	&= \frac{\kappa}{z^3} \left[1 +(2\alpha-1) \frac{z^2}{4\kappa} \Psi\left(1,2-\alpha; \frac{z^2}{4\kappa}\right) \right]~,  
\end{align}
where $\Psi$ denotes a confluent hypergeometric function, which is related to the incomplete gamma function. 

Instead of using the final result in \eqref{ex:lag.K.7}, it is more useful to take the first line and rewrite it as 
\begin{equation}
\label{ex:lag.K.8}
	K_\op(z) = \frac{\kappa}{z^3}- \frac{(2\alpha-1)}{8\Gamma(\alpha)} \int\limits_0^\infty \rmd x\,
	\e{-x} x^{\alpha-2}  \left[ \frac{1}{z+2i\sqrt{\kappa x}} + \frac{1}{z-2i\sqrt{\kappa x}} -\frac{2}{z} \right]~.
\end{equation}
This allows to perform the inverse Laplace transform using the general formula \eqref{gen:f.Laplace}, 
\begin{equation}
\label{ex:lag.K.9}
	K_\op(t) = \frac{\kappa}{2}t^2 + \frac{(\alpha-\frac12)}{\Gamma(\alpha)} \int\limits_0^\infty \rmd x\,
	\e{-x} x^{\alpha-2} \sin^2(\sqrt{\kappa x}t)~.
\end{equation}
The integral can be done using \cite[3.952.8]{Gradshteyn}, which yields the final result
\begin{align}
\label{ex:lag.K.10}
	 K_\op(t) &= \frac{\kappa}{2}t^2 + \frac{(2\alpha-1)}{4(\alpha-1)} \left[1- \e{\kappa t^2} \Phi\left(\frac32-\alpha,\frac12,\kappa t^2\right)\right]\\
	 &= \frac{\kappa}{2}t^2 + \frac{(2\alpha-1)}{4(\alpha-1)} \left[1- \Phi\left(\alpha-1,\frac12,-\kappa t^2\right)\right]~.
\end{align}

\begin{figure}[t]
	\includegraphics[width=0.48\textwidth]{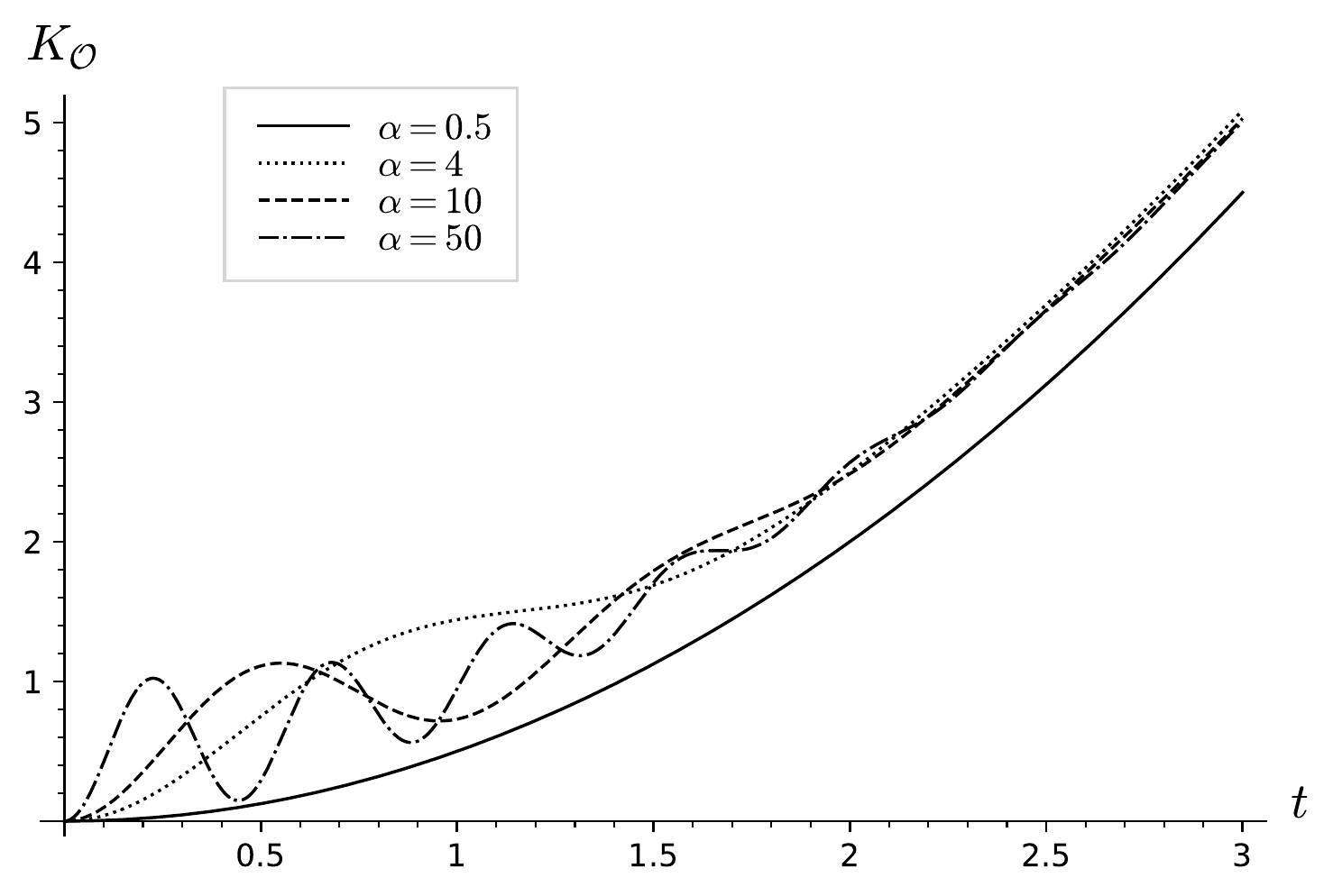}
	\hfill
	\includegraphics[width=0.48\textwidth]{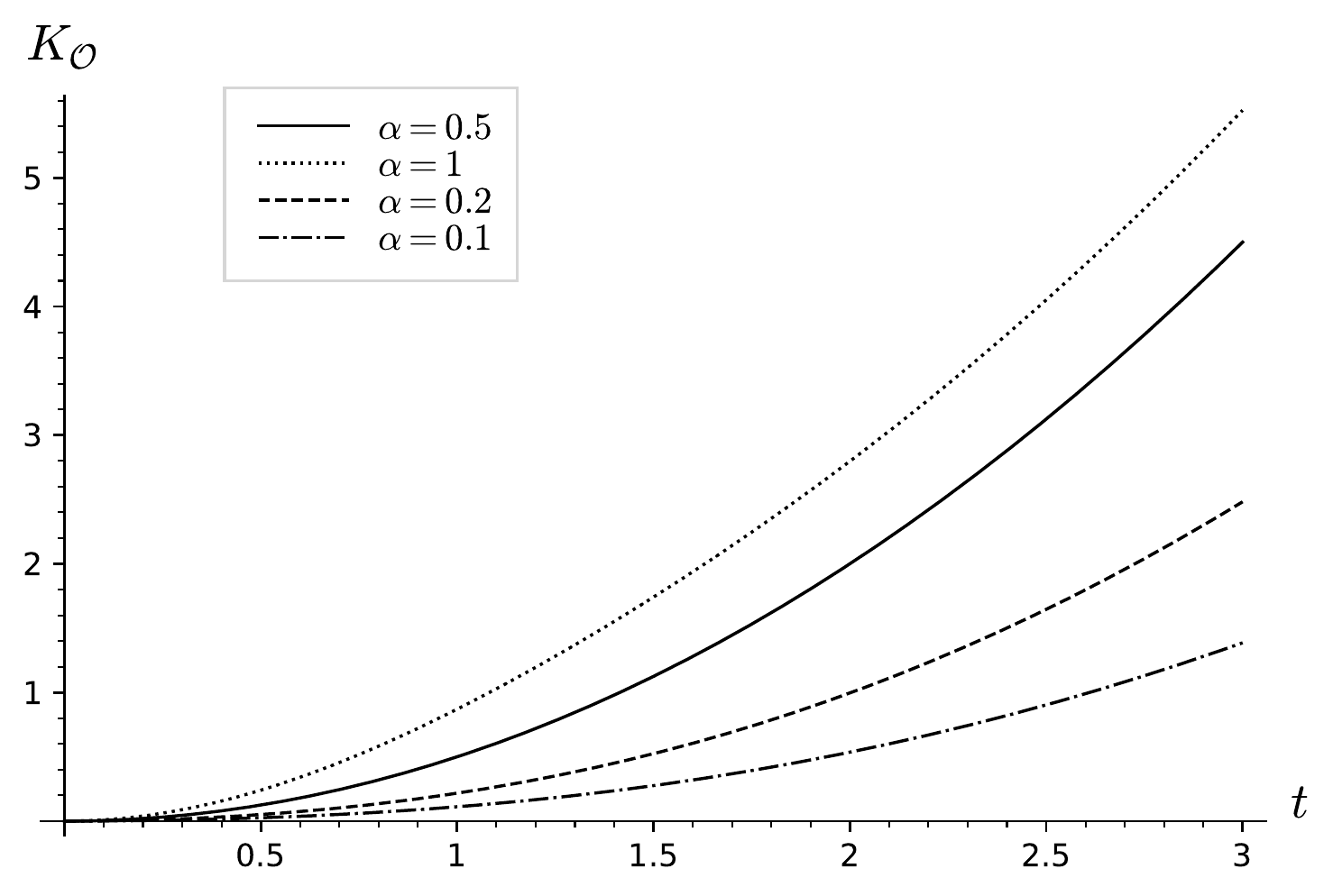}
\caption{Complexity \eqref{ex:lag.K.11} as a function of time for  various values of $\alpha$ ($\kappa=1$). The solid lines correspond to the Hermite polynomial case with $K_\op(t)=\frac12 \kappa t^2$. \label{ex:lag.K.fig}}  
\end{figure}

Formally, one can rewrite \eqref{ex:lag.K.10} as
\begin{equation}
\label{ex:lag.K.11}
	K_\op(t) = \alpha \kappa t^2 -\frac16 \alpha(2\alpha-1) \kappa^2 t^4 \operatorname{{}_2F_2}\left(\alpha+1,1;\frac52,3;-\kappa t^2\right)~,
\end{equation}
where $\operatorname{{}_2F_2}$ denotes a generalized hypergeometric function, expressing the complexity as a series in $t^2$. The special case $\alpha=\frac12$ reproduces the result \eqref{ex:herm.complex} of the Hermite polynomial case.
We show $K_\op(t)$ for a number of different parameters $\alpha$ in figure~\ref{ex:lag.K.fig}. 

In order to obtain the late-time complexity, we use the asymptotic behaviour of the confluent hypergeometric function in \eqref{ex:lag.K.10}. Hence,
\begin{equation}
\label{ex:lag.K.large.t}
	K_\op(t) = \frac12 \kappa t^2 +\frac{(2\alpha-1)}{4(\alpha-1)} \left[1 - \frac{\sqrt{\pi}}{\Gamma(\frac32 -\alpha)} (\kappa t^2)^{1-\alpha} +\cdots\right]~, 
\end{equation}
where the ellipses denote asymptotically vanishing terms. One must distinguish the following three cases:
\begin{equation}
\label{ex:lag.K.large.t1}
	K_\op(t) = \frac12 \kappa t^2 +
	\begin{cases}
		\frac{(2\alpha-1)}{4(\alpha-1)} +\cdots\qquad &(\alpha>1) \\
		\frac14 \ln (\kappa t^2) -\frac14 \psi\left(\frac12\right) +\cdots\qquad &(\alpha=1) \\
		\frac{\sqrt{\pi}(\kappa t^2)^{1-\alpha}}{2(\alpha-1)\Gamma(1/2-\alpha)} 
		- \frac{(2\alpha-1)}{4(1-\alpha)}  +\cdots\qquad &(0<\alpha<1)~.
	\end{cases}
\end{equation}

It is evident from figure~\ref{ex:lag.K.fig} that, for $\alpha<\frac12$, $K_\op(t)<\frac12 \kappa t^2$, whereas $\alpha>\frac12$ leads to $K_\op(t)>\frac12 \kappa t^2$. Moreover, for large enough $\alpha$ the complexity undergoes some oscillations before approaching the asymptotic late-time behaviour.

\section{Examples with polynomials of the Hahn class}
\label{hahn}

\subsection{Charlier polynomials}
\label{hahn:charlier}

Charlier polynomials \cite{Chihara:1978} provide an interesting example of a system with an unbounded discrete spectrum. Allowing for only one parameter, they are the simplest polynomials of the Hahn class. In the Askey scheme \cite{NIST:DLMF}, they are at the same level as the Laguerre polynomials.

The monic Charlier polynomials can be defined by their generating function\footnote{We use the monic polynomials of \cite{Chihara:1978}, which are related to the polynomials in the notation of \cite{NIST:DLMF} by $C_n^{(a)}(x) =(-a)^{-n} C_n(x;a)$.}
\begin{equation}
\label{hahn:ch.ge.func}
	\sum\limits_{n=0}^\infty C_n^{(a)}(x) \frac{w^n}{n!} = \e{-aw} (1+w)^x~.
\end{equation}
Explicit forms are
\begin{equation}
\label{hahn:ch.expl}
	 C_n^{(a)}(x) = n! L_n^{(x-n)}(a) = \sum\limits_{k=0}^n \binom{n}{k}\binom{x}{k} k! (-a)^{n-k} 
	 = (-a)^n \genhypF{2}{0}{-n,-x;-;-\frac1a}~.
\end{equation}
They are orthogonal with respect to the Poisson distribution,
\begin{equation}
\label{hahn:ch.orth}
	\sum\limits_{x=0}^\infty C_n^{(a)}(x) C_m^{(a)}(x) \frac{\e{-a}a^x}{x!} = n!a^n \,\delta_{mn}
\end{equation}
and satisfy the three-term recurrence relation
\begin{equation}
\label{hahn:ch.three}
	x C_n^{(a)}(x) = C_{n+1}^{(a)}(x) + (n+a) C_n^{(a)}(x) +an C_{n-1}^{(a)}(x)~.
\end{equation}

To insert the Charlier polynomials in the context of the recursion method, consider the three-term recurrence relation for the even polynomials \eqref{La:rec.even}. Obviously, \eqref{hahn:ch.three} has the form of \eqref{La:rec.even} with 
\begin{equation}
\label{hahn:ch.Delta}
	x=\Liou^2~,\qquad \Delta_{2n}=n~,\qquad \Delta_{2n+1} =a~.
\end{equation}
For simplicity, we shall not rescale the energy in this example. Hence, we identify
\begin{equation}
\label{hahn:ch.even}
	\bar{P}_n(\Liou^2) = C_n^{(a)}(x)~,\qquad \bar{w}_x = \frac{\e{-a}a^x}{x!}~,\qquad \bar{h}_n=h_{2n} = n!a^n~.
\end{equation}
The easiest way to obtain the odd polynomials is to consider their measure, $\dbar{w}_x=\frac{x}a \bar{w}_x$, which gives 
\begin{equation}
\label{hahn:ch.odd}
	\dbar{P}_n(\Liou^2) = C_n^{(a)}(x-1)~,\qquad \dbar{w}_x = \frac{\e{-a}a^{x-1}}{(x-1)!}~,\qquad h_{2n+1}= a \dbar{h}_n = n!a^{n+1}~.
\end{equation}

In order to calculate the complexity, we shall need the functions of the second kind. Consider first the even polynomials. The functions of the second kind associated with the polynomials \eqref{hahn:ch.even} can be elegantly found by applying the functional transformation \eqref{La:Q.def} directly to the generating function \eqref{hahn:ch.ge.func}. This yields the generating function for the functions of the second kind
\begin{align}
\label{hahn:ch.even.Q.gen.func}
	\bar{Q}(z;w) = \sum\limits_{n=0}^\infty \bar{Q}_n(z) \frac{w^n}{n!} 
	&= \sum\limits_{x=0}^\infty  \bar{w}_x\frac{e^{-aw}(1+w)^x}{z-x} \\
\label{hahn:ch.even.Q.gen.func1}
	&= \e{-(1+w)a} \sum\limits_{x=0}^\infty \frac{[(1+w)a]^x}{(z-x)x!}~.
\end{align} 
After writing 
\begin{equation}
\label{hahn:ch.trick}
	\frac{1}{z-x}= -\frac{(-z)_x}{(-z)(1-z)_x}~,
\end{equation}
\eqref{hahn:ch.even.Q.gen.func1} gives immediately
\begin{align}
\notag
	\bar{Q}(z;w) &= \frac{\e{-(1+w)a}}{z} \genhypF{1}{1}{-z;1-z;a(1+w)}\\
\label{hahn:ch.even.Q}
	&=	\frac1{z} \genhypF{1}{1}{1;1-z;-a(1+w)}~.
\end{align}
Recall that $\genhypF{1}{1}{}$ is just the Kummer confluent hypergeometric function $\Phi()$. 
The functions $\bar{Q}_n(z)$ can by found according to \eqref{hahn:ch.even.Q.gen.func} by differentiating \eqref{hahn:ch.even.Q} with respect to $w$,
\begin{equation}
\label{hahn:ch.even.Qn}
	\bar{Q}_n(z) = -\frac{n!(-a)^{n}}{(-z)_{n+1}} \genhypF{1}{1}{n+1;n+1-z;-a}~.
\end{equation}

Similarly, for the functions of the second kind associated with the the odd polynomials \eqref{hahn:ch.odd}, one obtains
\begin{align}
\label{hahn:ch.odd.Q}
	\dbar{Q}(z;w) &= - \frac{1}{1-z} \genhypF{1}{1}{1;2-z;-a(1+w)}~,\\
\label{hahn:ch.odd.Qn}
	\dbar{Q}_n(z) &= -\frac{n!(-a)^{n}}{(1-z)_{n+1}} \genhypF{1}{1}{1+n;n+2-z;-a}~.
\end{align}

Let us calculate the complexity. Recall from \eqref{La:K.laplace2} that
\begin{equation}
\label{hahn:ch.K.laplace2}
	K_\op(z) = \frac{2i}{z^2} \sum\limits_{n=0}^\infty \frac{(-1)^{n}}{h_n} \int\rmd \mu(\Liou) P_n(\Liou) Q_{n}(iz-\Liou) (\Delta_{n+1}-\Delta_n)~,
\end{equation}
but we also have, from \eqref{La:K.laplace.formal},
\begin{equation}
\label{hahn:ch.K.laplace.formal}
	K_\op(z) = i\sum\limits_{n=0}^\infty \frac{(-1)^nn}{h_n} \int\rmd \mu(\Liou) P_n(\Liou) Q_n(iz-\Liou)~,
\end{equation}
and, from \eqref{La:K.gen.func.laplace3},
\begin{equation}
\label{hahn:ch.K.unity}
	\frac1z = i\sum\limits_{n=0}^\infty \frac{(-1)^n}{h_n} \int\rmd \mu(\Liou) P_n(\Liou) Q_n(iz-\Liou)~.
\end{equation}
After substituting the coefficients \eqref{hahn:ch.Delta} into \eqref{hahn:ch.K.laplace2}, we use \eqref{hahn:ch.K.laplace.formal} and \eqref{hahn:ch.K.unity} to eliminate all the terms involving the even polynomials. This yields
\begin{align}
\notag
	(z^2+1)	K_\op(z) &=\frac{2a}z -i \sum\limits_{n=0}^\infty \frac{1}{h_{2n+1}} \int\rmd \mu(\Liou) P_{2n+1}(\Liou) Q_{2n+1}(iz-\Liou) 
 	(4n-4a+3)\\
\label{hahn:ch.K.only.odd}
 	&= \frac{2a}z -i\frac{a^2}{2} \sum\limits_x \dbar{w}_x  \sum\limits_{n=0}^\infty\frac{4n-4a+3}{h_{2n+1}\sqrt{x}} \dbar{P}_n(x)
    \left[\dbar{Q}_n(y_-) - \dbar{Q}_n(y_+) \right]~, 
\end{align}
where we have substituted the relevant relations from subsection~\ref{La:spectrum} and abbreviated 
\begin{equation}
\label{hahn:ch.y.def}
	y_\mp = (iz\mp \sqrt{x})^2~.
\end{equation} 
First, consider the combination
\begin{align}
\notag 
	(PQ)(x,y) &\equiv \sum\limits_{n=0}^\infty\frac{4n-4a+3}{h_{2n+1}} \dbar{P}_n(x)\dbar{Q}_n(y) \\ 
	&= 
	- \sum\limits_{n=0}^\infty\frac{(4n-4a+3)a^{n-1}}{(1-y)_{n+1}} 
	\genhypF{2}{0}{-n,1-x;-;-\frac1a}\genhypF{1}{1}{1+n;2-y+n;-a}~.
\end{align}
After writing out the generalized hypergeometric series, 
\begin{equation}
\notag 
	(PQ)(x,y) =- \sum\limits_{n=0}^\infty \sum\limits_{k=0}^n \sum\limits_{l=0}^\infty
	\frac{(4n-4a+3)(n-k+1)_k(1-x)_k (1+n)_l (-1)^l a^{n+l-k-1}}{k!l!(1-y)_{n+l+1}}~,
\end{equation}
we rearrange the sums over $n$ and $k$,
\begin{align}
\notag 
	(PQ)(x,y) &=- \sum\limits_{k=0}^\infty \sum\limits_{n=0}^\infty \sum\limits_{l=0}^\infty 
	\frac{(4n+4k-4a+3)(n+k+l)! (1-x)_k (-1)^l a^{n+l-1}}{k!l!n!(1-y)_{n+l+k+1}}~.
\end{align}
A further sum rearrangement introducing $m=n+l$ gives
\begin{align}
\notag
	(PQ)(x,y) &= -\sum\limits_{k=0}^\infty \sum\limits_{m=0}^\infty \binom{m+k}{m} \frac{(1-x)_k a^{m-1}}{(1-y)_{m+k+1}}
	\sum\limits_{l=0}^m \binom{m}{l}(-1)^l [4(m+k-l-a) +3] \\
\label{hahn:ch.PQ1}
	&=  -\sum\limits_{k=0}^\infty \left[ \frac{(4k-4a+3)(1-x)_k}{a(1-y)_{k+1}} +\frac{4(k+1)(1-x)_k}{(1-y)_{k+2}} \right]~,
\end{align}
where, on the second line, we have used the sums
\begin{equation}
\label{hahn:ch.sums}
	\sum\limits_{l=0}^m (-1)^l \binom{m}{l} = \delta_{m,0}~, \qquad \sum\limits_{l=0}^m l (-1)^l \binom{m}{l} = -\delta_{m,1}~.
\end{equation} 
The sum in \eqref{hahn:ch.PQ1} results in
\begin{align}
\notag
	(PQ)(x,y) &= -\frac{4}{a(1-y)} \hypF{2,1-x;2-y;1} +\frac{1+4a}{a(1-y)} \hypF{1,1-x;2-y;1} \\
\notag 
&\quad 
	-\frac{4}{(1-y)(2-y)} \hypF{2,1-x;3-y;1}  \\
\label{hahn:ch.PQ2}
	&=  \frac{4}{x-y+1} +\frac{1-4x}{a(x-y)} +\frac{4(x-1)}{a(x-y-1)}~.
\end{align}

At this point, using $x-y_\pm = z^2 \mp 2iz\sqrt{x}$ from \eqref{hahn:ch.y.def}, we calculate the combination
\begin{equation}
\label{hahn:ch.PQ.comb}
	\frac{(PQ)(x,y_-)-(PQ)(x,y_+)}{\sqrt{x}} = -\frac{4iz}a \left[ \frac{4a}{(z^2+1)^2+4z^2x} + \frac{1-4x}{z^2(z^2+4x)} 
	+ \frac{4(x-1)}{(z^2-1)^2+4z^2x} \right]~, 
\end{equation}
which appears in \eqref{hahn:ch.K.only.odd}. After substituting this result, \eqref{hahn:ch.K.only.odd} becomes 
\begin{align}
\label{hahn:ch.K}
	 K_\op(z) &= \frac{2a}{z(z^2+1)} - 2 \sum\limits_{x=0}^\infty \frac{\e{-a}a^{x}}{(x-1)!} \left\{ \frac{8az}{[(z^2+1)^2+4z^2x](z^2+1)} 
	- \frac{4x-1}{z(z^2+4x)(z^2+1)}\right\}~, 
\end{align}
where the term arising from the third term in the brackets of \eqref{hahn:ch.PQ.comb} has been combined with the first one after shifting the summation index. 

It is possible to carry out the summation in \eqref{hahn:ch.K}, but the result is not very enlightening. It is more instructive to rewrite 
$$  \frac{8z}{[(z^2+1)^2+4z^2x](z^2+1)} =\frac1x \left\{\frac{2}{z(z^2+1)} 
	- \frac{1-\sqrt{\frac{x}{x+1}}}{z[z^2+(\sqrt{x+1}-\sqrt{x})^2]}
	- \frac{1+\sqrt{\frac{x}{x+1}}}{z[z^2+(\sqrt{x+1}+\sqrt{x})^2]} \right\}
$$
and 
$$ \frac{4x-1}{z(z^2+4x)(z^2+1)} = \frac{1}{z(z^2+1)} - \frac{1}{z(z^2+4x)}~,
$$
so that
\begin{align}
\label{hahn:ch.K2}
	 K_\op(z) &= 2 \sum\limits_{x=0}^\infty \frac{\e{-a}a^{x+1}}{x!} \left\{ 
	 \frac{1-\sqrt{\frac{x}{x+1}}}{z[z^2+(\sqrt{x+1}-\sqrt{x})^2]}
	 + \frac{1+\sqrt{\frac{x}{x+1}}}{z[z^2+(\sqrt{x+1}+\sqrt{x})^2]} -\frac{1}{z(z^2+4x+4)}\right\}~. 
\end{align}

Finally, it is straightforward to compute the inverse Laplace transform of \eqref{hahn:ch.K2},
\begin{align}
\label{hahn:ch.Kt1}
	 K_\op(\tau) &= \sum\limits_{x=0}^\infty \frac{\e{-a}a^{x+1}}{(x+1)!} \left[ 
	 \frac{4\sin^2\left(\frac{\sqrt{x+1}-\sqrt{x}}2 \tau\right)}{1-\sqrt{\frac{x}{x+1}}}
	 +\frac{4\sin^2\left(\frac{\sqrt{x+1}+\sqrt{x}}2 \tau\right)}{1+\sqrt{\frac{x}{x+1}}}   
	 - \sin^2 \left(\sqrt{x+1}\tau\right) \right]~. 
\end{align}

\begin{figure}[t]
	\includegraphics[width=0.49\textwidth]{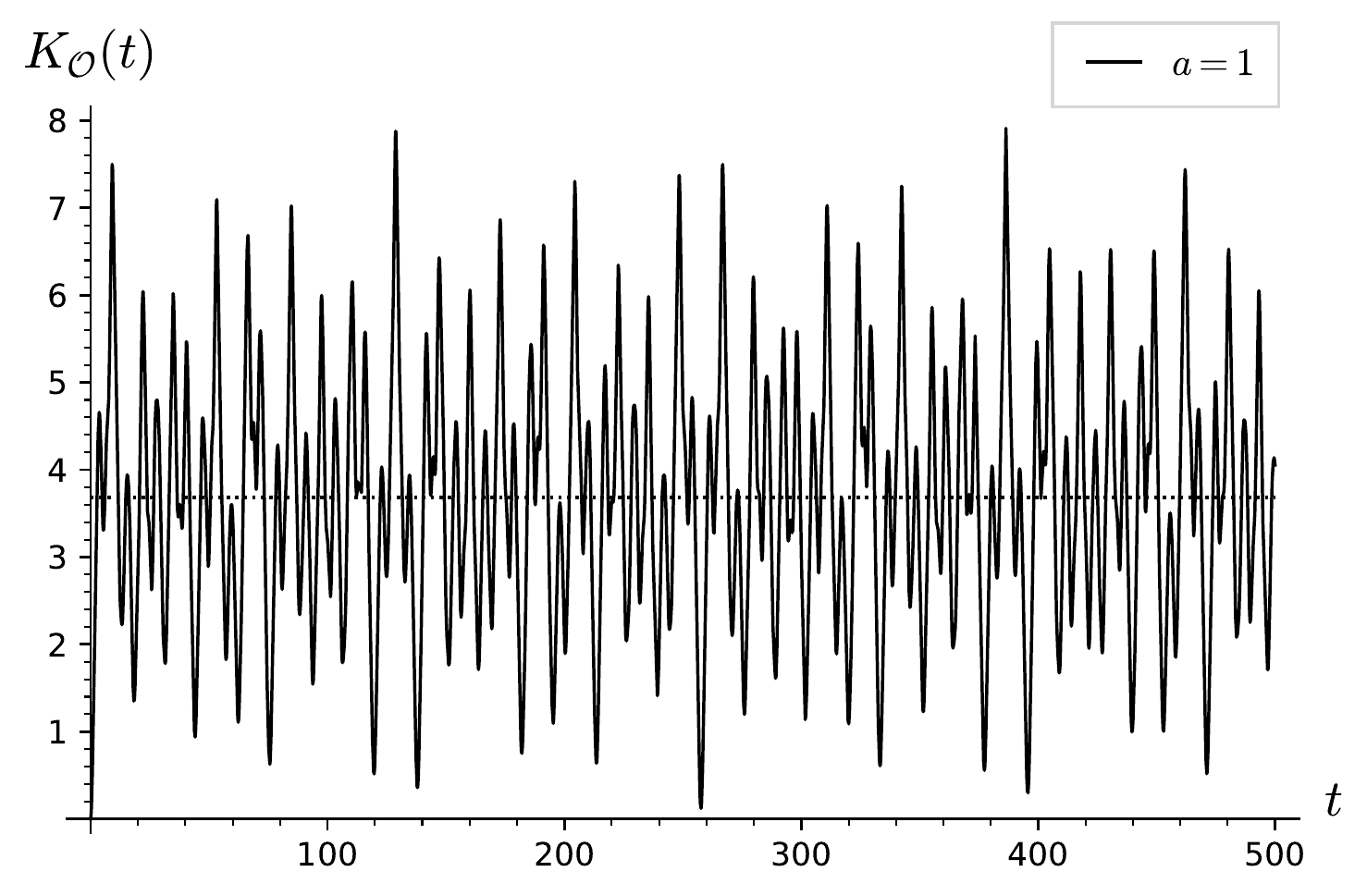}
	\hfill
	\includegraphics[width=0.49\textwidth]{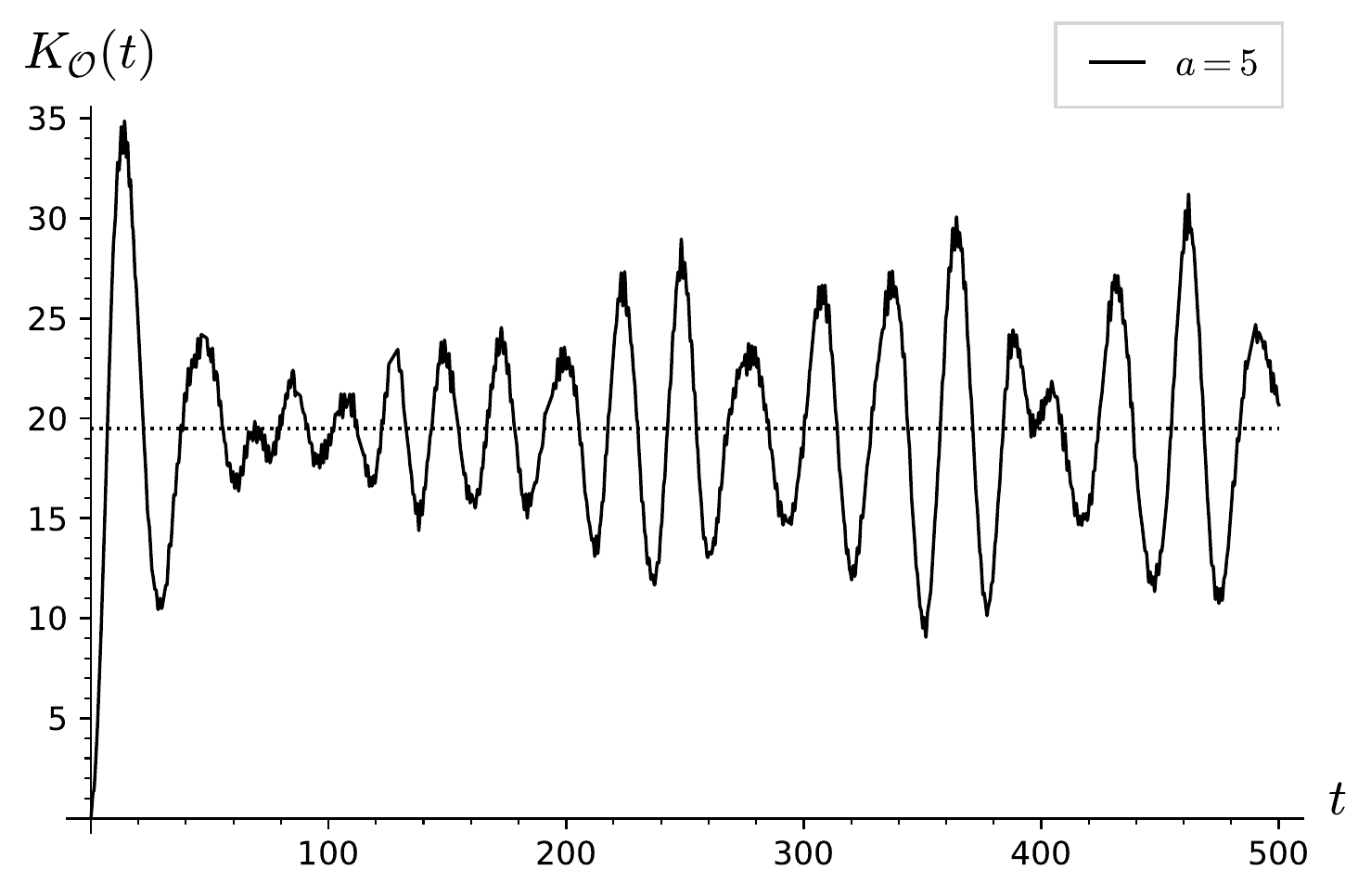}\\
	\includegraphics[width=0.49\textwidth]{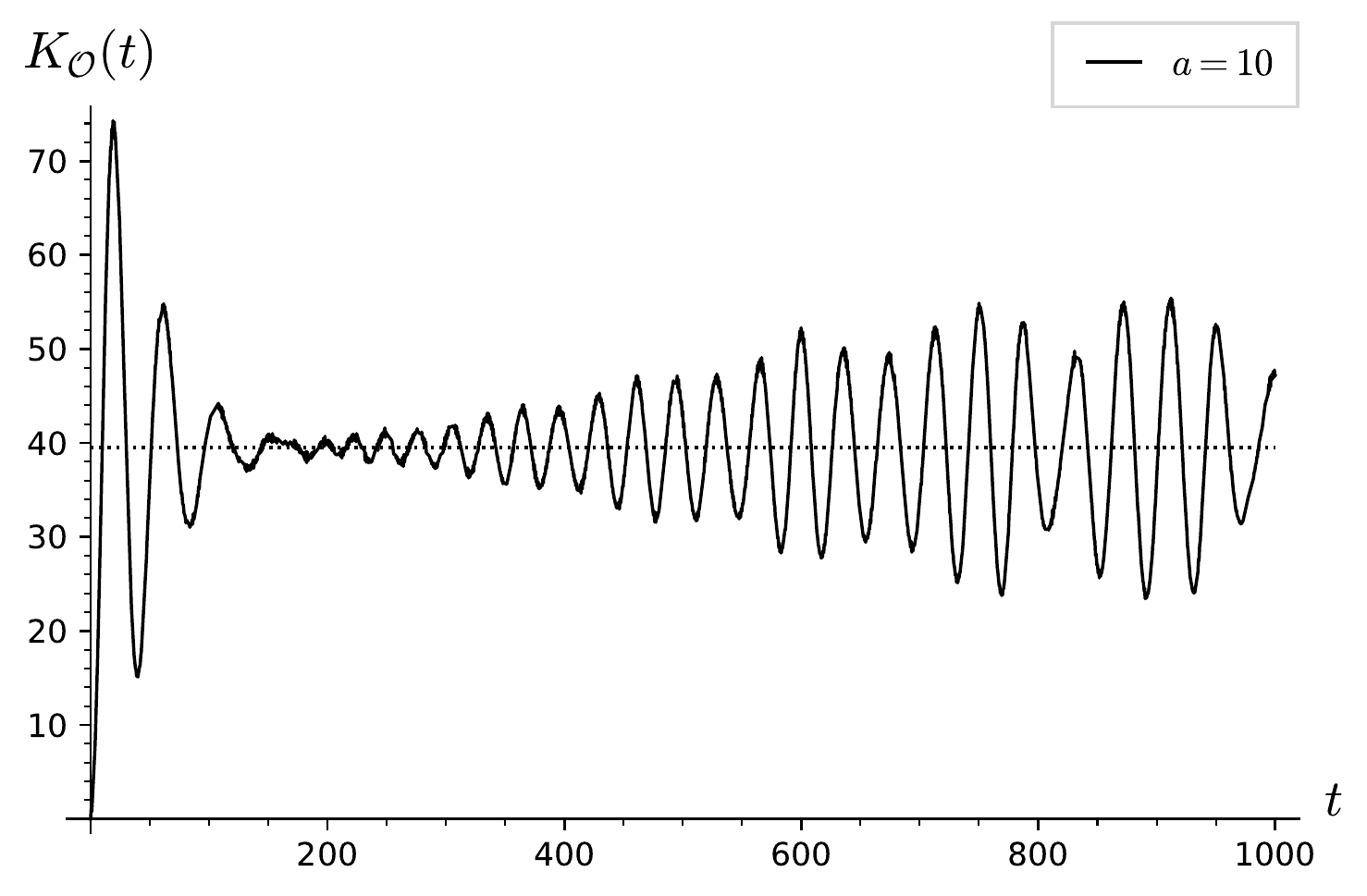}
	\hfill
	\includegraphics[width=0.49\textwidth]{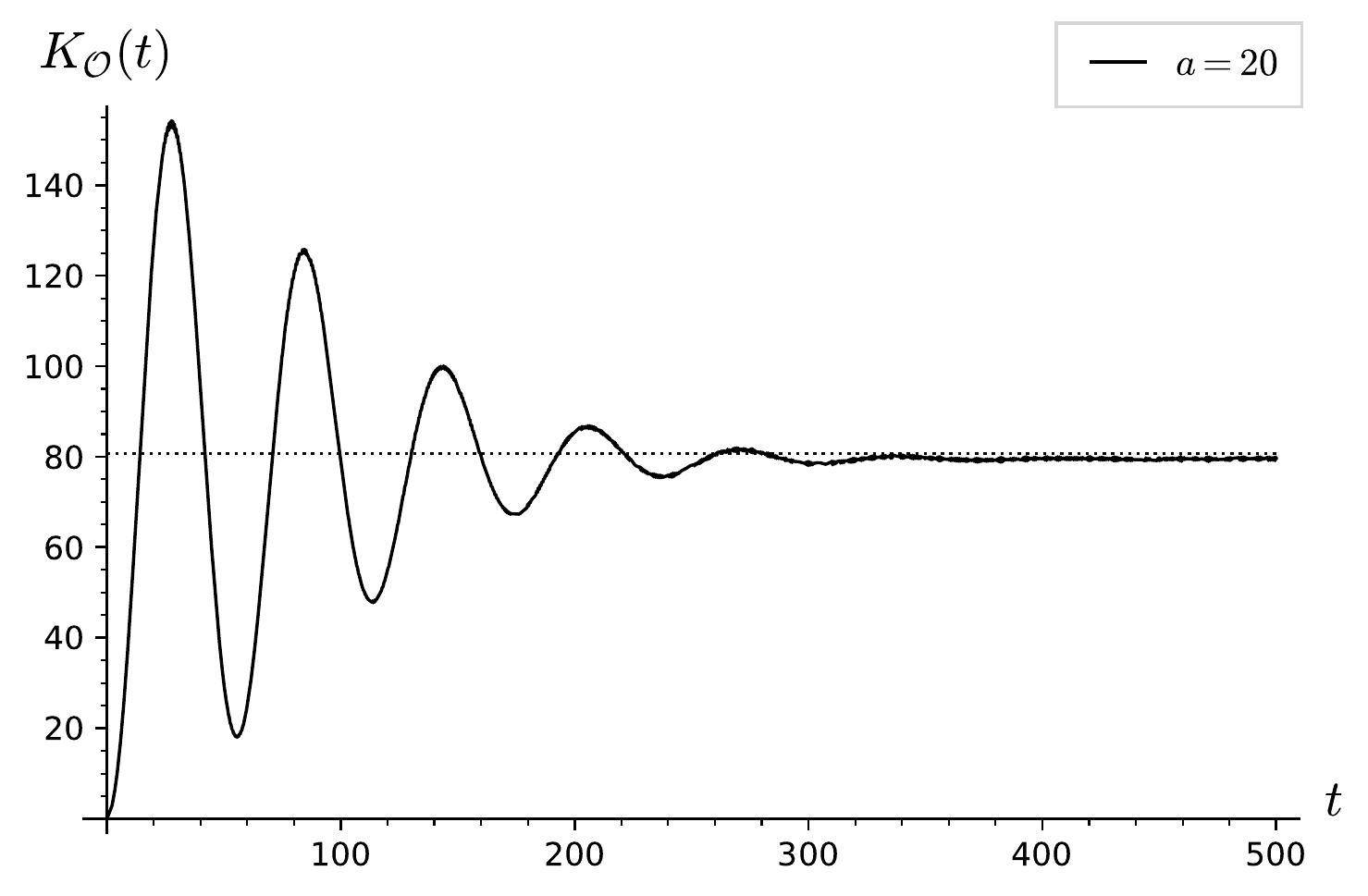}\\
	\includegraphics[width=0.49\textwidth]{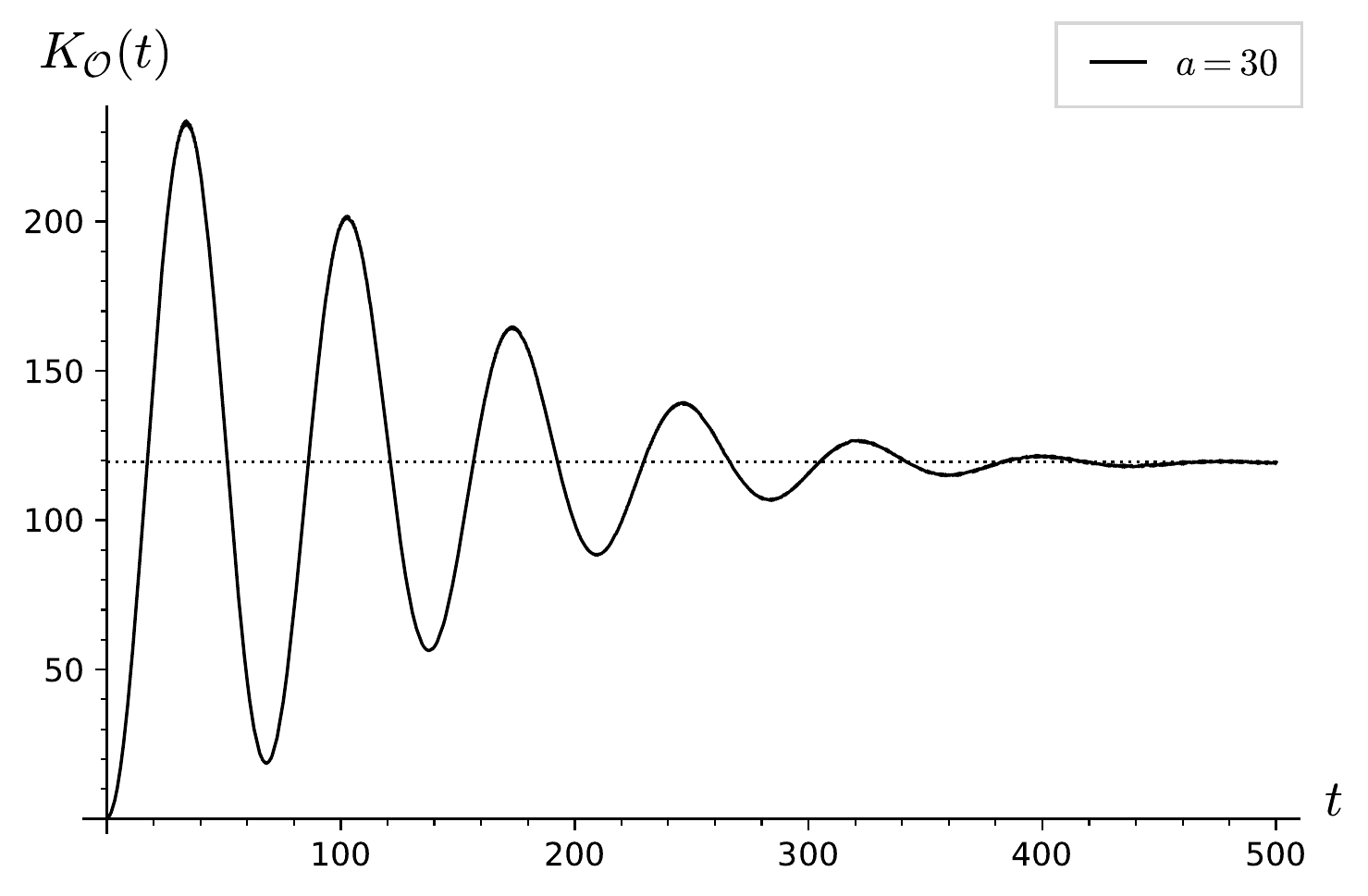}
	\hfill
	\includegraphics[width=0.49\textwidth]{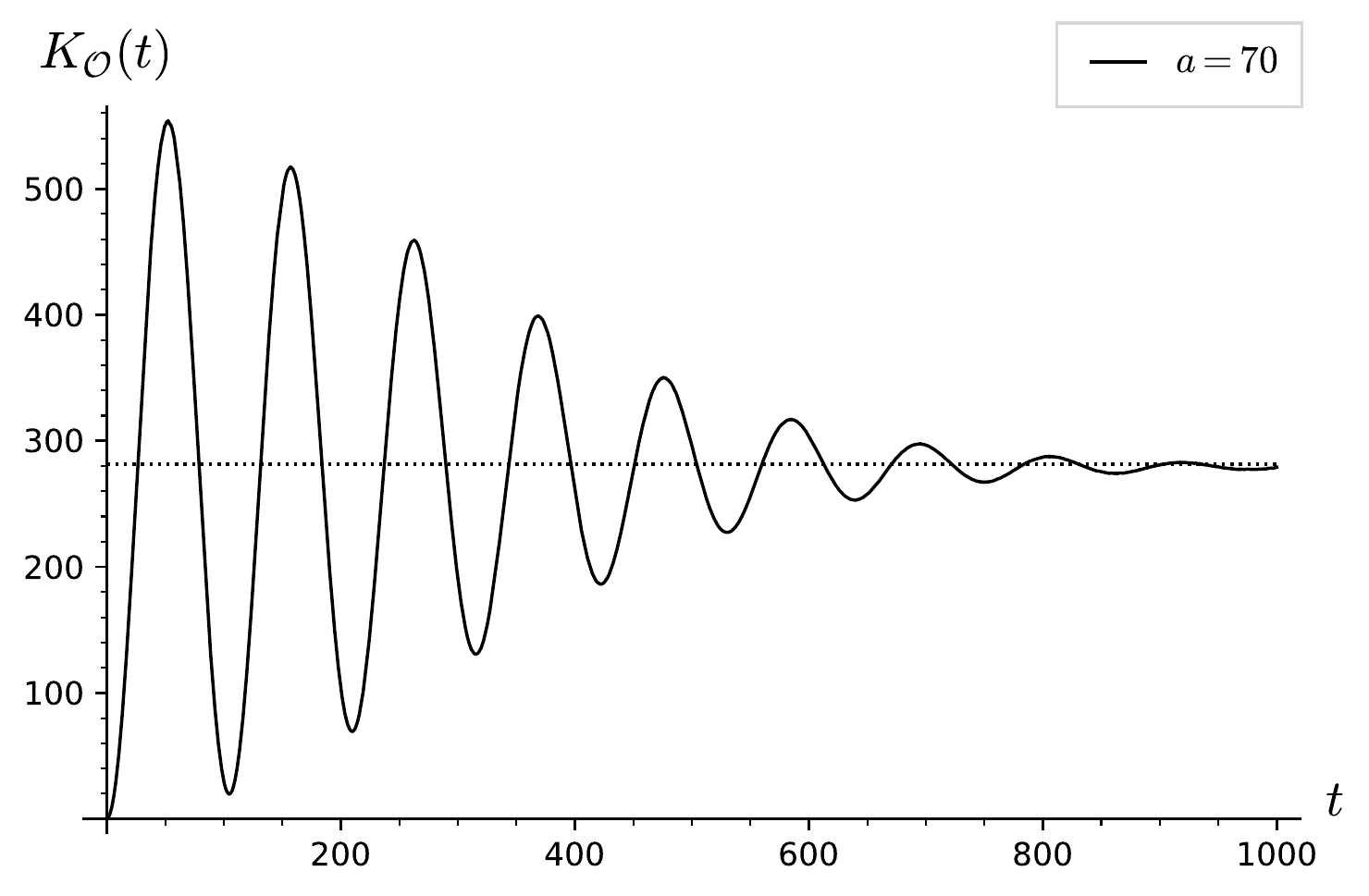}
\caption{Complexity \eqref{hahn:ch.Kt1} as a function of time for a variety of values of $a$. The dashed line represents $K_\op^\ast$ \eqref{hahn:ch.Kast}. Note the different time scales. For $a=20,30,70$, it is likely that the complexity restarts oscillating for larger times than shown. \label{hahn:ch.K.fig}}  
\end{figure}

The complexity \eqref{hahn:ch.Kt1} is shown in figure~\ref{hahn:ch.K.fig} for a number of values of the parameter $a$. 
The behaviour of the complexity at late times is quite surprising. Naively, because the $\Delta_{2n}$ grow linearly and the $\Delta_{2n+1}$ are constant, from the results of subsections \ref{ex:Chebyshev.U} and \eqref{ex:hermite} one would expect a growth between linear and quadratic. In fact, the complexity does not grow at late times, but oscillates more or less erratically (depending on $a$) around the mean value 
\begin{equation}
\label{hahn:ch.Kast}
	K_\op^\ast = 4a - \frac12 \left(1-\e{-a}\right)~,
\end{equation}
which is found from \eqref{hahn:ch.Kt1} by replacing the sine squares by $\frac12$. The same value can be obtained by looking at the coefficient of the $1/z$ term in \eqref{hahn:ch.K2}.

The initial oscillations are easily understood when $a$ is large. In that case, the Poisson weight $\e{-a} a^{x+1}/(x+1)!$ can be approximated by a Gaussian distribution peaked at $x=a-1$. Hence, the dominant oscillation is 
$$ 
	\sin^2 \left(\frac{\sqrt{a}-\sqrt{a-1}}2\tau\right) \approx \sin^2 \frac{\tau}{4\sqrt{a}}~, 
$$
with all the oscillations with nearby frequencies starting in sync at $\tau=0$. As time passes, they get more and more out of sync, which produces the late-time average \eqref{hahn:ch.Kast}. 

A complexity that approaches a constant is characteristic of a finite chain of wave functions. Indeed, the Charlier polynomials can be obtained as a limit from the Krawtchouk polynomials, which have a finite spectrum. The limit is \cite{NIST:DLMF} 
\begin{equation}
\label{hahn:kr.to.ch}
	\lim\limits_{N\to\infty} K_n(x;N^{-1}a,N) = (-a)^n C_n^{(a)}(x)~.
\end{equation}
We will discuss the Krawtchouk polynomials in the next subsection.

\subsection{Krawtchouk polynomials}
\label{hahn:kraw}

The Krawtchouk polynomials \cite{NIST:DLMF} provide an example of a finite spectrum. They are a special case of the Hahn polynomials, and the limit \eqref{hahn:kr.to.ch} gives the Charlier polynomials.   
We shall work with the monic Krawtchouk polynomials, which are given explicitly by\footnote{In the notation of \cite{NIST:DLMF}, the Krawtchouk polynomials are $K_n(x;p,N) = \hypF{-n,-x;-N;p^{-1}}$.}
\begin{equation}
\label{hahn:kr.expl}
	K^{(p,N)}_n(x) = (-p)^n n!\binom{N}{n} \hypF{-n,-x;-N;p^{-1}}~.
\end{equation}
They may be defined via their generating function
\begin{equation}
\label{hahn:kr.genfunc}
	\sum\limits_{n=0}^N K^{(p,N)}_n(x) \frac{w^n}{n!} = [1+(1-pw)]^x (1-pw)^{N-x}
\end{equation}
and are orthogonal with respect to the binomial distribution,
\begin{equation}
\label{hahn:kr.ortho}
	\sum\limits_{x=0}^N K^{(p,N)}_m(x)K^{(p,N)}_n(x) \binom{N}{x} p^x(1-p)^{N-x} = [p(1-p)]^n \binom{N}{n} (n!)^2\,\delta_{mn}~.
\end{equation}
Moreover, they satisfy the three-term recurrence relation
\begin{equation}
\label{hahn:kr.three.term}
	x K_n^{(p,N)}(x) = K_{n+1}^{(p,N)}(x) + [p(N-n) +n(1-p)] K_n^{(p,N)}(x) +p(1-p)n (N-n+1)K_{n-1}^{(p,N)}(x)~.
\end{equation}

We have already met the Krawtchouk polynomials in the special case $p=\frac12$ of the simple finite system \eqref{gen:Kraw.P}. Here, we shall provide a different example. For this purpose, we identify \eqref{hahn:kr.three.term} with \eqref{La:rec.even} by setting  
\begin{equation}
\label{hahn:kr.Deltas}
		x=\Liou^2~,\qquad \Delta_{2n}=n(1-p)~,\qquad \Delta_{2n+1} =p(N-n)
\end{equation}
with $x,n\in \{0,1,\ldots,N\}$. Therefore, for the even polynomials, we have
\begin{equation}
\label{hahn:kr.even}
\begin{aligned}
\bar{P}_n(\Liou^2) &= K_n^{(p,N)}(x)~,\qquad 
	\bar{w}_x = \binom{N}{x} p^x (1-p)^{N-x}~,\\
	\bar{h}_n&=h_{2n} = [p(1-p)]^n \binom{N}{n} (n!)^2~.
\end{aligned}	
\end{equation}
For the odd polynomials, one easily gets from \eqref{La:rec.odd} 
\begin{equation}
\label{hahn:kr.odd}
\begin{aligned}
	\dbar{P}_n(\Liou^2) &= K_n^{(p,N)}(x-1)~,\qquad 
	\dbar{w}_x = \binom{N-1}{x-1} p^{x-1} (1-p)^{N-x}~,\\ 
	h_{2n+1}&= pN \dbar{h}_n = pN [p(1-p)]^n \binom{N-1}{n} (n!)^2~.
\end{aligned}
\end{equation}

The functions of the second kind can be calculated using the generating function \eqref{hahn:kr.genfunc}. 
\begin{align}
\label{hahn:kr.even.Q.gen.func}
	\bar{Q}(z;w) &= \sum\limits_{n=0}^N \bar{Q}_n(z) \frac{w^n}{n!} = \sum\limits_{x=0}^N  \bar{w}_x\frac{[1+(1-p)w]^x(1-pw)^{N-x}}{z-x}~.
\end{align} 
Using \eqref{hahn:ch.trick}, one obtains
\begin{align}
\notag 
	\bar{Q}(z;w) &= \frac{[(1-p)(1-pw)]^N}{z} \hypF{-N,-z;1-z;-\frac{p[1+(1-p)w]}{(1-p)(1-pw)}} \\
\label{hahn:kr.even.Q.gen.func2}
	&= \frac1z \hypF{-N,1;1-z;p+p(1-p)w}~,
\end{align} 
and \eqref{hahn:kr.even.Q.gen.func} gives
\begin{equation}
\label{hahn:kr.even.Qn}
	\bar{Q}_n(z) = -\frac{[p(1-p)]^n(-N)_n n!}{(-z)_{n+1}} \hypF{n-N,n+1;n+1-z;p}~.
\end{equation}

Similarly, for the odd part, one finds
\begin{align}
\label{hahn:kr.odd.Q.gen.func2}
	\dbar{Q}(z;w) &= -\frac1{1-z} \hypF{1-N,1;2-z;p+p(1-p)w}~,\\
\label{hahn:kr.odd.Qn}
	\dbar{Q}_n(z) &= -\frac{[p(1-p)]^n(1-N)_n n!}{(1-z)_{n+1}} \hypF{n+1-N,n+1;n+2-z;p}~.
\end{align} 

We have now all the ingredients that we need to calculate the complexity. We shall proceed as in the previous section, cf.\ \eqref{hahn:ch.K.laplace2}--\eqref{hahn:ch.K.only.odd}. The expression corresponding to \eqref{hahn:ch.K.only.odd} is 
\begin{equation}
\label{hahn:kr.K.only.odd}
 	(z^2+1)	K_\op(z)= \frac{2pN}z -i\frac{(pN)^2}{2} \sum\limits_x \dbar{w}_x \sum\limits_{n=0}^{N-1} \frac{4n-4pN+3-2p}{h_{2n+1}\sqrt{x}} \dbar{P}_n(x)
 	\left[\dbar{Q}_n(y_-) - \dbar{Q}_n(y_+) \right]~,
\end{equation}
where $y_\pm$ are again given by \eqref{hahn:ch.y.def}. 

Consider the combination
\begin{align}
\notag 
	(PQ)(x,y) &\equiv \sum\limits_{n=0}^{N-1}\frac{4n-4pN+3-2p}{h_{2n+1}} \dbar{P}_n(x)\dbar{Q}_n(y) \\ 
	&= 
	- \sum\limits_{n=0}^{N-1}\frac{(4n-4pN+3-2p)(N-n)_n p^n}{pN (1-y)_{n+1}} \\
\notag
	&\quad \times
	\hypF{-n,1-x;1-N;p^{-1}}\hypF{n+1-N, n+1;n+2-y;p}~.
\end{align}
After writing out the generalized hypergeometric series, this becomes
\begin{equation}
\notag 
	(PQ)(x,y) =-\sum\limits_{n=0}^{N-1} \sum\limits_{k=0}^n \sum\limits_{l=0}^{N-1-n} 
	\frac{(4n-4pN+3-2p)(n+l)!(1-x)_k (N-1-k)! (-1)^l p^{n+l-k}}{pN k! l! (n-k)!(1-y)_{n+l+1} (N-1-n-l)!}~.
\end{equation}
We rearrange first the sums over $n$ and $k$,
\begin{align}
\notag 
	(PQ)(x,y) &=-\sum\limits_{k=0}^{N-1} \sum\limits_{n=0}^{N-1-k} \sum\limits_{l=0}^{N-1-n-k} 
	\frac{(n+l+k)!(1-x)_k (N-1-k)! (-1)^l p^{n+l}}{pN k! l! n!(1-y)_{n+l+k+1} (N-1-n-l-k)!}\\
\notag &\quad \times 
	[4(n+k)-4pN+3-2p]~,
\end{align}
and then the sums over $n$ and $l$, setting $m=n+l$,
\begin{align}
\notag 
	(PQ)(x,y) &=-\sum\limits_{k=0}^{N-1} \sum\limits_{m=0}^{N-1-k} 
	\frac{(m+k)!(1-x)_k (N-1-k)! p^m}{pN k! m! (1-y)_{m+k+1} (N-1-m-k)!}\\
\notag
	&\quad \times 	
	\sum\limits_{l=0}^{m} \binom{m}{l} (-1)^l [4(m+k-l)-4pN+3-2p]~.
\end{align}
The sum over $l$ can be performed using \eqref{hahn:ch.sums}, which yields 
\begin{equation}
\notag 
	(PQ)(x,y) =-\frac1{pN}\sum\limits_{k=0}^{N-1} \left[ (4k-4pN+3-2p)\frac{(1-x)_k}{(1-y)_{k+1}} 
	+ 4p(k+1)(N-k-1) \frac{(1-x)_k}{(1-y)_{k+2}}\right]~.
\end{equation}
At this point, the sum over $k$ can be extended to infinity, because $x\in\{1,2,\ldots,N\}$ in \eqref{hahn:kr.K.only.odd}, so that $(1-x)_k=0$ for $k\geq N$. The summation can then be carried out, with the result
\begin{align}
\notag 
	(PQ)(x,y) &=-\frac1{pN} \left[ \frac4{1-y} \hypF{2,1-x;2-y;1} -\frac{4pN+1+2p}{1-y}\hypF{1,1-x;2-y;1} \right.\\
\notag &\quad \left.
	\frac{4p(N+1)}{(1-y)(2-y)} \hypF{2,1-x;3-y;1} -\frac{8p}{(1-y)(2-y)}\hypF{3,1-x;3-y;1} \right]\\
\label{hahn:kr.PQ}
	&= \frac1{pN} \left[ \frac{(1-4x)(1-2p)}{x-y} + \frac{4(1-p)(x-1)}{x-y-1} + \frac{4p(N-x)}{x-y+1} \right]~.
\end{align}
It is a nice check that \eqref{hahn:kr.PQ} reduces to \eqref{hahn:ch.PQ2} in the limit $pN=a, N\to \infty$.

To continue, we use $x-y_\mp = z^2 \pm 2iz\sqrt{x}$ from \eqref{hahn:ch.y.def} and calculate the combination
\begin{equation}
\label{hahn:kr.PQ.comb}
	\frac{(PQ)(x,y_-)-(PQ)(x,y_+)}{\sqrt{x}} = -\frac{4iz}{pN} \left[ \frac{4p(N-x)}{(z^2+1)^2+4z^2x} + \frac{(1-4x)(1-2p)}{z^2(z^2+4x)} 
	+ \frac{4(1-p)(x-1)}{(z^2-1)^2+4z^2x} \right]~. 
\end{equation}

We now return to \eqref{hahn:kr.K.only.odd}, substituting \eqref{hahn:kr.PQ.comb} and shifting $x$ by one in the term arising from the third term in brackets in \eqref{hahn:kr.PQ.comb}. This gives
\begin{align}
\label{hahn:kr.K}
	 K_\op(z) &= \frac{2pN}{z(z^2+1)} - 2 pN \sum\limits_{x} \binom{N-1}{x-1} p^{x-1}(1-p)^{N-x}
\\ \notag &\quad \times 
	\left\{ \frac{8p(N-x)z}{[(z^2+1)^2+4z^2x](z^2+1)} + \frac{(1-4x)(1-2p)}{z(z^2+4x)(z^2+1)}\right\} \\
\label{hahn:kr.K2}
	&= 2 pN \sum\limits_{x=0}^{N-1} \binom{N-1}{x} p^{x}(1-p)^{N-1-x} 
\\ \notag &\quad \times
		\left\{ \frac{(1-p)\left(1+\sqrt{\frac{x}{x+1}}\right)}{z[z^2+(\sqrt{x+1}+\sqrt{x})^2]}
		+ \frac{(1-p)\left(1-\sqrt{\frac{x}{x+1}}\right)}{z[z^2+(\sqrt{x+1}-\sqrt{x})^2]}
		- \frac{1-2p}{z(z^2+4x+4)}\right\}~,
\end{align}
where we have carried out similar steps as in the previous subsection. 
We can readily perform the inverse Laplace transform of \eqref{hahn:kr.K2}. This yields our final result for the complexity,
\begin{align}
\label{hahn:kr.Kt1}
	 K_\op(\tau) &= \sum\limits_{x=0}^{N-1} \binom{N}{x+1} p^{x+1} (1-p)^{N-(x+1)}\\
\notag &\quad \times
	 \left[ 
	 \frac{4(1-p)\sin^2\left(\frac{\sqrt{x+1}-\sqrt{x}}2 \tau\right)}{1-\sqrt{\frac{x}{x+1}}}
	 +\frac{4(1-p)\sin^2\left(\frac{\sqrt{x+1}+\sqrt{x}}2 \tau\right)}{1+\sqrt{\frac{x}{x+1}}}   
	 - (1-2p)\sin^2 \left(\sqrt{x+1}\tau\right) \right]~. 
\end{align}

\begin{figure}[t]
	\includegraphics[width=0.49\textwidth]{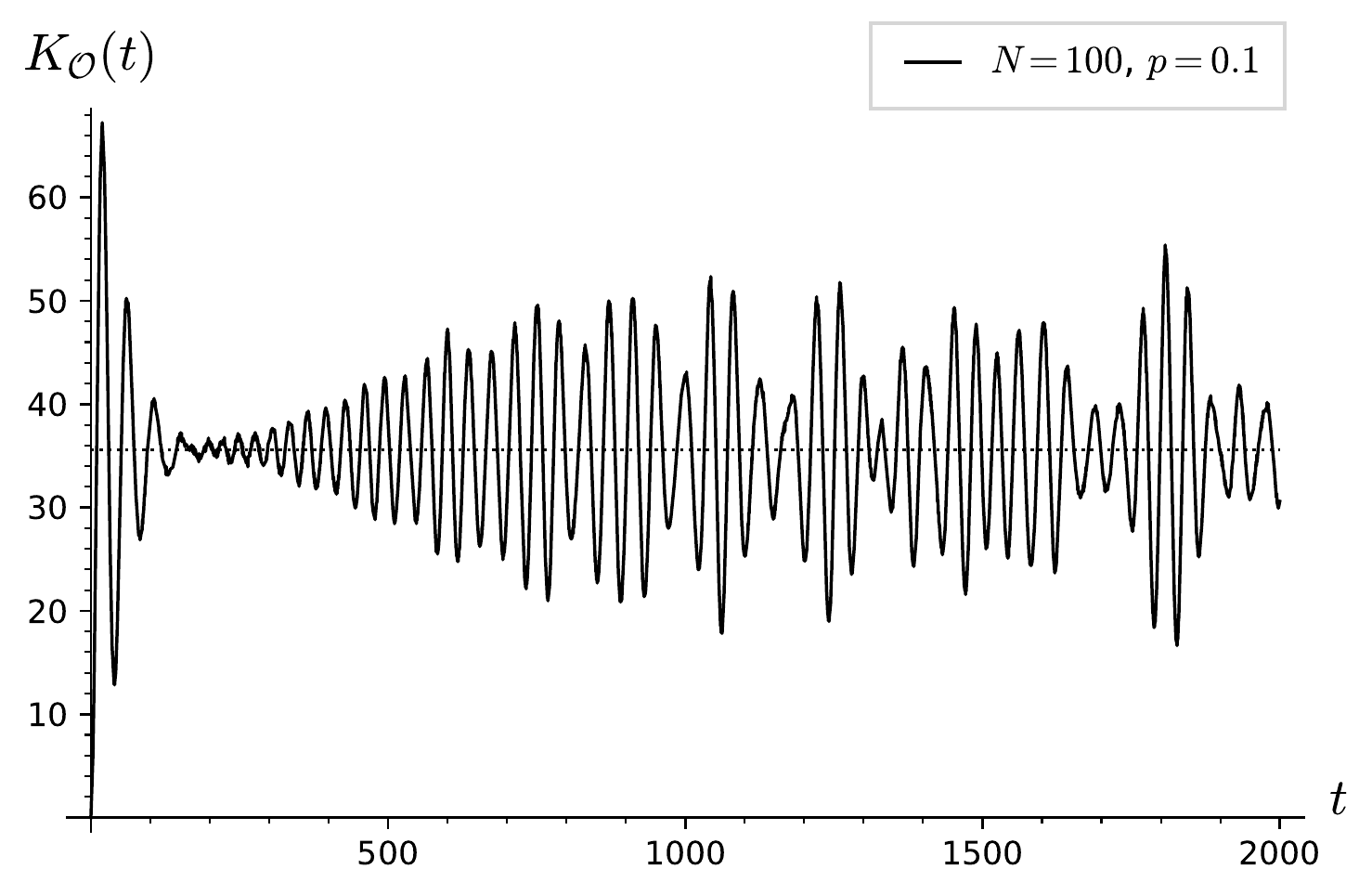}
	\hfill
	\includegraphics[width=0.49\textwidth]{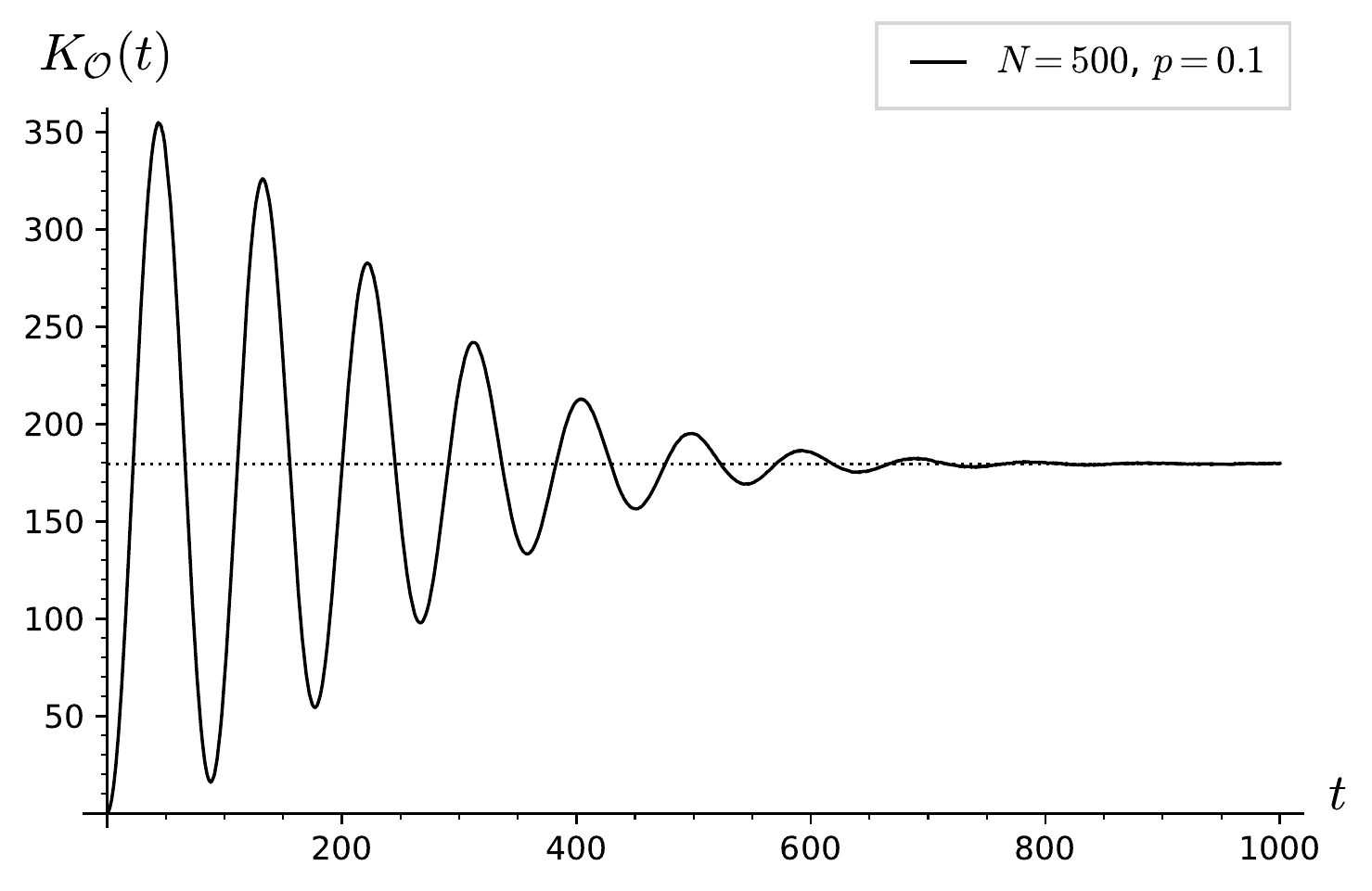}\\
	\includegraphics[width=0.49\textwidth]{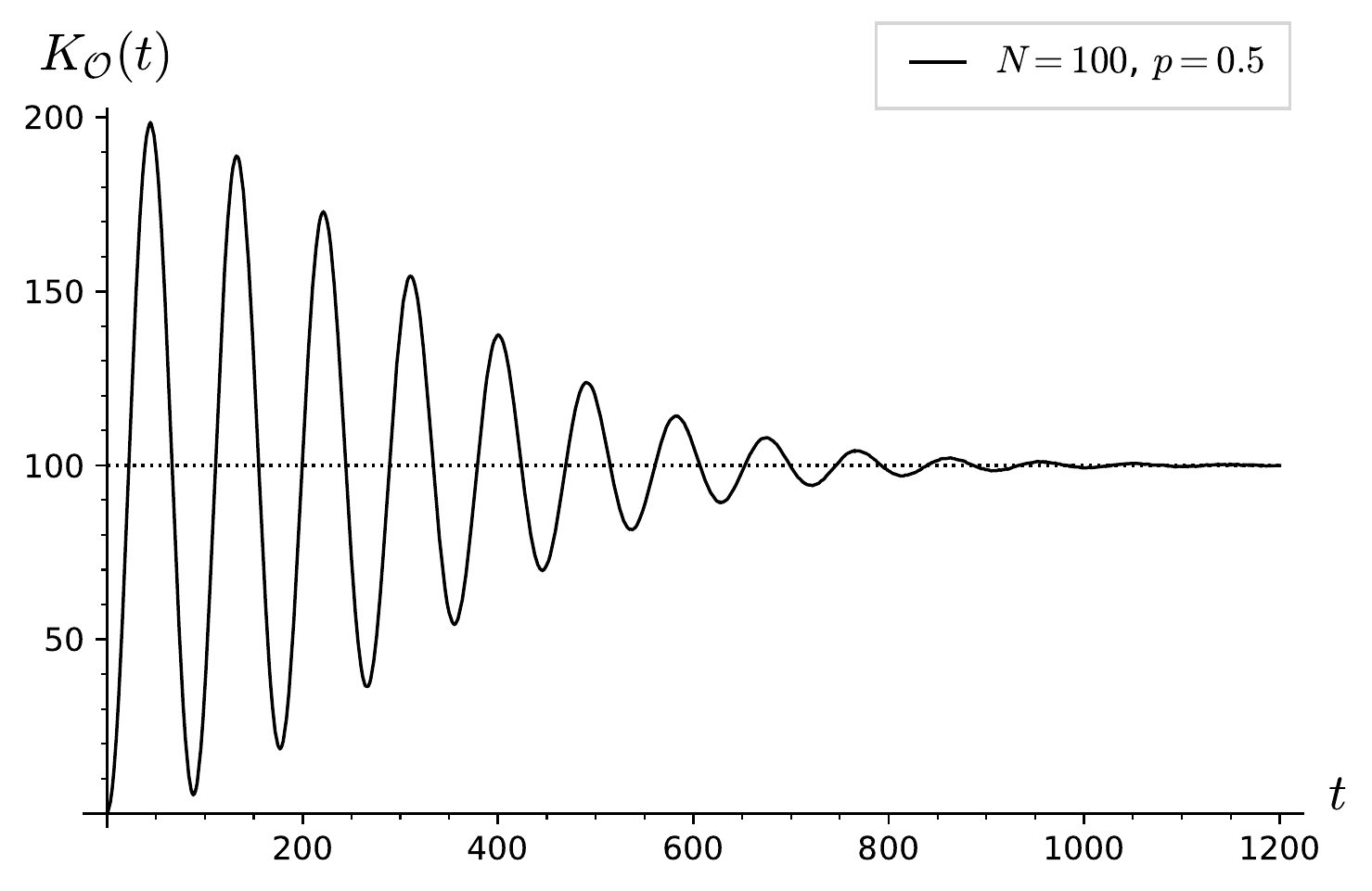}
	\hfill
	\includegraphics[width=0.49\textwidth]{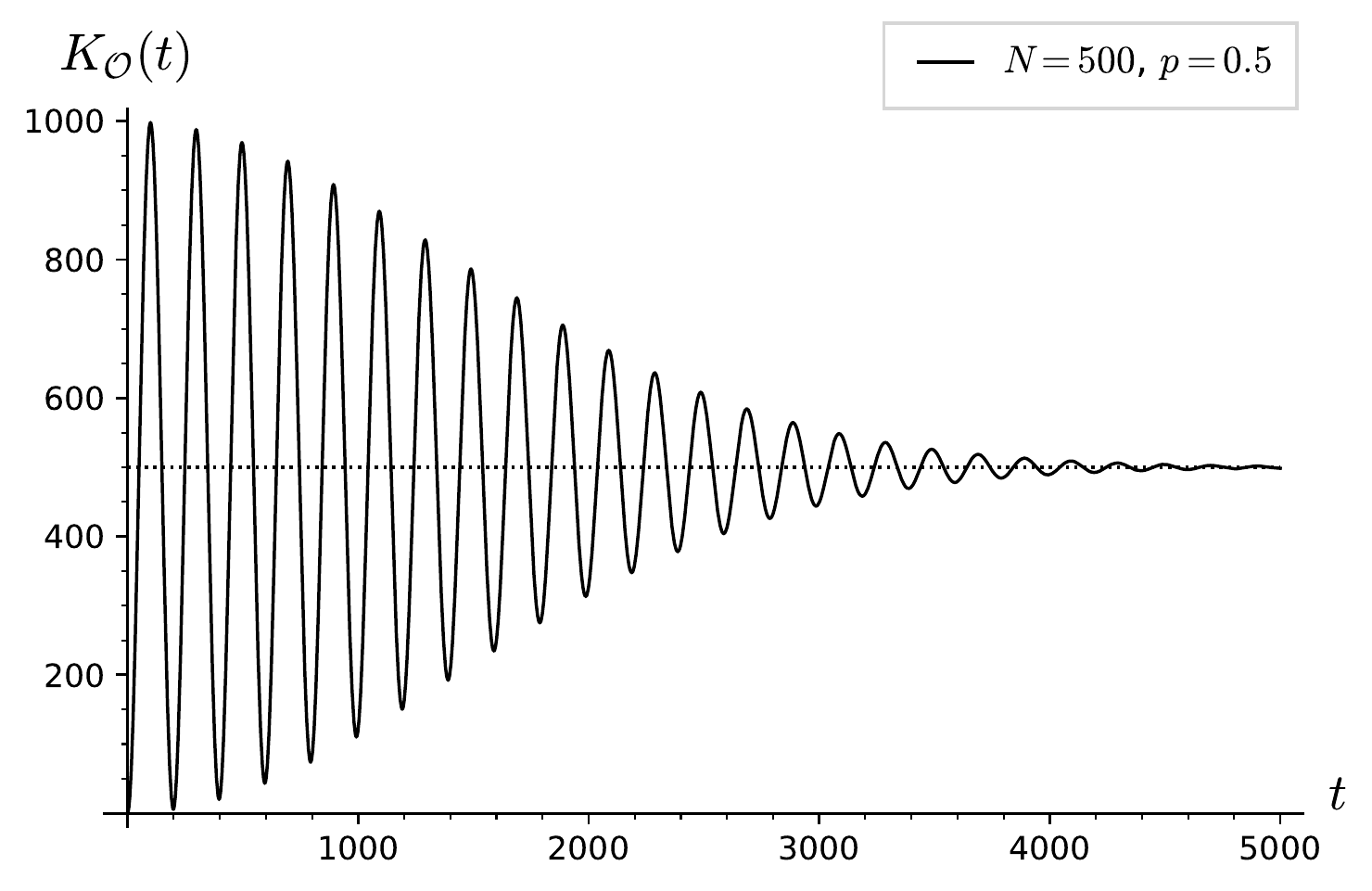}\\
	\includegraphics[width=0.49\textwidth]{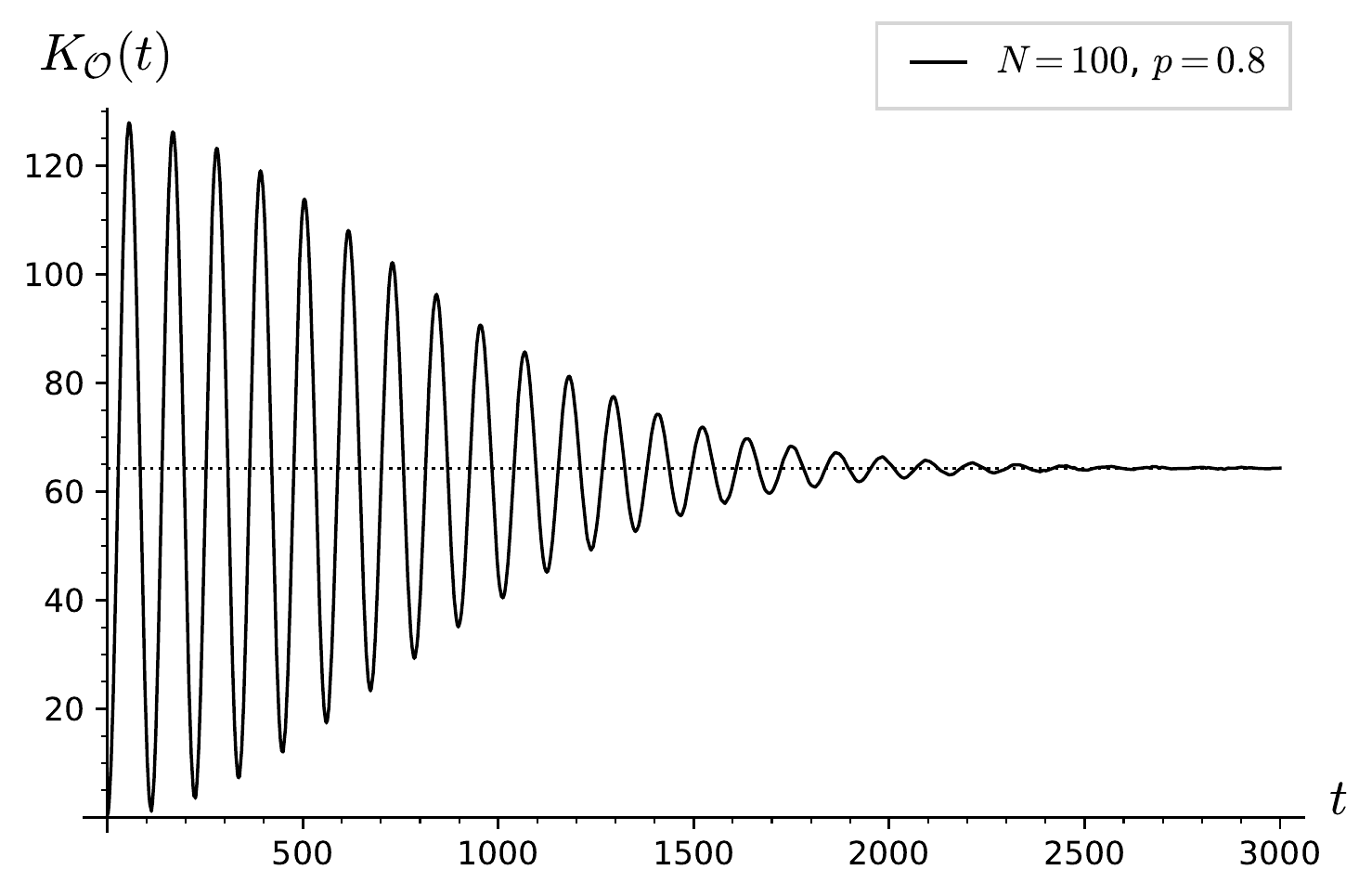}
	\hfill
	\includegraphics[width=0.49\textwidth]{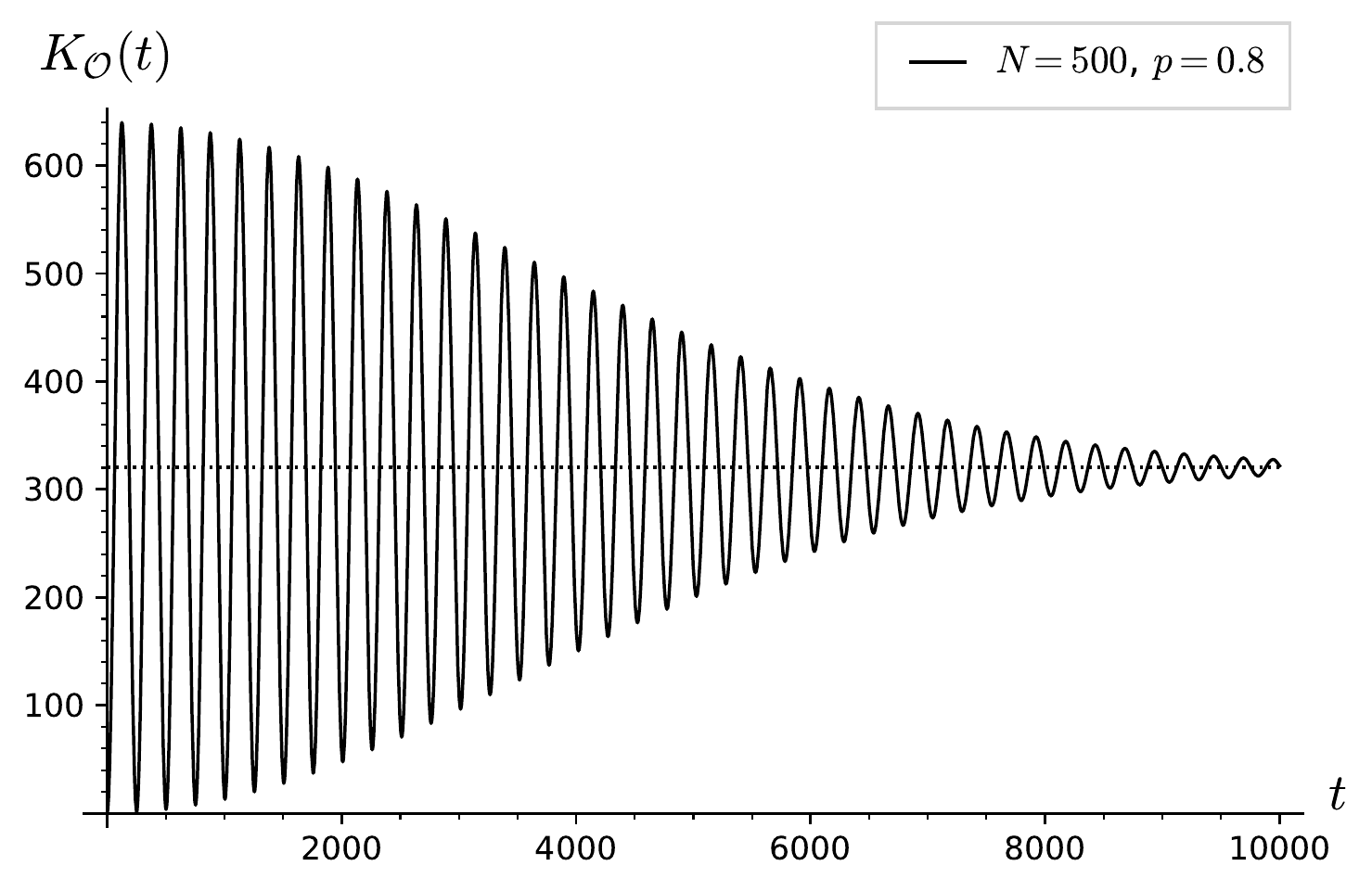}
\caption{Complexity \eqref{hahn:kr.Kt1} as a function of time for various combinations of $p$ and $N$. The dashed line represents the value $K_\op^\ast$ \eqref{hahn:kr.Kmean}. Note the different time scales. It is interesting to note that, for $p=0.5$, the first maximum nearly reaches the end of the chain ($2N$). \label{hahn:kr.K.fig}}  
\end{figure}

At late times, the complexity oscillates around the mean value 
\begin{equation}
\label{hahn:kr.Kmean}
	K_\op^\ast = 4Np(1-p) -\frac12 (1-2p)[ 1-(1-p)^N]~,
\end{equation}
which is found by setting the sine squares in \eqref{hahn:kr.Kt1} to $\frac12$. We illustrate the complexity for various values of $N$ and $p$ in figure~\ref{hahn:kr.K.fig}. It can be seen that the complexity is qualitatively quite similar to the case with the Charlier polynomials discussed in the previous subsection.

\subsection{Meixner polynomials}
\label{hahn:meix}

The Meixner polynomials \cite{NIST:DLMF} provide another example of an unbounded discrete spectrum. In the Askey scheme, possessing two parameters, they are at the same level as the Krawtchouk polynomials. Therefore, they are a special case of the Hahn polynomials, and reduce to the Charlier polynomials in a certain limit. 
   
As before, we shall work with the monic polynomials, which are given explicitly by\footnote{In the notation of \cite{NIST:DLMF}, the Meixner polynomials are $M_n(x;\beta,c) = \hypF{-n,-x;\beta;1-c^{-1}}$.}
\begin{equation}
\label{hahn:mx.expl}
	M^{(\beta,c)}_n(x) = (\beta)_n \left(\frac{-c}{1-c}\right)^n \hypF{-n,-x;\beta;1-c^{-1}}~,
\end{equation}
with $\beta>0$ and $0<c<1$. 

They may be defined via their generating function
\begin{equation}
\label{hahn:mx.genfunc}
	\sum\limits_{n=0}^N M^{(\beta,c)}_n(x) \frac{w^n}{n!} = \left(1+\frac{w}{1-c}\right)^x \left(1+\frac{cw}{1-c}\right)^{-x-\beta}
\end{equation}
and are orthogonal with respect to
\begin{equation}
\label{hahn:mx.ortho}
	\sum\limits_{x=0}^N M^{(\beta,c)}_m(x)M^{(\beta,c)}_n(x) \frac{(1-c)^\beta(\beta)_xc^x}{x!} 
	= n! (\beta)_n c^n (1-c)^{-2n} \,\delta_{mn}~.
\end{equation}
Moreover, they satisfy the three-term recurrence relation
\begin{equation}
\label{hahn:mx.three.term}
	x M_n^{(\beta,c)}(x) = M_{n+1}^{(\beta,c)}(x) + \frac{c(n+\beta)+ n}{1-c} M_n^{(\beta,c)}(x) +\frac{nc(n-1+\beta)}{(1-c)^2} M_{n-1}^{(\beta,c)}(x)~.
\end{equation}
The Meixner polynomials reduce to the Charlier polynomials in the the limit
\begin{equation}
\label{hahn:mx:mx.to.ch}
	\lim\limits_{\beta\to \infty} M^{(\beta, \frac{a}{a+\beta})}_n(x) = C_n^{(a)}(x)~. 
\end{equation}

In order to use the Meixner polynomials in the context of Krylov complexity, we identify \eqref{hahn:mx.three.term} with the recurrence relation of the even polynomials \eqref{La:rec.even} by setting  
\begin{equation}
\label{hahn:mx.Deltas}
		x=\Liou^2~,\qquad \Delta_{2n}=\frac{n}{1-c}~,\qquad \Delta_{2n+1} =\frac{c(n+\beta)}{1-c}
\end{equation}
with $x,n\in \{0,1,2\ldots\}$. Therefore, for the even polynomials, one has
\begin{equation}
\label{hahn:mx.even}
\begin{aligned}
\bar{P}_n(\Liou^2) &= M_n^{(\beta,c)}(x)~,\qquad 
	\bar{w}_x = \frac{(1-c)^\beta(\beta)_xc^x}{x!}~,\\
	\bar{h}_n&=h_{2n} = n! (\beta)_n c^n (1-c)^{-2n}~.
\end{aligned}	
\end{equation}
For the odd polynomials, one gets from \eqref{La:rec.odd} 
\begin{equation}
\label{hahn:mx.odd}
\begin{aligned}
	\dbar{P}_n(\Liou^2) &= M_n^{(\beta+1,c)}(x-1)~,\qquad 
	\dbar{w}_x = \frac{(1-c)^{\beta+1}(\beta+1)_{x-1}c^{x-1}}{(x-1)!}~,\\ 
	h_{2n+1}&= \Delta_1 \dbar{h}_n = n! (\beta)_{n+1} c^{n+1} (1-c)^{-2n-1}~.
\end{aligned}
\end{equation}

In order to find the functions of the second kind for the even part, we consider again the generating function
\begin{align}
\label{hahn:mx.even.Q.gen.func}
	\bar{Q}(z;w) &= \sum\limits_{n=0}^N \bar{Q}_n(z) \frac{w^n}{n!} 
	= \sum\limits_{x=0}^N  \bar{w}_x \frac{\left(1+\frac{w}{1-c}\right)^x\left(1+\frac{cw}{1-c}\right)^{-x-\beta}}{z-x}~.
\end{align} 
Using \eqref{hahn:ch.trick}, one finds
\begin{align}
\notag 
	\bar{Q}(z;w) &= \frac{(1-c)^{2\beta}}{z(1-c+cw)^\beta} \hypF{\beta,-z;1-z;\frac{c(1-c+w)}{1-c-cw)}} \\
\label{hahn:mx.even.Q.gen.func2}
	&= \frac1z \hypF{\beta,1;1-z;-\frac{c(1-c+w)}{(1-c)^2}}~,
\end{align} 
from which one gets the functions of the second kind
\begin{equation}
\label{hahn:mx.even.Qn}
	\bar{Q}_n(z) = -\frac{(\beta)_n n!\, (-c)^n}{(1-c)^{2n} (-z)_{n+1}} \hypF{\beta+n,n+1;n+1-z;-\frac{c}{1-c}}~.
\end{equation}

Similarly, for the odd part, one has
\begin{align}
\label{hahn:mx.odd.Q.gen.func2}
	\dbar{Q}(z;w) &= -\frac1{1-z} \hypF{1+\beta,1;2-z;-\frac{c(1-c+w)}{(1-c)^2}}~,\\
\label{hahn:mx.odd.Qn}
	\dbar{Q}_n(z) &= -\frac{(\beta+1)_n n!\, (-c)^n}{(1-c)^{2n} (1-z)_{n+1}} \hypF{\beta+n+1,n+1;n+2-z;-\frac{c}{1-c}}~.
\end{align} 

To calcultate the complexity, we proceed again as in subsection~\ref{hahn:charlier}, cf.\ \eqref{hahn:ch.K.laplace2}--\eqref{hahn:ch.K.only.odd}, with the result  
\begin{equation}
\label{hahn:mx.K.only.odd}
 	(z^2+1)	K_\op(z)= \frac{2c\beta}{(1-c)z} -\frac{i}{2} \left( \frac{c\beta}{1-c}\right)^2 \sum\limits_x \dbar{w}_x \sum\limits_{n=0}^{\infty} \frac{4n+1 -\frac{4c\beta-2}{1-c}}{h_{2n+1}\sqrt{x}} \dbar{P}_n(x)
 	\left[\dbar{Q}_n(y_-) - \dbar{Q}_n(y_+) \right]~,
\end{equation}
where $y_\pm$ are again given by \eqref{hahn:ch.y.def}. 

Consider the combination
\begin{align}
\notag 
	(PQ)(x,y) &\equiv \sum\limits_{n=0}^{\infty}\frac{4n+1-\frac{4c\beta-2}{1-c}}{h_{2n+1}} \dbar{P}_n(x)\dbar{Q}_n(y) \\ 
	&= 
	- \sum\limits_{n=0}^{\infty}\frac{\left(4n+1-\frac{4c\beta-2}{1-c}\right)(\beta+1)_n}{\beta(1-y)_{n+1}} 
	\left(\frac{c}{1-c}\right)^ {n-1}\\
\notag
	&\quad \times
	\hypF{-n,1-x;\beta+1;\frac{c-1}{c}}\hypF{n+1+\beta, n+1;n+2-y;-\frac{c}{1-c}}~.
\end{align}
The next steps are analogous to the calculations in the previous subsections. We write out the generalized hypergeometric series,
\begin{align}
\notag 
	(PQ)(x,y) &=-\sum\limits_{n=0}^{\infty} \sum\limits_{k=0}^n \sum\limits_{l=0}^{\infty} 
	\frac{(n+l)!(1-x)_k (\beta+1+k)_{n+l-k} (-1)^l}{\beta k! l! (n-k)!(1-y)_{n+l+1}}\\
\notag &\quad \times
	\left(4n+4k+1-\frac{4c\beta-2}{1-c}\right) \left(\frac{c}{1-c}\right)^{n+l-k-1}	
\end{align}
and rearrange first the sums over $n$ and $k$,
\begin{align}
\notag 
	(PQ)(x,y) &=-\sum\limits_{k=0}^{\infty} \sum\limits_{n=0}^{\infty} \sum\limits_{l=0}^{\infty} 
	\frac{(n+l+k)!(1-x)_k (\beta+1+k)_{n+l} (-1)^l}{\beta k! l! n!(1-y)_{n+l+k+1}} \\
\notag &\quad \times
	\left(4n+4k+1-\frac{4c\beta-2}{1-c}\right) \left(\frac{c}{1-c}\right)^{n+l-1}~.
\end{align}
Rearranging also the sums over $n$ and $l$, with $m=n+l$, yields
\begin{align}
\notag 
	(PQ)(x,y) &=-\sum\limits_{k=0}^{\infty} \sum\limits_{m=0}^{\infty} \binom{m+k}{m}
	\frac{(1-x)_k (\beta+1+k)_m}{\beta (1-y)_{m+k+1}} \left(\frac{c}{1-c}\right)^{m-1}\\
\notag
	&\quad \times 	
	\sum\limits_{l=0}^{m} \binom{m}{l} (-1)^l \left[4(m+k-l)+1 -\frac{4c\beta-2}{1-c}\right]~.
\end{align}
The sum over $l$ can be performed using \eqref{hahn:ch.sums}, so that we get
\begin{equation}
\notag 
	(PQ)(x,y) =-\sum\limits_{k=0}^{\infty} \left[\left(4k+1 -\frac{4c\beta-2}{1-c}\right) \frac{(1-c)(1-x)_k}{\beta c(1-y)_{k+1}} 
	+ \frac{4(k+1)(\beta+k+1)(1-x)_k}{\beta (1-y)_{k+2}}\right]~.
\end{equation}
Carrying out the remaining summation, we obtain
\begin{align}
\notag 
	(PQ)(x,y) &= -\frac{4(1-c)}{\beta c (1-y)} \hypF{2,1-x;2-y;1} +\frac{1-3c+4\beta c}{\beta c(1-y)}\hypF{1,1-x;2-y;1} \\
\notag &\quad 
	-\frac{4(\beta-1)}{\beta(1-y)(2-y)} \hypF{2,1-x;3-y;1} -\frac{8}{\beta(1-y)(2-y)}\hypF{3,1-x;3-y;1} \\
\label{hahn:mx.PQ}
	&= \frac1{\beta} \left[ \frac{(1-4x)(1+c)}{c(x-y)} + \frac{4(x-1)}{c(x-y-1)} + \frac{4(x+\beta)}{x-y+1} \right]~.
\end{align}
One can verify that \eqref{hahn:mx.PQ} reduces to \eqref{hahn:ch.PQ2} in the limit \eqref{hahn:mx:mx.to.ch}.

To continue, we use $x-y_\mp = z^2 \pm 2iz\sqrt{x}$ from \eqref{hahn:ch.y.def} and calculate the combination
\begin{equation}
\label{hahn:mx.PQ.comb}
	\frac{(PQ)(x,y_-)-(PQ)(x,y_+)}{\sqrt{x}} = -\frac{4iz}{\beta} \left[ \frac{4(x+\beta)}{(z^2+1)^2+4z^2x} + \frac{(1-4x)(1+c^{-1})}{z^2(z^2+4x)} 
	+ \frac{4c^{-1}(x-1)}{(z^2-1)^2+4z^2x} \right]~. 
\end{equation}

Returning to \eqref{hahn:mx.K.only.odd}, we substitute \eqref{hahn:mx.PQ.comb}. Shifting also $x$ by one in the term arising from the third term in brackets in \eqref{hahn:mx.PQ.comb}, we find
\begin{align}
\notag
	 K_\op(z) &= \frac{2c\beta}{(1-c)z(z^2+1)} - 2 \beta \sum\limits_{x} \frac{(1-c)^{\beta-1} c^{x+1} (\beta+1)_{x-1}}{(x-1)!} 
\\ \notag &\quad \times 
	\left\{ \frac{8(x+\beta)z}{[(z^2+1)^2+4z^2x](z^2+1)} + \frac{(1-4x)(1+c^{-1})}{z(z^2+4x)(z^2+1)}\right\} \\
\label{hahn:mx.K}
	&= 2 \beta \sum\limits_{x=0}^{\infty} \frac{(1-c)^{\beta-1} c^{x+1} (\beta+1)_x}{x!} 
\\ \notag &\quad \times
		\left\{ \frac{1+\sqrt{\frac{x}{x+1}}}{z[z^2+(\sqrt{x+1}+\sqrt{x})^2]}
		+ \frac{1-\sqrt{\frac{x}{x+1}}}{z[z^2+(\sqrt{x+1}-\sqrt{x})^2]}
		- \frac{1+c}{z(z^2+4x+4)}\right\}~,
\end{align}
where we have carried out similar steps as in the previous subsections. 

Finally, after performing the inverse Laplace transform, we obtain the complexity
\begin{align}
\label{hahn:mx.Kt1}
	 K_\op(\tau) &= \sum\limits_{x=0}^{\infty} \frac{(1-c)^{\beta-1} c^{x+1} (\beta)_{x+1}}{(x+1)!} \\
\notag &\quad \times
	 \left[ 
	 \frac{4\sin^2\left(\frac{\sqrt{x+1}-\sqrt{x}}2 \tau\right)}{1-\sqrt{\frac{x}{x+1}}}
	 +\frac{4\sin^2\left(\frac{\sqrt{x+1}+\sqrt{x}}2 \tau\right)}{1+\sqrt{\frac{x}{x+1}}}   
	 - (1+c)\sin^2 \left(\sqrt{x+1}\tau\right) \right]~. 
\end{align}

\begin{figure}[t]
	\includegraphics[width=0.49\textwidth]{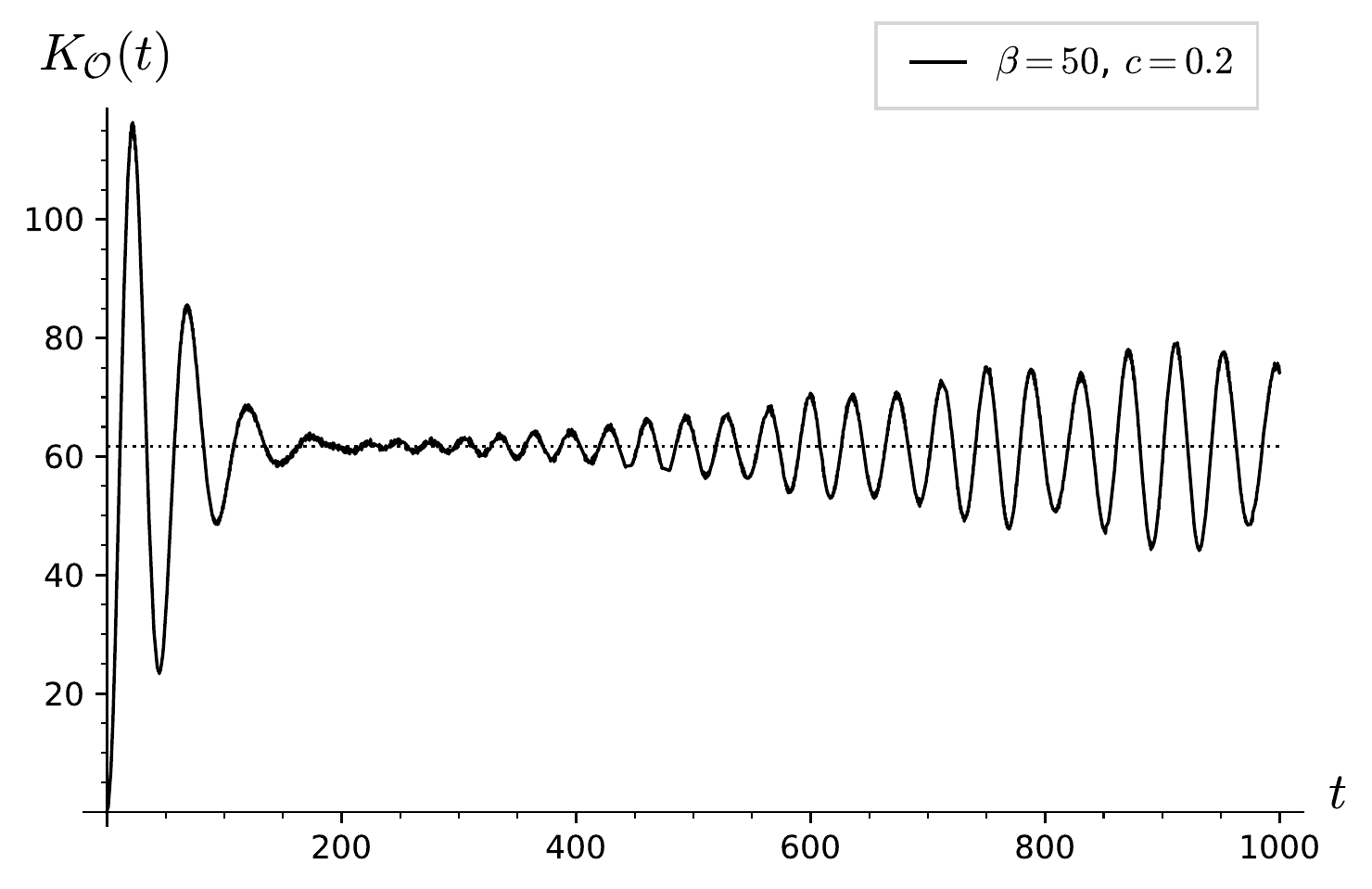}
	\hfill
	\includegraphics[width=0.49\textwidth]{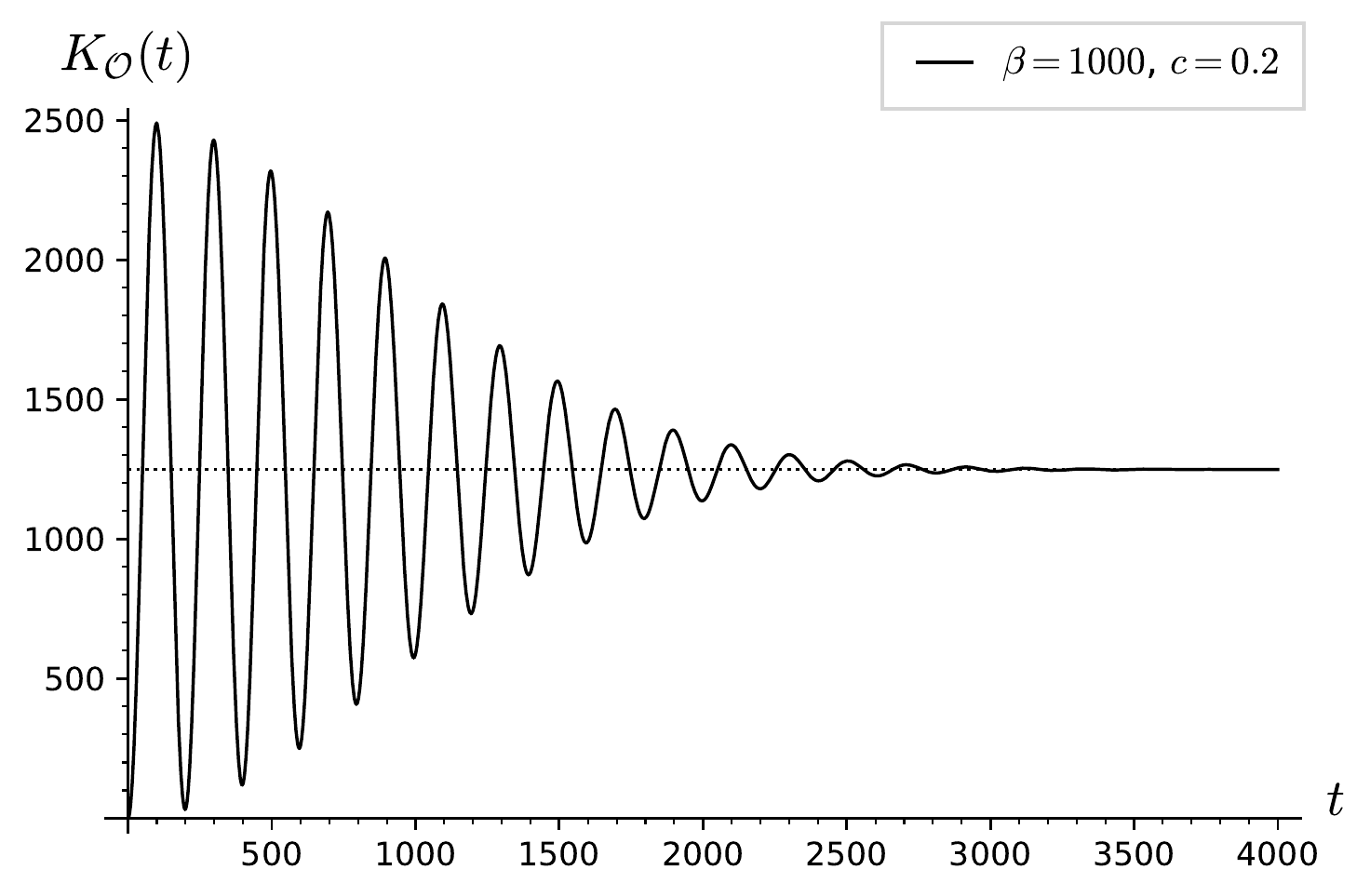}\\
	\includegraphics[width=0.49\textwidth]{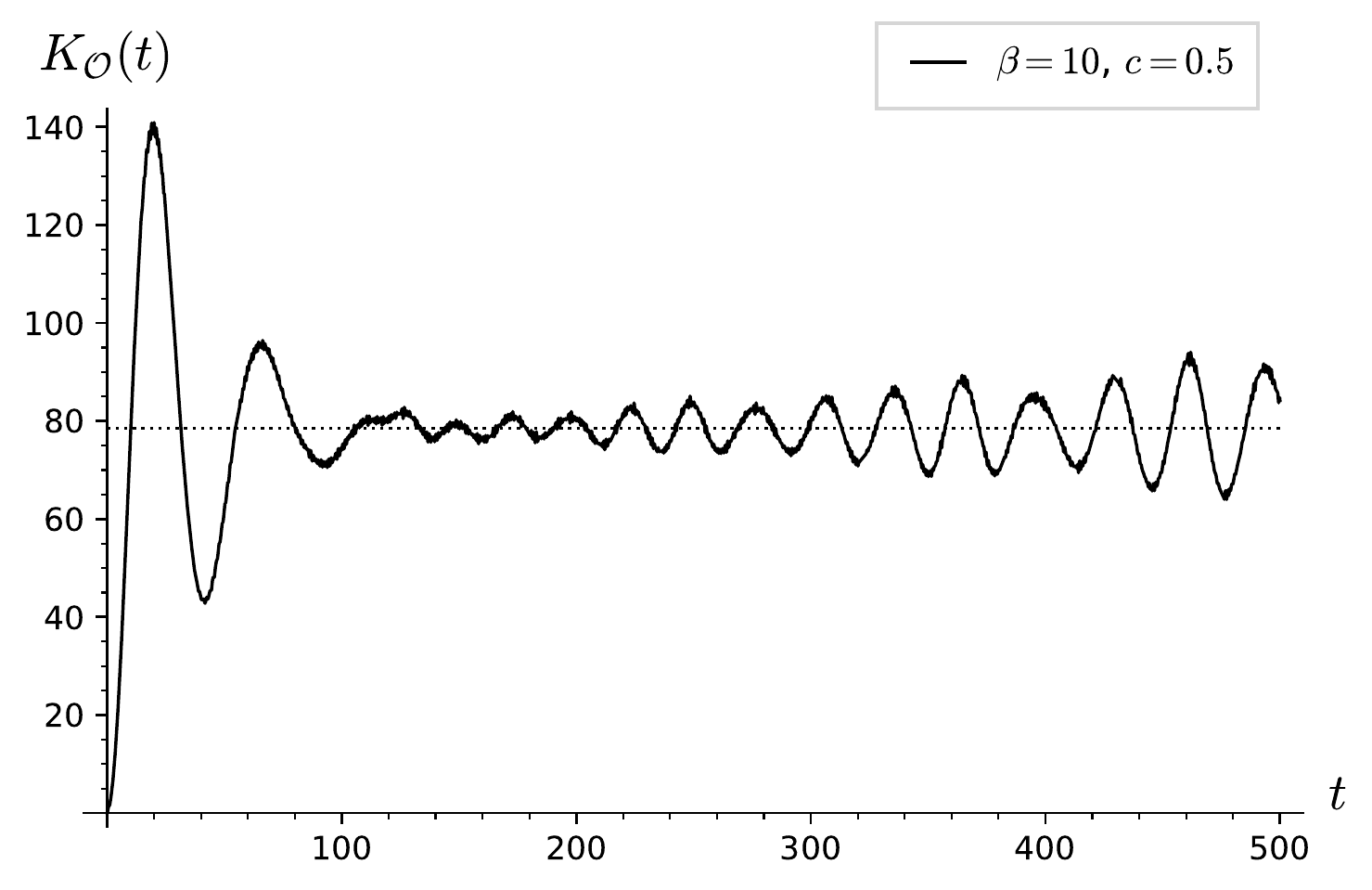}
	\hfill
	\includegraphics[width=0.49\textwidth]{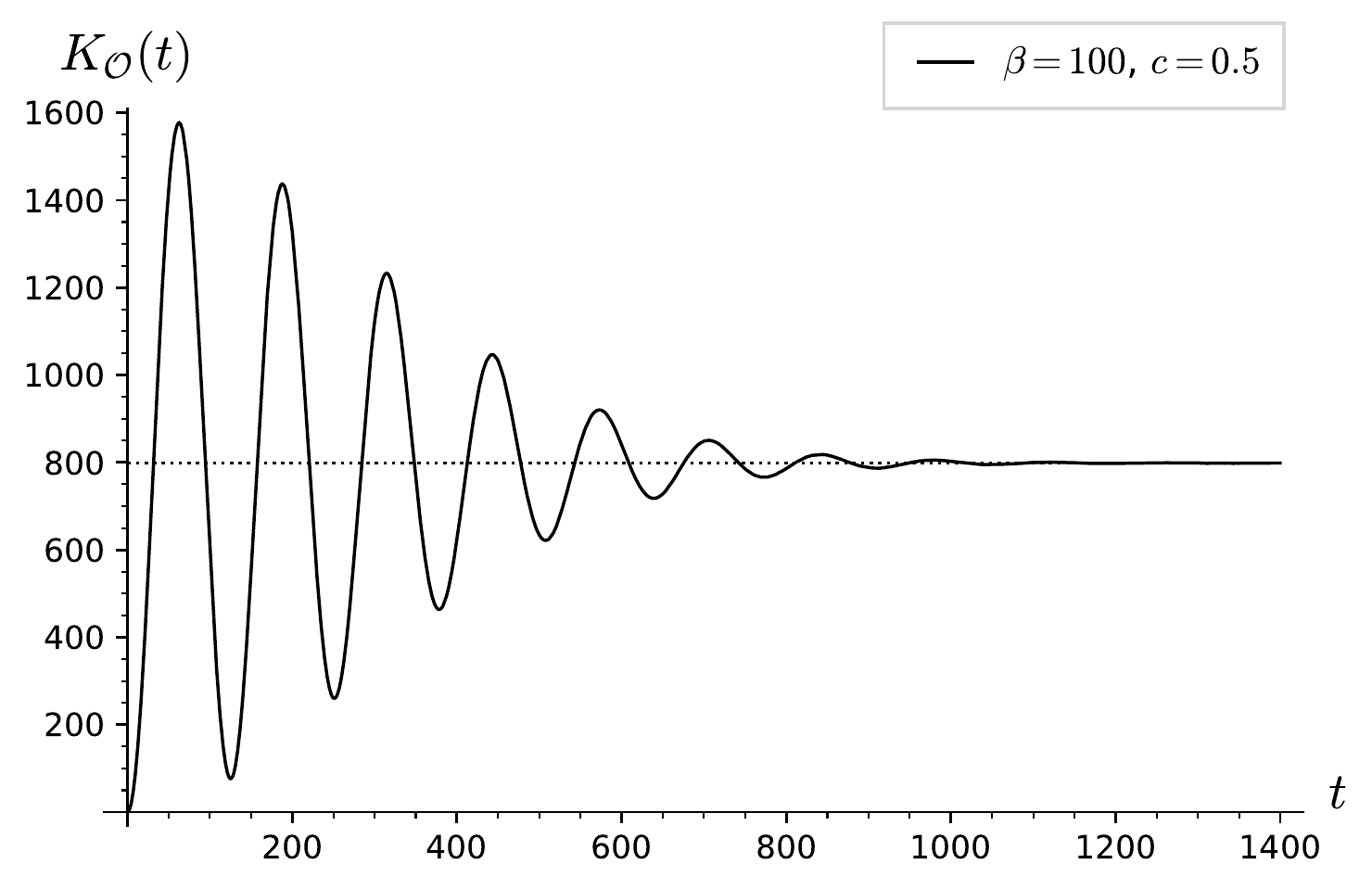}\\
	\includegraphics[width=0.49\textwidth]{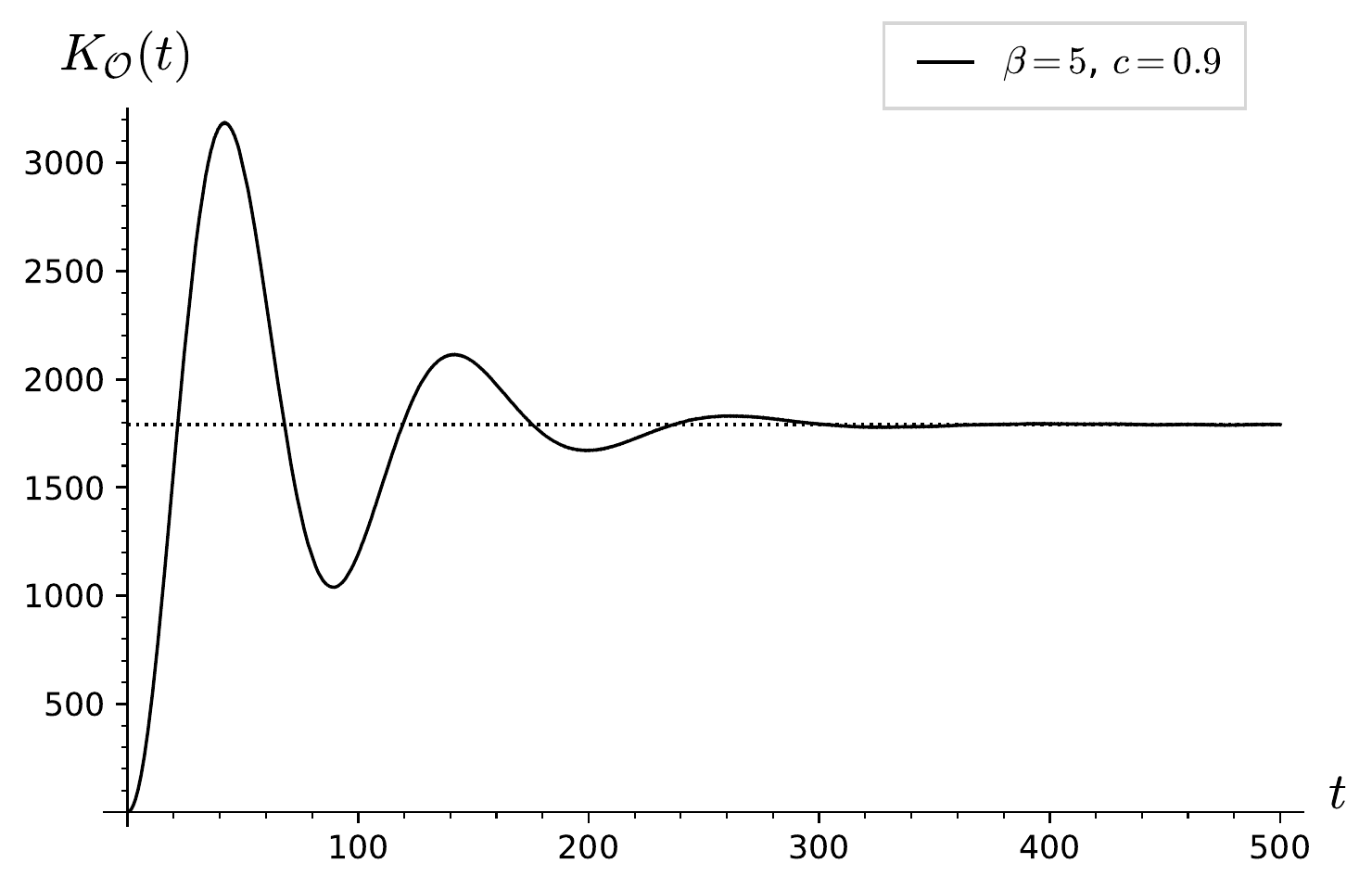}
	\hfill
	\includegraphics[width=0.49\textwidth]{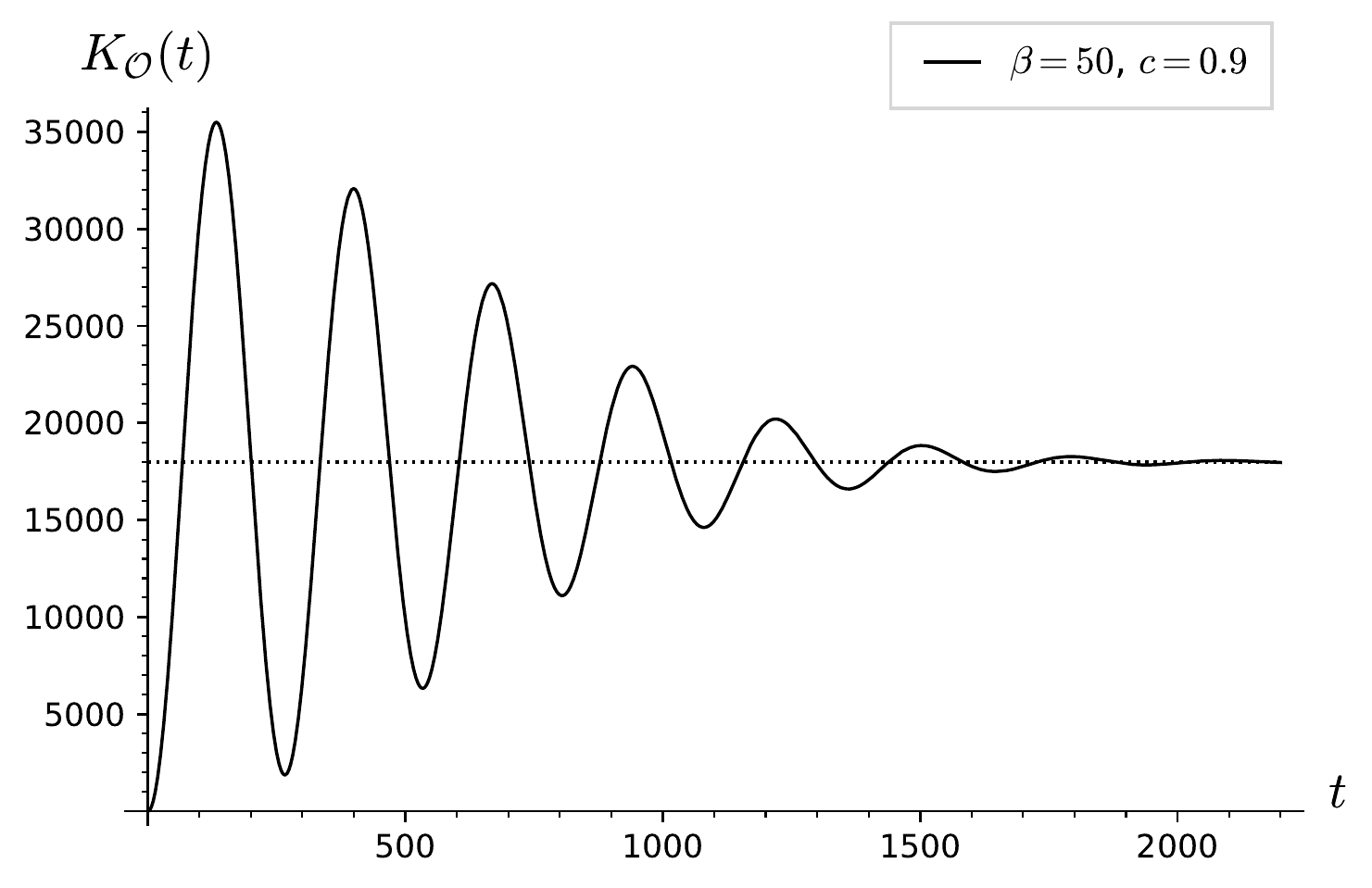}
\caption{Complexity \eqref{hahn:mx.Kt1} as a function of time for various combinations of $\beta$ and $c$. The dashed line represents the value $K_\op^\ast$ \eqref{hahn:mx.Kmean}. Note the different time scales. \label{hahn:mx.K.fig}}  
\end{figure}

The result \eqref{hahn:mx.Kt1} is quite similar to the results of the Charlier and Krawtchouk cases. At late times, the complexity oscillates (more or less erratically) around the mean value
\begin{equation}
\label{hahn:mx.Kmean}
	K_\op^\ast = \frac{4\beta c}{(1-c)^2} -\frac{1+c}{2(1-c)} \left[ 1-(1-c)^\beta\right]~,
\end{equation}
which is found by setting the sine squares in \eqref{hahn:mx.Kt1} to $\frac12$. The complexity is plotted in figure~\ref{hahn:mx.K.fig} for various values of $\beta$ and $c$.

\subsection{Other polynomials of the Hahn class}
\label{hahn:other}

The remaining polynomials of the Hahn class \cite{NIST:DLMF} are the Hahn polynomials, the Meixner-Pollaczek polynomials, and the continuous Hahn polynomials. We will not provide the full analysis of these cases, but limit ourselves to a few comments.

The Hahn polynomials are orthogonal with respect to a discrete measure on $x\in\{0,1,\ldots,N\}$ and have the explicit form (monic polynomials)
\begin{equation}
\label{hahn:hahn.expl}
	Q_n^{(\alpha,\beta,N)}(x) =\frac{(\alpha+1)_n(-N)_n}{(n+\alpha+\beta+1)_n} \genhypF{3}{2}{-n,n+\alpha+\beta+1,-x;\alpha+1,-N;1}~,
\end{equation}
with $\alpha,\beta>-1$. The following coefficents $\Delta_n$ give rise to Hahn polynomials:
\begin{subequations}
\begin{align}
\label{hahn:hahn.delta1}
	\Delta_{2n} &= \frac{n(n+\beta)(n+\alpha+\beta+N+1)}{(2n+\alpha+\beta+1)(2n+\alpha+\beta+2)}~,\\
\label{hahn:hahn.delta2}
	\Delta_{2n+1} &= \frac{(N-n)(n+\alpha+1)(n+\alpha+\beta+1)}{(2n+\alpha+\beta+1)(2n+\alpha+\beta+2)}~.
\end{align}
\end{subequations}
This is a generalization of the Krawtchouk and Meixner cases. The former is reproduced in the limit $\alpha=pt$, $\beta=(1-p)t$, $t\to \infty$, the latter by $\alpha=b-1$, $\beta=N(c^{-1}-1)$, $N\to \infty$.

The Meixner-Pollaczek polynomials are orthogonal with respect to a weight function with support on $(-\infty,\infty)$. They have two parameters, which places them at the same level as the Jacobi, Krawtchouk and Meixner polynomials in the Askey scheme \cite{NIST:DLMF}. In our context, one parameter is fixed by demanding that the spectrum is symmetric $(\phi=\frac{\pi}2)$. The resulting polynomials give rise to the $SU(1,1)$ case of \cite{Caputa:2021sib}, as discussed in subsection~\ref{gen:simple}.

A more general case with continuous support, $x \in(-\infty,\infty)$, is given by the continuous Hahn polynomials. 
These polynomials occur in relation with exactly solvable Quantum mechanics problems \cite{Bender:1987, Odake:2019bxl, Atakishiyev:1998md}. In general, continuous Hahn polynomials are parameterized by two complex coefficients $a$ and $b$, or, equivalently, by four real parameters. However, only three of these real parameters are free, because one parameter can be absorbed by a shift of the independent variable $x$. Moreover, demanding the weight function to be symmetric reduces the number of free parameters to two. 
There remain two possibilities, in both of which $P_n(x)\sim p_n(x;a,b,\bar{a},\bar{b})$ is a continuous Hahn polynomial of degree $n$.  First, one can choose $a$ and $b$ to be real and positive, in which case one has the coefficients
\begin{equation}
\label{hahn:conh.delta1}
	\Delta_n = \frac{n(2a+n-1)(2b+n-1)(2a+2b+n-2)}{4(2a+2b+2n-1)(2a+2b+2n-3)}~,\qquad (a,b>0)~.
\end{equation}
The special case $b=a+\frac12$ reduces to the Meixner-Pollaczek case. 
Second, for $b=\bar{a}$, $a+\bar{a}>0$, one has
\begin{equation}
\label{hahn:conh.delta2}
	\Delta_n = \frac{n|2a+n-1|^2(2a+2\bar{a}+n-2)}{4(2a+2\bar{a}+2n-1)(2a+2\bar{a}+2n-3)}~,\qquad (a+\bar{a}>0)~.
\end{equation}
In both cases, \eqref{hahn:conh.delta1} and \eqref{hahn:conh.delta2}, $\Delta_n$ grows as $n^2$ for large $n$. This implies maximal exponential growth of the complexity at late times, just as in the Meixner-Pollaczek case.

\section{Tricomi-Carlitz polynomials}
\label{Carlitz}

As an exotic example of non-classical orthogonal polynomials, we consider the Tricomi-Carlitz polynomials \cite{Tricomi:1951,Carlitz:1958,Chihara:1978}. They are also known as Karlin-McGregor polynomials, as they have been studied by Karlin and McGregor in the connection with random walks \cite{Karlin:1958}.  
In our context, the Tricomi-Carlitz polynomials are interesting, because their sequence of $\Delta_n$ converges to zero for $n\to \infty$. By the argument given in subsection~\ref{gen:spectra}, the spectrum must consist of countably many points, \ie it must be discrete.
As we shall see below, the spectrum is bounded, but countably infinite.

Tricomi \cite{Tricomi:1951} studied the polynomials 
\begin{equation}
\label{more:tri.pol.def}
	t^{(\alpha)}_n(z) = (-1)^n L_n^{(z-\alpha-n)}(z) = \sum\limits_{k=0}^n (-1)^k \binom{z-\alpha}{k} \frac{z^{n-k}}{(n-k)!}~,  
\end{equation}
where $L_n^{(\alpha)}$ denotes a Laguerre polynomial. Carlitz \cite{Carlitz:1958} noted that, if one sets 
\begin{equation}
\label{more:car.pol.def}
	f^{(\alpha)}_n(x) = x^n t_n^{(\alpha)}(x^{-2})
\end{equation}
with $\alpha>0$, then the polynomials $f^{(\alpha)}_n(x)$ constitute a set of orthogonal polynomials with positive definite norm. The monic polynomials \cite{Chihara:1978}
\begin{equation}
\label{more:car.pol.monic}
	F^{(\alpha)}_n(x) = \frac{n!}{(\alpha)_n} f^{(\alpha)}_n(x)
\end{equation}
satisfy the three-term recurrence relation
\begin{equation}
\label{more:car.three-term}
	x F^{(\alpha)}_n(x) = F^{(\alpha)}_{n+1}(x) + \frac{n}{(n+\alpha)(n+\alpha-1)} F^{(\alpha)}_{n-1}(x)~.
\end{equation}
Because the sequence of 
\begin{equation}
\label{more:car.delta.n}
	\Delta_n = \frac{n}{(n+\alpha)(n+\alpha-1)} \qquad (n\geq 1)
\end{equation}
approaches zero for large $n$, by the argument given in subsection~\ref{gen:spectra}, the spectrum must be discrete, as we mentioned above. Indeed, the orthogonality relation is 
\begin{equation}
\label{more:car.ortho}
	\int\limits_{-\infty}^{\infty}  F^{(\alpha)}_m(x)F^{(\alpha)}_n(x) \rmd \Psi^{(\alpha)}(x) 
	= \frac{n!}{(\alpha)_n(\alpha+1)_n} \delta_{mn}~,
\end{equation}
where $\Psi^{(\alpha)}$ is a step function with jumps 
\begin{equation}
\label{more:car.Psi.def}
	\rmd \Psi^{(\alpha)} = \frac{\alpha (k+\alpha)^{k-1}\e{-(k+\alpha)}}{2 k!} \qquad \text{at } x = \pm \frac1{\sqrt{k+\alpha}}~. 
\end{equation}
The generating function is 
\begin{equation}
\label{more:car.gen.func}
	F^{(\alpha)}(x;w) = \sum\limits_{n=0}^\infty F^{(\alpha)}_n(x) \frac{(\alpha)_n w^n}{n!} = \e{w/x} (1-xw)^{x^{-2}-\alpha}~.
\end{equation}

Furthermore, the Tricomi-Carlitz polynomials are related to the Hermite polynomials through the limit \cite{Lopez:1999}
\begin{equation}
\label{more:tri.to.hermit}
	\lim\limits_{\alpha\to\infty} f_n^{(\alpha)}\left(\frac{\sqrt{2}x}{\alpha} \right) = \frac{2^{-n/2}}{n!} H_n(x)~.
\end{equation}

From \eqref{more:car.delta.n} we see that $\Delta_n\sim n^{-1}$ for large $n$, which corresponds to the exponent $\lambda=-1$ in the continuum approximation of subsection~\ref{cont}. This would indicate a late time complexity that grows as $t^{2/3}$, but as we pointed out in that subsection, this result should be taken with a grain of salt, because the continuum approximation is not justified at large times, if $\lambda<0$.

Let us find the functions of the second kind \eqref{La:Q.def} for the Tricomi-Carlitz polynomials. This time, our approach is to calculate $Q_0(z)$ and then use the recurrence relation \eqref{La:Q.rec} to obtain the $Q_n(z)$ for $n>0$. Therefore, consider
\begin{align}
\notag
	Q_0(z) &= \int \rmd \Psi^{(\alpha)}(x) \frac{1}{z-x} \\
\notag
	&= \frac{\alpha}2 \sum\limits_{k=0}^\infty \frac{(k+\alpha)^{k-1} \e{-(\alpha+k)}}{k!} 
	\left( \frac1{z-\frac1{\sqrt{\alpha+k}}} + \frac1{z+\frac1{\sqrt{\alpha+k}}} \right)\\
\label{more:tri.Q0.1}
	&= \frac{\alpha}z \sum\limits_{k=0}^\infty \frac{(k+\alpha)^{k} \e{-(\alpha+k)}}{k! (\alpha+k-z^{-2})}~.
\end{align}
After expanding $(\alpha+k)^k =[(\alpha+k-z^{-2}) +z^{-2}]^k$, \eqref{more:tri.Q0.1} reads
\begin{equation}
\notag
	Q_0(z) = \frac{\alpha}z \sum\limits_{k=0}^\infty \sum\limits_{l=0}^k 
	\frac{(k+\alpha-z^{-2})^{l-1} (z^{-2})^{k-l} \e{-(\alpha+k)}}{l! (k-l)!}~,
\end{equation}
which, after a rearrangement of the summations, takes the form
\begin{equation}
\label{more:tri.Q0.2}
	Q_0(z) = \frac{\alpha}z \sum\limits_{k=0}^\infty \sum\limits_{l=0}^\infty 
	\frac{(k+l+\alpha-z^{-2})^{l-1} z^{-2k} \e{-(\alpha+k+l)}}{l! k!}~.
\end{equation}
The summation over $l$ can be done with the help of the identity\footnote{The identity \eqref{more:tri.sum.ident} has been used by Carlitz \cite{Carlitz:1958} to prove the orthogonality relation \eqref{more:car.ortho}. The proof of \eqref{more:tri.sum.ident} is sufficiently easy to leave it as an exercise for the reader.}
\begin{equation}
\label{more:tri.sum.ident}
	\sum\limits_{n=0}^\infty \frac{a(a+n)^{n-1}}{n!} (z\e{-z})^n = \e{az}~.
\end{equation}
Thus, \eqref{more:tri.Q0.2} becomes
\begin{align}
\notag
	Q_0(z) &= \frac{\alpha}z \sum\limits_{k=0}^\infty  
	\frac{z^{-2k} \e{-z^{-2}}}{k!(\alpha+k-z^{-2})}\\
\notag
	&= \frac{\alpha z^{-1}}{\alpha -z^{-2}} \e{-z^{-2}} \genhypF{1}{1}{\alpha-z^{-2};1+\alpha-z^{-2};z^{-2}}\\
\label{more:tri.Q0.3}
	&= \frac{\alpha z^{-1}}{\alpha -z^{-2}} \genhypF{1}{1}{1;1+\alpha-z^{-2};-z^{-2}}~,
\end{align}
where $\genhypF{1}{1}{}=\Phi()$ is the standard confluent hypergeometric function.

It takes a little effort to obtain
\begin{align}
\notag
	Q_1(z) &= z Q_0(z)-1 \\
\label{more:tri.Q1}
	&= \frac{z^{-2}}{(\alpha -z^{-2})_2} \genhypF{1}{1}{2;2+\alpha-z^{-2};-z^{-2}}~,
\end{align}
and little more to prove that  
\begin{equation}
\label{more:tri.Qn}
	Q_n(z) = \frac{\alpha n! z^{-(n+1)}}{(\alpha)_n (\alpha -z^{-2})_{n+1}} \genhypF{1}{1}{n+1;n+1+\alpha-z^{-2};-z^{-2}}
\end{equation}
satisfy the recurrence relation \eqref{La:Q.rec}. Written as a series, \eqref{more:tri.Qn} reads

\begin{equation}
\label{more:tri.Qn.series}
	Q_n(z) = \frac{\alpha z^{-(n+1)}}{(\alpha)_n} \sum\limits_{k=0}^\infty \frac{(n+k)!(-z^{-2})^k}{k!(\alpha-z^{-2})_{n+k+1}}~.
\end{equation}

Let us attempt to calculate the complexity. We shall use the generating function \eqref{La:K.gen.func.laplace3}, which now reads
\begin{equation}
\label{more:tri.Kz.general}
	K_\op(\lambda;z) = \sum\limits_{n=0}^\infty \frac{i(-\e{\lambda})^n}{h_n} \int\rmd \Psi^{(\alpha)}(x) F_n^{(\alpha)}(x) Q_n(iz-x)
\end{equation}
and consider first the combination 
\begin{equation}
\label{more:tri.FQ.1}
	(FQ)(x,y) = \sum\limits_{n=0}^\infty \frac{(-\e{\lambda})^n}{h_n} F_n^{(\alpha)}(x) Q_n(y)
\end{equation}
with $y=iz-x$. Substituting $h_n$, $F_n^{(\alpha)}$ and $Q_n$ from \eqref{more:car.ortho}, \eqref{more:car.pol.monic}, and \eqref{more:tri.Qn.series}, respectively, gives
\begin{equation}
\label{more:tri.FQ.2}
	(FQ)(x,y) = \sum\limits_{n=0}^\infty \frac{(n+\alpha)}{y} \left(-\e{\lambda}\frac{x}{y}\right)^n \sum\limits_{k=0}^n (-1)^k \binom{x^{-2}-\alpha}{k} \frac{(x^{-2})^{n-k}}{(n-k)!} \sum\limits_{l=0}^\infty \frac{(n+l)!(-y^{-2})^l}{l!(\alpha-y^{-2})_{n+l+1}}~.
\end{equation}
After rearranging the summations, this becomes
\begin{equation}
\label{more:tri.FQ.3}
	(FQ)(x,y) = \sum\limits_{m=0}^\infty \frac{m!(-y^{-2})^m}{y(\alpha-y^{-2})_{m+1}} 
		\sum\limits_{k=0}^m \binom{x^{-2}-\alpha}{k} \frac{(-\e{\lambda}xy)^k}{(m-k)!} 
		\sum\limits_{n=0}^{m-k} \binom{m-k}{n}(n+k+\alpha) \left(\frac{\e{\lambda}y}{x}\right)^n~.
\end{equation}
The sum over $n$ can be carried out now, which yields 
\begin{align}
\label{more:tri.FQ.4}
	(FQ)(x,y) &= \sum\limits_{m=0}^\infty \frac{m!(-y^{-2})^m}{(\alpha-y^{-2})_{m+1}} 
		\sum\limits_{k=0}^m \binom{x^{-2}-\alpha}{k} \frac{(-\e{\lambda}xy)^k}{(m-k)!} 
		\left(1+\frac{\e{\lambda}y}{x} \right)^{m-k-1} \\
\notag &\quad \times
	\left[ \e{\lambda} \frac{m+\alpha}{x} +  \frac{k+\alpha}{y}\right]~.
\end{align}
To proceed, let us define the following three sums,
\begin{subequations}
\begin{align}
	S_1 &= \sum\limits_{m=0}^\infty \frac{m!(-y^{-2})^m}{(\alpha-y^{-2})_{m+1}} 
			\sum\limits_{k=0}^m \binom{x^{-2}-\alpha}{k} \frac{(-\e{\lambda}xy)^k}{(m-k)!} 
			\left(1+\frac{\e{\lambda}y}{x} \right)^{m-k-1} \frac{m+\alpha}{x}~,\\
	S_2 &= \sum\limits_{m=0}^\infty \frac{m!(-y^{-2})^m}{(\alpha-y^{-2})_{m+1}} 
				\sum\limits_{k=0}^m \binom{x^{-2}-\alpha}{k} \frac{(-\e{\lambda}xy)^k}{(m-k)!} 
				\left(1+\frac{\e{\lambda}y}{x} \right)^{m-k-1} \frac{k+\alpha}{y}~,\\
	S_3 &= \sum\limits_{m=0}^\infty \frac{(m+1)!(-y^{-2})^{m+1}}{(\alpha-y^{-2})_{m+2}} 
				\sum\limits_{k=0}^m \binom{x^{-2}-\alpha}{k} \frac{(-\e{\lambda}xy)^k}{(m-k)!} 
				\left(1+\frac{\e{\lambda}y}{x} \right)^{m-k}~.
\end{align}
\end{subequations}
It is easy to see that they satisfy
\begin{equation}
\label{tri:S.rel1}
	x S_1 - y S_2 = S_3~.
\end{equation}
It is a bit more difficult to show that 
\begin{equation}
\label{tri:S.rel2}
	S_1 = \frac{1}{x+y\e{\lambda}} -\e{\lambda} S_2
\end{equation}
holds. To prove \eqref{tri:S.rel2}, take $S_1$, substitute the identity
\begin{equation}
\label{more:tri.sum.aux}
	\frac{m!}{(\gamma)_{m+1}} = \sum\limits_{l=0}^m \binom{m}{l} \frac{(-1)^l}{\gamma+l}~,
\end{equation}
and write $m+\alpha = (\alpha+l-y^{-2}) + y^{-2} + (m-l)$. For the first term, the sum over $l$ is non-zero only for $m=0$, which gives the first term on the right hand side of \eqref{tri:S.rel2}. The other two terms yield the second term on the right hand side of \eqref{tri:S.rel2} after some manipulations.

After solving the system \eqref{tri:S.rel1}--\eqref{tri:S.rel2} one obtains
\begin{equation}
\label{more:tri.FQ.5}
	(FQ)(x,y) = \e{\lambda}S_1 + S_2 = \frac{1}{y+x\e{\lambda}} \left[1 + (\e{2\lambda}-1) S_3\right]~.  
\end{equation}

Now, let us return to the generating function \eqref{more:tri.Kz.general}, which is 
\begin{equation}
\label{more:tri.Kz.general.2}
	K_\op(\lambda;z) = i \int\rmd \Psi^{(\alpha)}(x) (FQ)(x,iz-x)~.
\end{equation}
While setting $\lambda=0$ simply gives the completeness relation, to find the complexity we are interested in the term linear in $\lambda$. In \eqref{more:tri.FQ.5}, this term is  
\begin{align}
\label{more:tri:FQ.6}
	(FQ)'(x,y) &=-\frac{x}{(x+y)^2} + \frac2{x+y} \sum\limits_{m=0}^\infty \frac{(m+1)!(-y^{-2})^{m+1}}{(\alpha-y^{-2})_{m+2}} 
					\sum\limits_{k=0}^m \binom{x^{-2}-\alpha}{k} \frac{(-xy)^k}{(m-k)!} 
					\left(1+\frac{y}{x} \right)^{m-k}~.
\end{align}

The double sum in \eqref{more:tri:FQ.6} is known as a (Humbert) confluent hypergeometric series of two variables \cite{Humbert:1922}, \cite[9.261]{Gradshteyn},
so
\begin{align}
\label{more:tri:FQ.7}
	(FQ)'(x,y) &=-\frac{x}{(x+y)^2} - \frac2{y^2(x+y)(\alpha-y^{-2})_2} \Phi_1\left( 2,\alpha-x^{-2},\alpha-y^{-2}+2,-\frac{x}{y}, -\frac{x+y}{xy^2}\right)~,
\end{align}
which has the following integral representation \cite[3.385]{Gradshteyn} or \cite{Choi:2011},
\begin{align}
\label{more:tri:FQ.8}
	(FQ)'(x,y) &=-\frac{x}{(x+y)^2} - \frac2{y^2(x+y)} \int\limits_0^1 \rmd t\, t (1-t)^{\alpha-y^{-2}-1} \left(1+\frac{xt}{y} \right)^{x^{-2}-\alpha} \e{-\frac{x+y}{xy^2} t}~.
\end{align}

We were unable to make further progress analytically. One should carry out the summation over the spectrum in \eqref{more:tri.Kz.general.2}, setting $y=iz-x$. Obviously, odd terms in $x$ cancel in this sum, in particular, the terms that would behave as $1/z^2$. One such term is the first term on the right hand side of \eqref{more:tri:FQ.8}, but also the Humbert series hides such a term. It comes from $(\alpha-y^{-2})_{m+2}$ in the denominator in \eqref{more:tri:FQ.6}, which has a $1/z$ contribution that is odd in $x$ for $x=\pm x_k=(\alpha+k)^{-1/2}$. From the absence of $1/z^2$ terms we can conclude that the late time complexity grows less that linearly in time, but in order to conclude that it approaches a constant we should know whether the resulting expression is analytic in $z$. We suspect that it is not. The issue is complicated by the fact that it is not possible to consider the small $z$ limit in \eqref{more:tri:FQ.7} before summing over the spectrum, because the spectrum contains (countably) infinitely many values of $|x|$ that are smaller than any given, even if very small, $|z|$. We mention that the continuum approximation, which cannot be considered as valid in this case, would indicate a non-analytic behaviour $\sim z^{-5/3}$. There is no evidence of such a behaviour in our results.

\section{Operator inner products}
\label{iprod}

In this final section, we shall return to an important detail, which was left out at the beginning of subsection~\ref{La:rec.meth}. Given an operator $\op$ in a system with Hamiltonian $H$, its complexity $K_\op(t)$ will crucially depend on the choice of the inner product adopted before applying the Lanczos method. Other than that, there is no additional freedom or parameter to choose, except possibly for approximations such as a terminator function \cite{rec-method}. The choice of the inner product has a direct bearing on the results of the entire procedure, \ie the Lanczos coefficients, the measure $\mu(\Liou)$, the system of orthogonal polynomials, and the complexity. It is known that different choices may even lead to essentially different qualitative behaviours of the complexity including its late-time behaviour. We shall limit ourselves to a brief review of physical inner products that have been introduced in the literature.

\subsection{Frobenius inner product}
If the physical Hilbert space is finite, $\mathrm{dim}(\Hilb)=D$, then the easiest choice for the inner product between two operators $A$ and $B$ is the Frobenius (or Hilbert-Schmidt) inner product
\begin{equation}
\label{iprod:i1}
	(A|B) = \frac1D \Tr[A^\dagger B]~.
\end{equation}
The dimension of $\hat{\Hilb}$, which is the space of linear operators acting on $\Hilb$, is $D^2$, so the Lanczos algorithm necessarily terminates. The maximal dimension of the Krylov subspace $\mathcal{K} = \mathrm{span}\{\op_n\}$ is somewhat smaller than $D^2$, namely $\mathrm{dim}(\mathcal{K}) = N \leq D^2 -D+1$ \cite{Rabinovici:2020ryf}. 

For infinite-dimensional Hilbert spaces, besides writing $D=\Tr[1]$, one would need to define the trace with a suitable regularization. For example, on may consider \eqref{iprod:i1} as the infinite temperature limit of the thermal inner product, which we will discuss in the next subsection.

In some references, the ``connected'' inner product is used instead of \eqref{iprod:i1}, \ie
\begin{equation}
\label{iprod:i2}
	(A|B) = \frac1D \Tr[A^\dagger B] - \frac1{D^2} \Tr[A^\dagger] \Tr[B]~.
\end{equation}
The difference between the inner products \eqref{iprod:i1} and \eqref{iprod:i2} is irrelevant for the Lanczos procedure, because $\Tr[\Liou A]=\Tr[HA-AH]=0$.

\subsection{Thermal two-point functions}

In quantum thermodynamics, for a system with Hamiltonian $H$ and inverse temperature $\beta$, the thermal expectation value of an observable $A$ is defined by
\begin{equation}
\label{th:ev}
	\vev{A}_\beta = \frac1Z \Tr \left(\e{-\beta H} A\right)~,
\end{equation}
where $Z=\Tr \left(\e{-\beta H}\right)$ is the partition function.

The thermal (canonical) inner product between two observables $A$ and $B$ may be defined by  
\begin{equation}
\label{th:corr}
	\vev{A|B}_\beta = \frac1{\beta} \int\limits_0^\beta \rmd \lambda\, g(\lambda) \vev{\e{\lambda H} A^\dagger \e{-\lambda H} B}_\beta~,
\end{equation} 
where the weight function $g(\lambda)$ must satisfy 
\begin{equation}
\label{th:g.prop}
	g(\lambda)\geq 0~,\qquad g(\beta-\lambda) = g(\lambda)~,\quad \text{and} \quad  
	\int\limits_0^\beta \rmd \lambda\, g(\lambda) = \beta~. 
\end{equation}
Physically relevant choices of $g(\lambda)$ are $g(\lambda)=\frac\beta2 \left[\delta(\lambda)+\delta(\beta-\lambda)\right]$, $g(\lambda)=\beta \delta(\lambda-\frac\beta2)$ and $g(\lambda)=1$, corresponding to the ``standard'', the Wightman and the Kubo inner products, respectively \cite{rec-method, Parker:2018yvk}. In \cite{rec-method} the connected version of \eqref{th:corr} is used, in which $\vev{A^\dagger}_\beta \vev{B}_\beta$ is subtracted from the right hand side. However, our remark below \eqref{iprod:i2} applies here as well. 

\subsection{Microcanonical inner product}

In systems with an unbounded spectrum, the Krylov complexity calculated starting from the canonical inner product \eqref{th:corr} often obscures the telltale signatures of quantum chaos. In order to improve on this situation, a microcanonical inner product has been introduced in \cite{Kar:2021nbm}. The main argument of \cite{Kar:2021nbm} is that the mean energy between two energy eigenstates, $E=\frac12 (E_i+E_j)$, is a conserved quantity under the Liouvillian, and operator growth should be considered separately in each conserved sector. In what follows, we shall review the construction of the microcanonical inner product of \cite{Kar:2021nbm} with some generalizations. For simplicity, we will consider a system with discrete energy spectrum in the thermodynamic limit and take the ground state energy to be zero. 

If we define the density of states 
\begin{equation}
\label{th:nE}
	\nu(E) = \sum\limits_i \delta(E-E_i)~,
\end{equation}
where the sum is over all energy eigenstates, then the thermal expectation value \eqref{th:ev} takes the classical form
\begin{equation}
\label{th:evA}
	\vev{A}_\beta = \frac1Z \int\limits_0^\infty \rmd E\, \nu(E) \e{-\beta E} \vev{A}_E~, 
\end{equation}
where $\vev{A}_E$ denotes the expectation value of $A$ in the microcanonical ensemble of states with energy $E$, which we call $\mathcal{M}(E)$.\footnote{In the sense of the thermodynamic limit, the definition of the microcanonical ensemble includes an energy window $\epsilon\ll 1$, \ie $\mathcal{M}(E) = \{\ket{i}: |E_i-E|<\epsilon\}$, and one should accordingly smear the delta functions in \eqref{th:nE}.} Specifically, 
\begin{equation}
\label{th:A.E}
	\vev{A}_E = \frac1{N(E)} \sum\limits_{i \in \mathcal{M}(E)} \vev{i|A|i}~, \qquad N(E) = \sum\limits_{i \in \mathcal{M}(E)} 1 ~.
\end{equation}
Notice that our normalization is such that $\vev{1}_\beta=\vev{1}_E=1$.

The idea of \cite{Kar:2021nbm} is to rewrite the thermal correlator \eqref{th:corr} as a thermal expectation value, \ie in the form 
\begin{equation}
\label{th:corr.AB}
	\vev{A|B}_\beta = \frac1Z \int\limits_0^\infty \rmd E\, \nu(E) \e{-\beta E} \vev{A|B}_E~,
\end{equation}
where $\vev{A|B}_E$ is the microcanonical inner product we are looking for. To do this, one starts from
\begin{equation}
\label{th:AB.calc1}
	\vev{\e{\lambda H} A^\dagger \e{-\lambda H} B}_\beta = \frac1{Z} \sum_{i,j} \e{(\lambda-\beta)E_j -\lambda E_i} 
	\vev{i|A|j}^\ast \vev{i|B|j} 
\end{equation}
and introduces the pair density\footnote{Our definition of $\rho(E,\omega)$ differs from the definition of \cite{Kar:2021nbm} by the factor $\nu(E)$. This implies a different normalization also for derived quantities such as $\vev{A|B}_E$, cf.\ \eqref{th:AB.E}.}
\begin{equation}
\label{th:rho.def}
	\rho(E,\omega) = \frac1{\nu(E)} \sum_{i,j} \delta\left(E-\frac{E_i+E_j}2\right) \delta (\omega - E_i +E_j)~.
\end{equation}
Furthermore, let
\begin{equation}
\label{th:gamma.def}
	\gamma_\beta(\omega) = \frac1\beta \int\limits_0^\beta \rmd \lambda\, g(\lambda) \e{\omega\left(\frac\beta2-\lambda\right)}~.
\end{equation}
Then, \eqref{th:corr} takes the form \eqref{th:corr.AB} with the microcanonical inner product
\begin{equation}
\label{th:AB.E}
	\vev{A|B}_E = \int\limits_{-2E}^{2E} \rmd \omega\, \rho(E,\omega) \gamma_\beta(\omega) \vev{A|B}_{E,\omega}~,
\end{equation}
where
\begin{equation}
\label{th:AB.E.omega}
	\vev{A|B}_{E,\omega} = 
	\frac1{N(E+\frac{\omega}2) N(E-\frac{\omega}2)}  \sum\limits_{i \in \mathcal{M}(E+\frac{\omega}2)}
	\sum\limits_{j \in \mathcal{M}(E-\frac{\omega}2)} \vev{i|A|j}^\ast \vev{i|B|j}~. 
\end{equation}
For the Wightman inner product, for which $\gamma_\beta(\omega)=1$, \eqref{th:AB.E} reproduces the simple expression obtained in \cite{Kar:2021nbm}. In general, both $\rho(E,\omega)$ and $\gamma_\beta(\omega)$ are even functions of $\omega$.

In subsection~\ref{La:rec.meth} we assumed the start operator $|\op_0)$ of the Lanczos algorithm to be normalized. In the canonical or the infinite-temperature settings, the initial norm is irrelevant, because one can normalize the operator once and for all. In the microcanonical approach, the norm is energy-dependent and one should keep track of it. Hence, we write explicitly 
\begin{equation}
\label{th:op.norm}
	|\op_0)_E = \vev{\op|\op}_E^{-\frac12} |\op)_E~.
\end{equation}
The definition of the (normalized) measure \eqref{La:measure.def} now reads 
\begin{equation}
\label{th:measure.def}
	\int \rmd\mu_E(\Liou)\, f(\Liou) = \frac{\vev{\op|f(\Liou)\op}_E}{\vev{\op|\op}_E}~,
\end{equation} 
which, together with \eqref{th:AB.E}, leads to
\begin{equation}
\label{th:measure}
	\frac{\rmd\mu_E(\omega)}{\rmd \omega} = \rho(E,\omega) \gamma_\beta(\omega) \frac{\vev{\op|\op}_{E,\omega}}{\vev{\op|\op}_E}  ~, \qquad \omega\in [-2E,2E]~.
\end{equation} 
Clearly, the microcanonical inner product gives rise to a measure with bounded support. 

The wave functions \eqref{La:phi.n} become
\begin{equation}
\label{th:wave.functions}
	\phi_{E,n}(t) = \frac{i^n}{\sqrt{\vev{\op|\op}_E h_n}} \vev{\op(t)|\op_n}_E~.
\end{equation}
In particular, with \eqref{th:op.norm}, 
\begin{equation}
\label{th:wave.function.0}
	\phi_{E,0}(t) = \frac{\vev{\op(t)|\op}_E}{\vev{\op|\op}_E}~.
\end{equation}

Thermal Krylov complexity can then be defined as a weighted average of the microcanonical complexities $K_{\op,E}(t)$, which are defined as usual by \eqref{La:K.complexity} with the wave functions $\phi_{E,n}(t)$. We point out that the weight is not unique. Motivated by \eqref{th:wave.function.0}, the authors of \cite{Kar:2021nbm} chose to include the energy dependent operator norm into the weight,  
\begin{equation}
\label{th:th.complexity}
	K_{\op,\beta}(t) = \frac{\int_0^\infty \rmd E\, \nu(E) \e{-\beta E}  \vev{\op|\op}_E K_{\op,E}(t)}{\int_0^\infty \rmd E\, \nu(E) 
	\e{-\beta E}  \vev{\op|\op}_E}~.
\end{equation}
One may also choose to define thermal Krylov complexity as an ordinary thermal expectation value,
\begin{equation}
\label{th:th.complexity.standard}
	K_{\op,\beta}(t) = \frac1Z \int\limits_0^\infty \rmd E\, \nu(E) \e{-\beta E} K_{\op,E}(t)~.
\end{equation}
We remark that this is a choice of the unit length on the (energy dependent) chains of wave functions rather than a consequence of operator normalization.

Because the measure \eqref{th:measure} has bounded support with $|\omega|\leq 2E$, in the thermodynamic limit the Krylov complexities $K_{\op,E}(t)$ grow at most as $K_{\op,E}(t)\lesssim  Et$ at late times. Therefore, \eqref{th:th.complexity.standard} gives immediately 
\begin{equation}
\label{th:th.complexity.late.time}
	K_{\op,\beta}(t) \lesssim \vev{E}_\beta t~,
\end{equation} 
which resembles Lloyd's bound on complexity growth \cite{Lloyd:2000}.

\section*{Acknowledgements}
Most of the work described in this paper was carried out while Y.Y.\ was \emph{Visiting Researcher} at the University of Naples ``Federico II'', for which he is grateful. Y.Y.\ would also like to thank the Physics Department ``Ettore Pancini'' and the INFN, sezione di Napoli, for kind hospitality and generous financial support during his stay. 

The work of W.M.\ is partially supported by the INFN, research initiative STEFI. The work of Y.Y.\ is partially supported by the Ministry of Science and Technology (MOST 109-2918-I-009-005 and MOST 110-2112-M-A49-008), ROC.

\bibliographystyle{JHEPnotes}
\bibliography{krylov,citedlmf}

\end{document}